\documentclass[%
 reprint,
superscriptaddress,
 amsmath,amssymb,
 aps,
floatfix,
]{revtex4-2}
\usepackage{graphicx}
\usepackage{dcolumn}
\usepackage{bm}
\usepackage{bbm}
\usepackage{multirow}
\usepackage{tabularx}
\usepackage{xcolor}
\usepackage{hyperref}

\newcolumntype{C}{>{\centering\arraybackslash}X}

\begin{document}

\preprint{APS/123-QED}

\title{Nonlinear-response theory for lossy superconducting quantum circuits}

\author{V. Vadimov}
 \email{vasilii.1.vadimov@aalto.fi}
 \affiliation{
QCD Labs, QTF Centre of Excellence, Department of Applied Physics, Aalto University, P.O. Box 15100, FI-00076 Aalto, Finland}
\author{M. Xu}%
\affiliation{
Institute  for Complex Quantum Systems and IQST, Ulm University - Albert-Einstein-Allee 11, D-89069  Ulm, Germany}
\author{J. T. Stockburger}
\affiliation{
Institute  for Complex Quantum Systems and IQST, Ulm University - Albert-Einstein-Allee 11, D-89069  Ulm, Germany}
\author{J. Ankerhold}
\affiliation{
Institute  for Complex Quantum Systems and IQST, Ulm University - Albert-Einstein-Allee 11, D-89069  Ulm, Germany}
\author{M. M\"ott\"onen}
\affiliation{
QCD Labs, QTF Centre of Excellence, Department of Applied Physics, Aalto University, P.O. Box 15100, FI-00076 Aalto, Finland}
\affiliation{VTT Technical Research Centre of Finland Ltd, QTF Centre of Excellence, P.O. Box 1000, FI-02044 VTT, Finland}

\makeatletter
\newsavebox{\@brx}
\newcommand{\llangle}[1][]{\savebox{\@brx}{\(\m@th{#1\langle}\)}%
  \mathopen{\copy\@brx\kern-0.5\wd\@brx\usebox{\@brx}}}
\newcommand{\rrangle}[1][]{\savebox{\@brx}{\(\m@th{#1\rangle}\)}%
  \mathclose{\copy\@brx\kern-0.5\wd\@brx\usebox{\@brx}}}
\renewcommand{\imath}{\textrm{i}}
\makeatother

\date{\today}

\begin{abstract}
We introduce a numerically exact and  computationally feasible nonlinear-response theory developed for lossy superconducting quantum circuits based on a framework of quantum dissipation in a minimally extended state space.
Starting from the Feynman--Vernon path integral formalism for open quantum systems with the system degrees of freedom being the nonlinear elements of the circuit, we eliminate the temporally non-local influence functional of all linear elements by introducing auxiliary harmonic modes with complex-valued frequencies coupled to the 
nonlinear elements. In our work, we propose a concept of time-averaged observables, inspired by experiment, and provide an explicit formula for producing their quasiprobability distribution. We illustrate the consistency of our formalism with the well-established Markovian input-output theory by applying them the dispersive readout of a superconducting transmon qubit. For an important demonstration of our approach beyond weak coupling, we analyze the low-frequency linear response of a capacitively and resistively shunted Josephson junction and observe signatures of a much-debated quantum phase transition at a finite temperature. The developed framework enables a comprehensive fully quantum-mechanical treatment of nonlinear quantum circuits coupled to their environment, without the limitations of typical approaches to weak dissipation, high temperature, and weak drive. This versatile tool paves the way for accurate models of quantum devices and increased fundamental understanding of quanutm mechanics such as that of the quantum measurement. 
\end{abstract}

\maketitle


\section{Introduction}

Quantum systems invariably interact with their environments, which typically consist of a macroscopic number of degrees of freedom~\cite{gardiner2004quantum, breuer2002theory, weiss2012quantum}. These interactions give rise to a diverse range of phenomena, including decoherence, dissipation-induced phase transitions~\cite{leggett1987dynamics, schon1990quantum}, and emergence of classical physics in quantum systems~\cite{zurek2003decoherence} among many others~\cite{weiss2012quantum}. Understanding and accurately describing the dynamics of open quantum systems is of fundamental importance in various fields, ranging from quantum information science and quantum computing~\cite{nielsen2010quantum, averin2012macroscopic, beige2000quantum, verstraete2009quantum} to condensed matter physics~\cite{kamenev2023field} and quantum optics~\cite{gardiner2004quantum, scully1999quantum}. Recent technological advances~\cite{georgescu202025, preskill2018quantum, reiter2017dissipative, tan2017quantum, ilias2022criticality, ronzani2018tunable, hernandez2022autonomous, bouton2021quantum} have opened up new possibilities for experimental studies of open quantum systems, necessitating the development of accurate and computationally efficient theoretical models.

Traditional Markovian approaches~\cite{gardiner2004quantum, redfield1965theory, lindblad1976generators, gorini1976completely} for treating open quantum systems rely on assumptions of weak coupling and time-scale separation of the system and the bath, or substantially high temperature, which may not always hold in experimentally relevant setups leading to non-Markovian dynamics~\cite{liu2011experimental, smirne2011experimental, cialdi2014two, gessner2014local, tang2012measuring, white2020demonstration, andersson2019non}. Furthermore, traditional approaches may have limited applicability in the presence of structured reservoirs with gaps or other singularities in the spectral density, and they become overly complicated in the context of strong driving~\cite{dann18,wu22}. Non-perturbative treatments often involve discretization of the bath spectra~\cite{bulla2005numerical,bulla2008numerical,DeFilippis2020} or employ the Feynman--Vernon path integral formalism~\cite{feynman2000theory}. The latter includes techniques such as path integral Monte Carlo (PIMC)~\cite{suzuki1993quantum, egger1994low, egger2000path, muhlbacher2005nonequilibrium}, unraveling the time-nonlocal influence functional using auxiliary stochastic fields~\cite{stockburger1999stochastic, stockburger2002exact}, introducing a set of auxiliary density operators governed by a system of time-local hierarchical equations of motion (HEOM)~\cite{tanimura1989time, tanimura2020numerically}, and others~\cite{beck2000multiconfiguration, wang2003multilayer, makri1995tensor, makri1998quantum, gull2011continuous}. Recently, a broad variety of numerically exact methods has been unified within the framework of quantum dissipation in a minimally extended state space (QD-MESS)~\cite{QDMESS}, also developed independently under names of bexcitonic HEOM~\cite{Chen2024a} and dissipation-embedded quantum master equation in second quantization (DQME-SQ)~\cite{Li2024}, which provides high accuracy with relatively moderate computational resources.

In addition to the inevitable interaction with the natural environment, the study of open quantum systems is also motivated by the need to carry out measurements on real quantum systems~\cite{jacobs2014quantum}. In such cases, the measurement apparatus itself acts as an environment, leading to controllable decoherence.
In this setup, standard von~Neumann measurements~\cite{landau2013quantum} can be applied only on the macroscopic measurement device or its part~\cite{jacobs2014quantum}, hence a single measurement may yield very limited information about the system of interest. This has two consequences: first, no real measurement is instantaneous since the result usually comes from integration of continuous measurements, and second, even continuous measurements implemented with a specific device cannot provide access to arbitrary observables of the system of interest.

In many experimental setups it is common to probe the electromagnetic field~\cite{blais2021circuit} emitted from the measured system. To this end, the input-output (IO) theory~\cite{collett1984squeezing, gardiner1985input, yurke1984quantum} for quantum optical systems and superconducting quantum circuits was developed to calculate the properties of the emitted far-field radiation and its integral characteristics. A major advantage of the IO theory is the possibility to generalize it for quantum networks~\cite{Combes2017}, where the output from one part of the system may serve as the input for another part. These theories are generally accurate in quantum optical systems but rely on the Born--Markov approximation, which may break down, e.g., in the strong-coupling regime. In fact, non-Markovian effects may be more pronounced in superconducting microwave circuits~\cite{white2020demonstration, andersson2019non} which are among the leading platforms for quantum-computing applications~\cite{blais2021circuit, kjaergaard2020superconducting, arute2019quantum}. Effects of retardation and correlations between the system and its environment are critical for the design and optimization of superconducting qubits, quantum gates, initialization, and readout protocols, especially in circuits with tunable dissipative elements~\cite{tan2017quantum, vadimov2022single, silveri2019broadband, esteve2018quantum, aiello2022quantum} and distributed-element qubits~\cite{hyyppa2022unimon}. Thus, a rigorous and computationally efficient IO theory that goes beyond the limitations of Born--Markov approximations is essential for accurately characterizing and harnessing the dynamics of superconducting circuits.

The first non-Markovian IO theory was presented in Ref.~\cite{diosi2012non} based on temporally non-local Heisenberg equations of motion for the degrees of freedom of the system and the field operators in the case of a one-dimensional chiral field propagating in one direction. However, solving these equations may be technically involved for nonlinear systems. Later, this approach was generalized for quantum networks~\cite{zhang2013non} and squeezed input fields~\cite{gross2022master}. 
In other studies~\cite{link2022non, Propp2022}, the IO formalism was developed for atoms in a cavity which is coupled to a Markovian bath.
Furthermore, due to the approximation of the chiral field, it is infeasible to properly account for thermal effects. This approximation introduces negative-frequency modes, which do not have a thermal state at a positive temperature. It is a reasonable approximation in quantum optics, since the characteristic frequency scale is much higher than the temperature scale, especially if the back-scattering can be neglected. This is not the case for the superconducting quantum circuits, where thermal effects are important and reflections from the qubits and resonators are usually strong.

The main goal of this paper is to develop a general formalism for dissipative superconducting quantum circuits which goes beyond the Born--Markov approximation, utilizing modern techniques for open quantum systems. We refer to this formalism as nonlinear-response (NLR) theory. In superconducting circuits, the sources and measurement devices of the microwave field are typically connected to other components by dispersion-free coaxial or coplanar waveguides. These transmission lines act as dissipation channels, in addition to dielectric losses~\cite{oliver2013materials, martinis2005decoherence} or quasiparticles in the superconductors~\cite{gordon2022environmental, wang2014measurement, barends2011minimizing, catelani2022using, rainis2012majorana}. 

In our formalism, we utilize the QD-MESS approach to account for the non-Markovian environment, which we transform to a finite number of intrinsically damped bosonic modes~\cite{xu2022taming, QDMESS} interacting with the nonlinear degrees of freedom of the circuit. By incorporating these auxiliary modes, we can capture the full non-Markovian dynamics induced by the system-environment interaction including memory effects and correlations, while re-casting this dynamics in a time-local form. Our formalism provides a systematic framework for studying the input-output characteristics of the circuit, taking into account the interplay between the dissipation, the nonlinearity of the circuit, and the external driving fields. The suggested approach is similar to the black-box quantization technique~\cite{nigg2012black, solgun2014blackbox}, but it is free from the limitation of weak nonlinearity and dissipation~\cite{nigg2012black}, and the need of a discretization of the environmental spectra~\cite{solgun2014blackbox}.  

The structure of the paper is as follows: In Sec.~\ref{sec:lagrangian-formalism}, we introduce the classical Lagrangian formalism for lossless lumped-element circuits and transmission lines. We define the input and output fields for each of the transmission lines and introduce the generating functional, which contains the full information about the statistics of the output field.
Section~\ref{sec:nlr-theory} contains the key enabling result of our paper, namely, a detailed derivation of the NLR theory. We eliminate the linear degrees of freedom of the circuit and unravel the resulting time-nonlocal influence functional by introducing auxiliary bosonic modes. We split these auxiliary degrees of freedom into two classes: dynamical, which have a purely classical origin and describe dissipative properties of the circuit, and fluctuation modes, which provide a proper quantum statistics of the noise. Finally, we derive a time-local QD-MESS equation that governs the dynamics of the dissipative superconducting quantum circuit. Furthermore, we define observables as time averages of the output field and provide an explicit formula for their quasiprobability distribution through the generating functional. In Sec.~\ref{sec:transmon-readout}, we illustrate our theory by applying it to the problem of dispersive readout of a superconducting transmon qubit.  Comparison to standard, linearized Markovian IO theory for such a system reproduces the known salient features of the scattering coefficient while providing a more accurate picture of the non-resonant background. Furthermore, we compute the generating function for the two quadratures for the readout of a hot transmon and its associated quasiprobability without the need for a stochastic sampling algorithms, typically used for such calculations~\cite{Wiseman1993, Wiseman2009}.
In Sec.~\ref{sec:rsj}, we present an analysis of a strongly dissipative system, namely a resistively shunted Josephson junction. Finally, we summarize our findings and provide our conclusions in Sec.~\ref{sec:conclusions}.

\section{Lagrangian formalism for superconducting circuits}

\label{sec:lagrangian-formalism}

\begin{figure}[t]
    \includegraphics[width=1\linewidth]{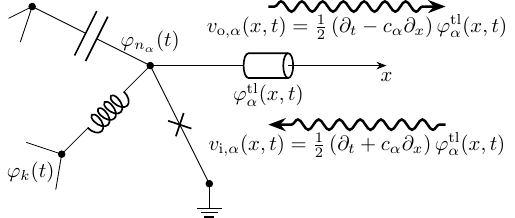}
    \caption{Transmission line $\alpha$ terminated at a node $n_\alpha$ which is connected to other nodes via capacitors, inductors, and Josephson junctions. The electromagnetic field in the line is split into input and output fields which propagate towards and away from the coupling node.}
    \label{fig:node}
\end{figure}

Although many parts of this theory are general and agnostic to the physical system in question, we motivate our studies by a particular class of open quantum systems, namely, superconducting quantum circuits. We consider circuits formed by lumped linear elements such as capacitors, inductors, and ohmic resistors, and nonlinear Josephson junctions. Due to their inherent nonlinearity, the latter are the key components of every superconducting qubit. The control and the measurement of the circuit is provided by coupling to semi-infinite transmission lines, which inevitably introduce additional dissipation channels in the system.

Such systems indeed fall into the Caldeira--Leggett~\cite{caldeira1983path} class of systems. However, from the experimental point of view, not all of its observables can be probed directly. The only experimentally accessible quantities are the output fields in the transmission lines, which belong rather to the environment. As we further discuss below, the division between the system and the environment may indeed be ambiguous. Therefore, there is a need of a theory which connects the classical drive field with the experimentally observable output field.

Throughout this paper, we use the following notation: Lowercase bold characters denote vectors and uppercase bold characters denote matrices. A bold capital character with subscripts denotes a certain block of the matrix denoted by the corresponding character. Functions of time and frequency denoted by the same character are related to each other by Fourier transformation:
\begin{gather}
    X(\omega) = \int\limits_{-\infty}^{+\infty} X(t) \mathrm e^{\imath \omega t}\;\mathrm dt, \\
    X(t) = \frac{1}{2\pi} \int\limits_{-\infty}^{+\infty} X(\omega) \mathrm e^{-\imath \omega t}\;\mathrm dt.
\end{gather}
The operators in the Hilbert space are emphasized by $\hat \cdot$ accents (caret accents), the operators in the extended Liouville space are denoted by $\check \cdot$ (caron or h\`a\v{c}ek) accents.

We consider a superconducting circuit, consisting of various dissipation-free lumped elements such as capacitors, inductors, and Josephson junctions, coupled to one or several transmission lines, which introduce dissipation to the system. We study the case of lumped resistors in the end of this section. We assume, that the studied circuit has $N+1$ nodes and $K$ transmission lines coupled to it. We number the nodes from $0$ to $N$, where node $0$ corresponds to the ground at zero potential. Following Refs.~\cite{burka04,girvi09} (see also~\cite{peika74}), we associate a flux $\varphi_k(t)$ to each node $k = 1, \ldots, N$ and a flux $\varphi^\mathrm{tl}_\alpha(x, t)$ to each transmission line $\alpha = 1,\ldots, K$. The flux of the ground node $\varphi_0$ is identically zero. We numerate  nodes  by Latin letters and transmission lines by Greek letters. For each line $\alpha$ there is a node $n_\alpha$, to which it is connected, see Fig.~\ref{fig:node}. The classical Lagrangian of the circuit reads as
\begin{multline}
    \mathcal L\left[\dot{\bm \varphi}, \bm \varphi, \dot{\bm \varphi}^\mathrm{tl}, \bm \varphi^\mathrm{tl}\right] = \mathcal L_\mathrm{lin}(\dot{\bm \varphi}, \bm \varphi) + \mathcal L_\mathrm J(\bm \varphi)\\  + \sum\limits_{\alpha=1} ^K\mathcal L ^\mathrm{tl}_\alpha\left[\dot \varphi^\mathrm{tl}_\alpha, \varphi^\mathrm{tl}_\alpha, \varphi_{n_\alpha}\right],
\end{multline}
where we have introduced Lagrangian of the lumped linear reactive elements
\begin{multline}
    \mathcal L_\mathrm{lin}(\dot{\bm \varphi}, \bm \varphi) = \sum\limits_{kk'=0}^N \left\{\frac{C_{kk'}\left(\dot \varphi_k - \dot \varphi_{k'}\right)^2}{2}  \right. \\ -  \left. \frac{\left(\varphi_k - \varphi_{k'}\right)^2}{2L_{kk'}}\right\},
\end{multline}
the Lagrangian of Josephson junctions
\begin{equation}
    \mathcal L_\mathrm{J}(\bm \varphi) = 
    \sum\limits_{kk'=0}^N E_{\mathrm J,kk'} \cos\left[\frac{2\pi}{\Phi_0} \left(\varphi_k - \varphi_k' - \Phi_{kk'}\right)\right], 
\end{equation}
and the Lagrangians of the transmission lines
\begin{multline}
    \mathcal L^\mathrm{tl}_\alpha\left[\dot \varphi^\mathrm {tl}_\alpha, \varphi^\mathrm{tl}_\alpha, \varphi_{n_\alpha}\right] = -\frac{\left(\varphi_{n_\alpha} - \left.\varphi^\mathrm{tl}_\alpha\right|_{x=0}\right)^2}{2 L_\varepsilon} \\ + 
    \int\limits_{0}^{+\infty} \left[
         \frac{C_{\ell,\alpha} \left(\dot \varphi^\mathrm{tl}_\alpha\right)^2}{2} - \frac{\left(\partial_{x} \varphi^\mathrm{tl}_\alpha\right)^2}{2 L_{\ell,\alpha}}
         \right]\;\mathrm dx.
\end{multline}
Here $\bm \varphi = \begin{bmatrix} \varphi_1 & \ldots & \varphi_N \end{bmatrix}^\mathsf T$ is a vector of temporally dependent fluxes at all the nodes and $\bm \varphi_\mathrm{tl} = \begin{bmatrix} \varphi^\mathrm{tl}_1 & \ldots  \varphi^\mathrm{tl}_K \end{bmatrix}^\mathsf T$ is vector of temporally dependent flux functions along all the transmission lines. The characteristics of the lumped capacitors, inductors, and Josephson junctions connecting nodes $k < k'$ are specified by $C_{kk'}$, $L_{kk'}$, and $E_{\mathrm J, kk'}$, respectively, and $\Phi_{kk'}$ denote the static flux bias of the junction between nodes $k$ and $k'$. If a pair of nodes $k$ and $k'$ is not connected by a capacitor, we use $C_{kk'} = 0$. Similarly for nodes not coupled by inductors or Josephson junctions we insert $L_{kk'} \to \infty$ or $E_{\mathrm J,kk'} = 0$, respectively. Each transmission line $\alpha$ is specified by the capacitance per the unit length $C_{\ell,\alpha}$ and the inductance per the unit length $L_{\ell,\alpha}$. The parameter $L_\varepsilon$ is an infinitesimally small positive inductance, which provides the boundary condition $\left.\varphi^\mathrm{tl}_\alpha\right|_{x=0} = \varphi_{n_\alpha}$ for the electromagnetic field in the transmission line.

\begin{figure}[t]
    \mbox{ }\\
    \begin{center}
    \includegraphics[width=1\linewidth]{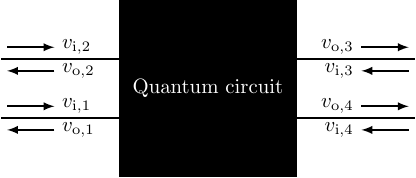}
    \end{center}
    \caption{Superconducting quantum circuit as a black box with four transmission lines, which provide the control and measurement signals of the circuit. }
    \label{fig:black-box}
\end{figure}

The field in the transmission line $\alpha$ can propagate in two directions: towards the node $n_\alpha$ and away from it. We refer to these electromagnetic fields as the input and the output, respectively. The voltages, associated with these fields, can be found as
\begin{equation}
    \begin{gathered}
        v_{\mathrm i, \alpha}(x, t) = \frac{1}{2} \left(\partial_t + c_\alpha \partial_{x}\right)\varphi_\alpha^\mathrm{tl}(x, t),\\
        v_{\mathrm o, \alpha}(x, t) = \frac{1}{2} \left(\partial_t - c_\alpha \partial_{x}\right)\varphi_\alpha^\mathrm{tl}(x, t),
    \end{gathered}
    \label{eq:io-field}
\end{equation}
where $c_\alpha = 1 / \sqrt{L_{\ell, \alpha} C_{\ell, \alpha}}$ is the speed of light in the transmission line $\alpha$. The main goal of this work is the development of a theoretical formalism for the calculation of the probabilistic characteristics of measurements performed on the output field, emitted to the transmission lines for a given input field. The input field in the transmission line $\alpha$ is specified by its voltage at the connecting node $v_{\mathrm i, \alpha}(t) \equiv v_{\mathrm i, \alpha}(x, t)|_{x=0}$. Then we measure the output field at the distance $x_0>0$ from the node connected to the line $v_{\mathrm o, \alpha}(t) \equiv v_{\mathrm o, \alpha}(x,t)|_{x=x_0}$. The full statistics of the output field is contained in the generating functional $\mathcal Z[\bm v_\mathrm i, \bm \eta_\mathrm o]$, where $\bm v_\mathrm i(t)$ is the vector of all the input fields $v_{\mathrm i, \alpha}(t)$ in all the transmission lines, and $\bm \eta_\mathrm o(t)$ is a vector of auxiliary functions $\eta_{\mathrm o, \alpha}(t)$, one function per transmission line. These function are typically called quantum drives~\cite{kamenev2023field} or counting fields~\cite{breuer2002theory, Bednorz2008, Landi2024}. The generating functional is similar to the characteristic function of probability theory, the averages and the arbitrary order correlators of the output fields are expressed through the variational derivatives of the generating functional over the respective counting fields
\begin{equation}
    \begin{gathered}
        \frac{\imath}{\hbar} \langle v_{\mathrm o, \alpha}(t)\rangle = \left.\frac{\delta \mathcal Z[\bm v_\mathrm i, \bm \eta_\mathrm o]}{\delta \eta_{\mathrm o,\alpha}(t)}\right|_{\bm \eta_\mathrm o = 0},\\
        \frac{\imath^2}{\hbar^2}\langle v_{\mathrm o, \alpha}(t)v_{\mathrm o, \alpha'}(t')\rangle = \left. \frac{\delta^2 \mathcal Z[\bm v_\mathrm i, \bm \eta_\mathrm o]}{\delta \eta_{\mathrm o,\alpha}(t) \delta \eta_{\mathrm o,\alpha'}(t')}\right|_{\bm \eta_\mathrm o = 0}, \\
        \frac{\imath^k}{\hbar^k}\left\langle 
        \prod\limits_{j=1}^k v_{\mathrm o, \alpha_j}(t_j)
        \right\rangle =
        \prod\limits_{j=1}^k \left.\frac{\delta}{\delta\eta_{\mathrm o, \alpha_j}(t_j)} \mathcal Z[\bm v_\mathrm i, \bm \eta_\mathrm o]\right|_{\bm \eta_\mathrm o = 0}.
    \end{gathered}
\end{equation}
We emphasize that quantum drive $\bm \eta_\mathrm o(t)$ has no direct physical meaning, it is an auxiliary field introduced to calculate statistical properties of the observables.
The goal of the paper is to provide an explicit way of evaluation of generating functional for arbitrary $\bm v_\mathrm i$ and $\bm \eta_\mathrm o$ starting from its general path integral expression. Moreover, the statistics of the output field is the only information which can be computed using the introduced NLR theory. The circuit itself is reduced to a black box, see Fig.~\ref{fig:black-box}, and the microscopic quantum state of the circuit remains inaccessible.

To construct the generating functional, we employ the Feynman--Vernon path integral formalism~\cite{feynman2000theory} for the treatment of the open quantum systems in question. We assume that at a time moment $t^\mathrm i$ in a distant past the circuit and the transmission lines were in a product state, with the transmission lines being prepared in the thermal coherent states~\cite{OzVogt1991}, corresponding to electromagnetic waves arriving at the coupling nodes. The particular state of the circuit at $t^\mathrm i$ is not important, since owing to the dissipation, the circuit relaxes to the thermal equilibrium state and any information on the initial state will be lost.

\section{Nonlinear-response theory for quantum circuits}

\label{sec:nlr-theory}

\subsection{Path integral formalism}

The generating functional can be expressed through the following path integral~\cite{weiss2012quantum,kamenev2023field}:
\begin{widetext}
    \begin{multline}
        \mathcal Z[\bm v_\mathrm i, \bm \eta_\mathrm o] = 
        \int \mathcal D\left[\bm \varphi_+, \bm \varphi^\mathrm {tl}_+, \bm \varphi_-, \bm \varphi^\mathrm{tl}_-\right]
        W^\mathrm i\left[
            \bm \varphi_+\left(t^\mathrm i\right),
            \bm \varphi_+^\mathrm{tl}\left(x, t^\mathrm i\right),
            \bm \varphi_-\left(t^\mathrm i\right),
            \bm \varphi_-^\mathrm{tl}\left(x, t^\mathrm i\right), \bm v_\mathrm i(t)
        \right] \\ \times
        \exp\left\{
        \frac{\imath}{\hbar} \int\limits_{t^\mathrm i}^{t^\mathrm f}
        \left[
        \mathcal L\left[\dot{\bm \varphi}_+, \bm \varphi_+, \dot{\bm \varphi}_+^\mathrm{tl}, \bm \varphi_+^\mathrm{tl}\right]
        -
            \mathcal L\left[\dot{\bm \varphi}_-, \bm \varphi_-, \dot{\bm \varphi}_-^\mathrm{tl}, \bm \varphi_-^\mathrm{tl}\right]
        +\frac{1}{2}\sum\limits_{\alpha=1}^K
        \eta_{\mathrm o, \alpha}(t) \left[v_{+,\mathrm o, \alpha}(x_0, t) + v_{-,\mathrm o, \alpha}(x_0, t)\right]\right]\;\mathrm dt
        \right\},
        \label{eq:generating-functional-0}
    \end{multline}
\end{widetext}
where forward and backward-propagating trajectories are denoted with subscripts `$+$' and `$-$', respectively, output voltage for the forward and backward-propagating trajectories is defined in accordance with Eq.~\eqref{eq:io-field}, $W^\mathrm i\left[\bm \phi_+, \bm \phi_+^\mathrm{tl}, \bm \phi_-, \bm \phi_-^\mathrm{tl}, \bm v_\mathrm i\right]$ is the full initial density matrix of the circuit, it's first four arguments $\bm \phi_\pm$ and $\bm \phi_{\pm}^\mathrm{tl}$ are $N$-dimensional vectors and are $K$-dimensional vector-valued functions of coordinate $x$, respectively, $t^\mathrm i$ and $t^\mathrm f$ are initial and final time moments. The path integral~\eqref{eq:generating-functional-0} is taken over trajectories which fulfil the boundary conditions $\bm \varphi_+(t^\mathrm f) = \bm \varphi_-(t^\mathrm f)$ and $\bm \varphi_+^\mathrm{tl}(x, t^\mathrm f) = \bm \varphi_-^\mathrm{tl}(x, t^\mathrm f)$.
The coherent displacements of the initial thermal coherent states of the transmission lines are such that under time evolution, the quantum-mechanical average input voltage at the connection nodes $\langle \hat {\bm v}_\mathrm i(x) |_{x=0}\rangle_{W_t[\ldots]}$ coincides with given classical input field $\bm v_\mathrm{i}(t)$.
This can be fulfilled by imposing
\begin{equation}
    \left\langle \hat v_{\mathrm i, \alpha}\left(x\right)\right\rangle_{W^\mathrm i[\ldots]}  = v_{\mathrm i, \alpha} \left(t + \frac{x}{c_\alpha}\right).
    \label{eq:coherent-displacement}
\end{equation}
Here, the quantum mechanical input field operator is defined by its action on arbitrary density matrix $W[\ldots]$ as
\begin{multline}
    \hat v_{\mathrm i,\alpha}(x) W\left[\ldots\right] = -\frac{1}{2} \left[c_\alpha \partial_x \phi_{+,\alpha}^\mathrm{tl}(x) \phantom{\frac{\delta}{\phi^\mathrm{tl}_+}}\right. \\ \left.+\imath \hbar C_{\ell, \alpha}\frac{\delta}{\delta \phi_{+,\alpha}^\mathrm{tl}(x)} \right] W\left[\ldots\right],
\end{multline}
quantum mechanical averaging of an operator $\hat A$ over the density matrix $W[\ldots]$ is given by
\begin{equation}
\langle \hat A \rangle_{W[\ldots]} = \int \mathcal D\left[\bm \phi^\mathrm{tl}\right]\;\mathrm d^N \bm \varphi\; \hat A W\left[\bm \phi, \bm \phi^\mathrm{tl}, \bm \phi, \bm \phi^\mathrm{tl}, \bm v_\mathrm i\right],
\end{equation}
and $W_t[\ldots]$ is the density matrix of the full system at time $t$:
\begin{multline}
W_{t'}\left[\bm \phi_+, \bm \phi_+^\mathrm{tl}, \bm \phi_-, \bm \phi_-^\mathrm {tl}\right] = \int \mathcal D\left[\bm \varphi_+, \bm \varphi^\mathrm {tl}_+, \bm \varphi_-, \bm \varphi^\mathrm{tl}_-\right] \\
        W^\mathrm i\left[
            \bm \varphi_+\left(t^\mathrm i\right),
            \bm \varphi_+^\mathrm{tl}\left(x, t^\mathrm i\right),
            \bm \varphi_-\left(t^\mathrm i\right),
            \bm \varphi_-^\mathrm{tl}\left(x, t^\mathrm i\right), \bm v_\mathrm i(t)
        \right] \\ \times
        \exp\left\{
        \frac{\imath}{\hbar} \int\limits_{t^\mathrm i}^{t'}
        \sum\limits_{s=\pm}
        s \mathcal L\left[\dot{\bm \varphi}_s, \bm \varphi_s, \dot{\bm \varphi}_s^\mathrm{tl}, \bm \varphi_s^\mathrm{tl}\right]
        \;\mathrm dt\right\},
\end{multline}
where trajectories of the path integral are fixed $\bm \varphi_\pm(t') = \bm \phi_{\pm}$ and $\bm \varphi_\pm^\mathrm{tl}(x, t') = \bm \phi_{\pm}^\mathrm{tl}(x)$.
We notice, that condition~\eqref{eq:coherent-displacement} is enough for unambiguous specification of the thermal coherent initial states of each transmission line. An explicit expression for the density matrix is presented in Appendix~\ref{sec:initial-condition}.

Throughout the paper we will use so-called classical and quantum trajectories, which are obtained by applying Keldysh rotation to forward and backward propagating trajectories:
\begin{equation}
    \begin{gathered}
        \bm \varphi_\mathrm c(t) = \frac{\bm \varphi_+(t) + \bm \varphi_-(t)}{2},\\
        \bm \varphi_\mathrm q(t) = \bm \varphi_+(t) - \bm \varphi_-(t),\\
        \bm \varphi_\mathrm c^\mathrm{tl}(x, t) = \frac{\bm \varphi_+^\mathrm{tl}(x, t) + \bm \varphi_-^\mathrm{tl}(x, t)}{2},\\
        \bm \varphi_\mathrm q^\mathrm{tl}(x, t) = \bm \varphi_+^\mathrm{tl}(x, t) - \bm \varphi_-^\mathrm{tl}(x, t).\\
    \end{gathered}
\end{equation}
Doing so significantly simplifies working with sources, such as input fields $\bm v_\mathrm i$ and counting fields~$\bm \eta_\mathrm o$. In Keldysh formalism, they play role of classical and quantum variables, respectively.
The path integral over the transmission line degrees of freedom has Gaussian form, hence, can be explicitly evaluated~\cite{weiss2012quantum}.
Elimination of the environmental degrees of freedom results in the emergence of time-nonlocal terms in the action for the degrees of freedom of the circuit. We take limits $t^\mathrm i \to -\infty$, $t^\mathrm f \to +\infty$, and $x_0 \to +0$, and obtain an expression for the generating functional in the form
\begin{equation}
    \mathcal Z[\bm v_\mathrm i, \bm \eta_\mathrm o] = \int \mathcal D[\bm \varphi_\mathrm c, \bm \varphi_\mathrm q] \exp \left\{\frac{\imath}{\hbar} S_\mathrm{circ}[\bm \varphi_\mathrm c, \bm v_\mathrm i, \bm \varphi_\mathrm q, \bm \eta_\mathrm o]\right\},
    \label{eq:generating-functional}
\end{equation}
where $\bm \varphi_\mathrm c$ and $\bm \varphi_\mathrm q$ are vectors of classical and quantum dynamical degrees of freedom~\cite{kamenev2023field} of the circuit, and the temporally non-local action of the circuit is given by
\begin{multline}
    S_\mathrm{circ}[\bm \varphi_\mathrm c, \bm v_\mathrm i, \bm \varphi_\mathrm q, \bm \eta_\mathrm o] = 
    S_\mathrm{lin}[\bm \varphi_\mathrm c, \bm \varphi_\mathrm q] + S_\mathrm{J}[\bm \varphi_\mathrm c, \bm \varphi_\mathrm q]\\-\sum\limits_{\alpha=1}^K
    F_\alpha^\mathrm{tl}\left[\varphi_{\mathrm c, n_\alpha}, v_{\mathrm i, \alpha}, \varphi_{\mathrm q, n_\alpha}, \eta_{\mathrm o , \alpha}\right].   
    \label{eq:effective-action-general}
\end{multline}
The first two terms are the actions of the linear reactive elements and of the Josephson junctions, respectively:
\begin{equation}
    \begin{gathered}
        S_\mathrm{lin}[\bm \varphi_\mathrm c, \bm \varphi_\mathrm q] =
        \int\limits_{-\infty}^{+\infty} \mathcal L_\mathrm{lin}\left(\dot{\bm \varphi}_\mathrm c, \dot{\bm \varphi}_\mathrm q, \bm \varphi_\mathrm c, \bm \varphi_\mathrm q\right)\;\mathrm dt, \\
        S_\mathrm J[\bm \varphi_\mathrm c, \bm \varphi_\mathrm q] = 
        \int\limits_{-\infty}^{+\infty} \mathcal L_\mathrm{J}\left(\bm \varphi_\mathrm c, \bm \varphi_\mathrm q\right)\;\mathrm dt,
    \end{gathered}
\end{equation}
where the two-trajectory Lagrangians read as
\begin{multline}
    \mathcal L_\mathrm{lin}\left(\dot{\bm \varphi}_\mathrm c, \dot{\bm \varphi}_\mathrm q, \bm \varphi_\mathrm c, \bm \varphi_\mathrm q\right) = \mathcal L_\mathrm{lin}\left(\dot{\bm  \varphi}_\mathrm c + \frac{\dot {\bm  \varphi}_\mathrm q}{2}, {\bm  \varphi}_\mathrm c + \frac{{\bm  \varphi}_\mathrm q}{2}\right)\\ - \mathcal L_\mathrm{lin}\left(\dot {\bm  \varphi}_\mathrm c - \frac{\dot {\bm  \varphi}_\mathrm q}{2}, {\bm  \varphi}_\mathrm c - \frac{{\bm  \varphi}_\mathrm q}{2}\right),
\end{multline}
\begin{equation}
    \mathcal L_\mathrm{J}\left(\bm \varphi_\mathrm c, \bm \varphi_\mathrm q\right) = \mathcal L_\mathrm{J}\left({\bm  \varphi}_\mathrm c + \frac{{\bm  \varphi}_\mathrm q}{2}\right)\\ - \mathcal L_\mathrm{J}\left({\bm  \varphi}_\mathrm c - \frac{{\bm  \varphi}_\mathrm q}{2}\right).
\end{equation}
Here, we use the same symbols for the for the two-path Lagrangians and for the ordinary Lagrangians, the former being distinguished by having twice the number of arguments.
The temporally non-local terms in~\eqref{eq:effective-action-general} are given by
\begin{multline}
    F_\alpha^\mathrm{tl}[\varphi_{\mathrm c, n_\alpha}, v_{\mathrm i, \alpha}, \varphi_{\mathrm q, n_\alpha}, \eta_{\mathrm o, \alpha}] \\ =
    \int_{-\infty}^{+\infty} \begin{bmatrix}
        \varphi_{\mathrm c, n_\alpha}^\ast(\omega) &
        v_{\mathrm i, \alpha}^\ast(\omega) &
        \varphi_{\mathrm q, n_\alpha}^\ast(\omega) &
        \eta_{\mathrm o, \alpha}^\ast(\omega)
    \end{bmatrix} \\ \times
    \begin{bmatrix}
        \bm 0 & \bm \Sigma_\alpha^\mathrm A(\omega) \\
        \bm \Sigma_\alpha^\mathrm R(\omega) & \bm \Sigma_\alpha^\mathrm K(\omega)
    \end{bmatrix}
    \begin{bmatrix}
        \varphi_{\mathrm c, n_\alpha}(\omega) \\
        v_{\mathrm i, \alpha}(\omega)\\
        \varphi_{\mathrm q, n_\alpha}(\omega) \\
        \eta_{\mathrm o, \alpha}(\omega)
    \end{bmatrix}\frac{d\omega}{4\pi},
    \label{eq:tl-nonlocal}
\end{multline}
a Feynman--Vernon influence functional extended by the inclusion of the counting fields.
Here, the retarded, advanced, and Keldysh components of the self-energies are given by
\begin{equation}
    \begin{gathered}
        \bm \Sigma^\mathrm R_\alpha(\omega) = \begin{bmatrix}
            -\imath \omega Z_\alpha^{-1} & -2 Z_\alpha^{-1} \\
            \imath \omega & 1
        \end{bmatrix},~\bm \Sigma^\mathrm A_\alpha(\omega) = \left[\bm \Sigma^\mathrm R_\alpha(\omega^\ast)\right]^\dagger,\\
        \bm \Sigma^\mathrm K_\alpha(\omega) = -\imath \omega
        \coth\left(\frac{\hbar \omega}{2 k_\mathrm B T}\right) \begin{bmatrix}
            Z_\alpha^{-1} & -\frac{1}{2} \\
            -\frac{1}{2} & \frac{1}{4} Z_\alpha
        \end{bmatrix},
    \end{gathered}
    \label{eq:tl-self-energy}
\end{equation}
where $Z_\alpha = \sqrt{L_{\ell,\alpha} / C_{\ell,\alpha}}$ is the impedance of the transmission line $\alpha$, and $T$ is the temperature of the transmission lines. The detailed derivation of the effective action~\eqref{eq:effective-action-general} from~\eqref{eq:generating-functional-0} is given in Appendix~\ref{sec:influence-functional}.

In the absence of the input field, the transmission line $\alpha$ is equivalent to an Ohmic shunt of resistance $Z_\alpha$ between the node $n_\alpha$ and the ground node with zero potential. Using this equivalence, we can phenomenologically introduce a term in the effective action which describes the lumped resistor of resistance $R_{kk'}$ between nodes $k$ and $k'$:\begin{multline}
    F_\mathrm{R}[\bm \varphi_{\mathrm c}, \bm \varphi_{\mathrm q}] = 
    \sum\limits_{kk'=0}^N\int \limits_{-\infty}^{+\infty} 
    \begin{bmatrix}
        \psi_{\mathrm c, kk'}^\ast (\omega) &
        \psi_{\mathrm q, kk'}^\ast (\omega)
    \end{bmatrix} \\ \times
    \begin{bmatrix}
        0 & 1 \\
        -1 & -\coth \left(\frac{\hbar \omega}{2 k_\mathrm B T}\right)
    \end{bmatrix}
    \begin{bmatrix}
        \psi_{\mathrm c, kk'}^\ast (\omega)\\
        \psi_{\mathrm q, kk'}^\ast (\omega)
    \end{bmatrix}
    \frac{\imath \omega}{R_{kk'}} \frac{\mathrm d\omega}{4\pi},
\end{multline}
where $\psi_{\mathrm c/\mathrm q,kk'} = \varphi_{\mathrm c/\mathrm q, k} - \varphi_{\mathrm c/\mathrm q, k'}$ is the flux difference between the nodes $k$ and $k'$.
The circuit action in the presence of lumped resistors becomes
\begin{multline}
S_\mathrm{circ}[\bm \varphi_\mathrm c, \bm v_\mathrm i, \bm \varphi_\mathrm q, \bm \eta_\mathrm o] = 
    S_\mathrm{lin}[\bm \varphi_\mathrm c, \bm \varphi_\mathrm q] + S_\mathrm{J}[\bm \varphi_\mathrm c, \bm \varphi_\mathrm q]\\-\sum\limits_{\alpha=1}^K
    F_\alpha\left[\varphi_{\mathrm c, n_\alpha}, v_{\mathrm i, \alpha}, \varphi_{\mathrm q, n_\alpha}, \eta_{\mathrm o , \alpha}\right] - F_\mathrm{R}[\bm \varphi_\mathrm c, \bm \varphi_\mathrm q].
    \label{eq:effective-action-0}
\end{multline}

\subsection{Core subsystem}

\begin{figure}[t]
    \begin{center}
        \includegraphics[width=0.9\linewidth]{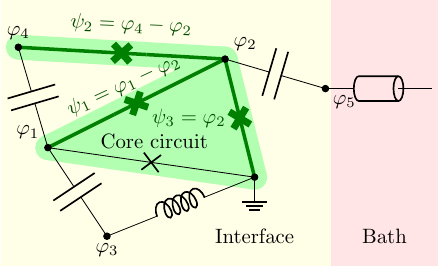}
    \end{center}
    \caption{Example of a circuit with core subsystem highlighted with green. The quantities $\psi_{1}$, $\psi_2$, and $\psi_3$ present new dynamical variables. The Josephson junction between node~1 and the ground node does not have its dedicated degree of freedom, since the flux difference across it can be expressed through $\psi_1$ and $\psi_3$.}
    \label{fig:core-subsystem}
\end{figure}
We proceed our analysis with further simplification of the circuit action~\eqref{eq:effective-action-0}. We notice, that in the most general form it can be expressed as
\begin{multline}
    S_\mathrm{circ}\left[\bm \varphi_\mathrm c, \bm v_\mathrm i, \bm \varphi_\mathrm q, \bm \eta_\mathrm o\right] = S_\mathrm J[\bm \varphi_\mathrm c, \bm \varphi_\mathrm q] \\ -\int\limits_{-\infty}^{+\infty} \begin{bmatrix}
    \bm \varphi_\mathrm c^\dagger(\omega) &
    \bm v_\mathrm i^\dagger(\omega) &
    \bm \varphi_\mathrm c^\dagger(\omega) &
    \bm \eta_\mathrm o^\dagger(\omega)
    \end{bmatrix} \\ \times
    \bm \Sigma_\mathrm{circ}(\omega)
    \begin{bmatrix}
        \bm \varphi_\mathrm c(\omega)\\
        \bm v_\mathrm i(\omega) \\
        \bm \varphi_\mathrm c(\omega) \\
        \bm \eta_\mathrm o(\omega)
    \end{bmatrix}\frac{\mathrm d\omega}{4\pi},
\end{multline}
where we incorporated both temporally local and non-local Gaussian contributions of the action into the self-energy $\bm \Sigma_\mathrm{circ}(\omega)$ possessing standard causality structure
\begin{equation}
    \bm \Sigma_\mathrm{circ}(\omega) = \begin{bmatrix}
        \bm 0 & \bm \Sigma_\mathrm{circ}^\mathrm A(\omega) \\
        \bm \Sigma^\mathrm R_\mathrm{circ}(\omega)& \bm \Sigma^\mathrm K_\mathrm{circ}(\omega)
    \end{bmatrix}.
    \label{eq:inv-greens-function}
\end{equation}

For completely linear systems, the path integral~\eqref{eq:generating-functional} can be evaluated exactly. However, in the presence of nonlinear elements such as Josephson junctions, this is no longer true. Note that the Lagrangian which describes Josephson junctions, depends only on the flux differences between the nodes connected by the respective junctions
\begin{multline}
    \mathcal L_\mathrm J\left(\bm \varphi_\mathrm c, \bm\varphi_\mathrm q\right) = -2 \sum\limits_{kk'=0}^N
    E_{\mathrm J, kk'} \\ \times \sin \left[\frac{2\pi \left(\psi_{\mathrm c, kk'} - \Phi_{kk'}\right)}{\Phi_0}\right]
    \sin \left[\frac{\pi  \psi_{\mathrm q, kk'}}{\Phi_0}\right],
\end{multline}
therefore it is natural to use $\psi_{\mathrm c/\mathrm q, kk'}$ as the new dynamical variables. However, in the presence of loops of Josephson junctions, these variables are linearly dependent, and hence we pick maximal linearly independent subsets among $\psi_{\mathrm c, kk'}$ and $\psi_{\mathrm q, kk'}$, and numerate them as $\psi_{\mathrm c, j}$ and $\psi_{\mathrm q, j}$, $j=1,\ldots,N_\mathrm{core}$, where $N_\mathrm{core}$ is the size of this maximal linearly independent subset. We refer to the subcircuit, described by these degrees of freedom, as a \textit{core circuit}, see an example in Fig.~\ref{fig:core-subsystem}.

\begin{figure}[t]
    \begin{center}
        \includegraphics[width=0.9\linewidth]{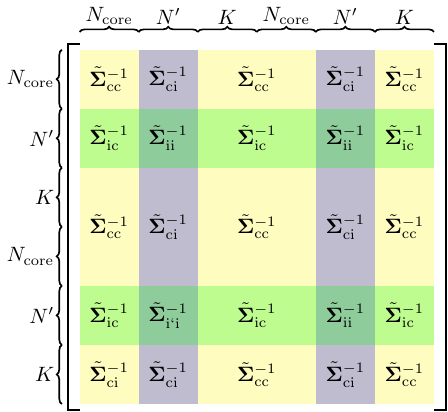}
    \end{center}
    \caption{Block structure of self-energy of the circuit~$\tilde{\bm \Sigma}_\mathrm{circ}(\omega)$. The yellow blocks correspond to the core circuit block and driving fields, green rows and blue columns correspond to the degrees of freedom $\bm \psi'_{\mathrm c/\mathrm q}$ of the interface. Here, $N' = N - N_\mathrm{core}$.}
    \label{fig:block-structure}
\end{figure}

The core dynamical variables $\bm \psi_{\mathrm c}$ and $\bm \psi_{\mathrm q}$ should be complemented by $\bm \psi'_{\mathrm c}$ and $\bm \psi'_\mathrm q$ in such a way, that there is an reversible linear transformation between the node fluxes $\bm \varphi_{\mathrm c/ \mathrm q}$ and the new degrees of freedom $\bm \psi_{\mathrm c/\mathrm q}$ and $\bm \psi'_{\mathrm c/\mathrm q}$. We denote this complementary linear subsystem as an \textit{interface} of the core system. We emphasize, that the choice of $\psi'_{\mathrm c/\mathrm q}$ is not unique and can be made in many ways. For the example shown in Fig.~\ref{fig:core-subsystem}, one of the suitable transformations is given by
\begin{equation}
    \begin{bmatrix}
        \varphi_{\mathrm c/\mathrm q, 1} \\
        \varphi_{\mathrm c/\mathrm q, 2} \\
        \varphi_{\mathrm c/\mathrm q, 3} \\
        \varphi_{\mathrm c/\mathrm q, 4} \\
        \varphi_{\mathrm c/\mathrm q, 5}
    \end{bmatrix} = 
    \begin{bmatrix}
        1 & 0 & 1 & 0 & 0 \\
        0 & 0 & 1 & 0 & 0 \\
        0 & 0 & 0 & 1 & 0 \\
        0 & 1 & 1 & 0 & 0 \\
        0 & 0 & 0 & 0 & 1
    \end{bmatrix}
    \begin{bmatrix}
        \psi_{\mathrm c/\mathrm q, 1} \\
        \psi_{\mathrm c/\mathrm q, 2} \\
        \psi_{\mathrm c/\mathrm q, 3} \\
        \psi'_{\mathrm c/\mathrm q, 1} \\
        \psi'_{\mathrm c/\mathrm q, 2}
    \end{bmatrix}.
\end{equation}
In the general case, we can define a transformation through its matrix~$\bm Q$ with unit determinant:
\begin{equation}
    \bm \varphi_{\mathrm c/\mathrm q} = \bm Q \begin{bmatrix}
        \bm \psi_{\mathrm c /\mathrm q} \\
        \bm \psi'_{\mathrm c/ \mathrm q}
    \end{bmatrix}.
    \label{eq:core-interface-splitting}
\end{equation}
Note that the action becomes quadratic with respect to the $\bm \psi'$ degrees of freedom, and hence they can be integrated out. By doing this, we shift the boundary between the system and its environment and consider the interface as an environment for the core system. Therefore, the generating functional can be expressed as a path integral over core subsystem degrees of freedom
\begin{equation}
\label{eq:generatingfunc}
    \mathcal Z[\bm v_\mathrm i,\bm \eta_\mathrm o] = \int \mathcal D[\bm \psi_\mathrm c, \bm \psi_\mathrm q] \exp\left\{\frac{\imath}{\hbar}S_\mathrm{core}\left[\bm \psi_\mathrm c, \bm v_\mathrm i,\bm \psi_\mathrm q, \bm \eta_\mathrm o\right]\right\},
\end{equation}
where the action of the core system arising from integrating out the interface degrees of freedom is given by
\begin{multline}
    S_\mathrm{core}[\bm \psi_\mathrm c, \bm v_\mathrm i, \bm \psi_\mathrm q, \bm \eta_\mathrm o] = S_\mathrm J[\bm \psi_\mathrm c, \bm \psi_\mathrm q]\\
    -\int\limits_{-\infty}^{+\infty}
    \begin{bmatrix}
        \bm \psi_\mathrm c^\dagger(\omega) &
        \bm v_\mathrm i^\dagger(\omega) &
        \bm \psi_\mathrm q^\dagger(\omega) &
        \bm \eta_\mathrm o^\dagger(\omega)
    \end{bmatrix} \\ \times
    \bm \Sigma_\mathrm{core}(\omega)
    \begin{bmatrix}
        \bm \psi_\mathrm c(\omega) \\
        \bm v_\mathrm i(\omega) \\
        \bm \psi_\mathrm q(\omega) \\
        \bm \eta_\mathrm o(\omega)
     \end{bmatrix}
    \frac{\mathrm d\omega}{4\pi}
    \label{eq:effective-action},
\end{multline}
where 
the core self-energy becomes
\begin{multline}
    \bm \Sigma_\mathrm{core}(\omega) = \begin{bmatrix}
        \bm 0 & \bm \Sigma_\mathrm{core}^\mathrm A(\omega) \\[0.5em]
        \bm \Sigma_\mathrm{core}^\mathrm R(\omega) & \bm \Sigma_\mathrm{core}^\mathrm K(\omega)
    \end{bmatrix} \\
    = \tilde{\bm \Sigma}^{-1}_\mathrm{cc} (\omega)- \tilde{\bm \Sigma}_{\mathrm{ci}}(\omega)
    \tilde{\bm \Sigma}_{\mathrm{ii}}^{-1}(\omega)
    \tilde{\bm \Sigma}_{\mathrm{ic}}(\omega),
    \label{eq:core-self-energy}
\end{multline}
where the matrices $\tilde{\bm \Sigma}_\mathrm{cc}$, $\tilde{\bm \Sigma}_\mathrm{ci}$, $\tilde{\bm \Sigma}_\mathrm{ic}$, and~$\tilde{\bm \Sigma}_\mathrm{ii}$ are blocks of the circuit self-energy $\tilde{\bm \Sigma}_\mathrm{circ}$ after the transformation to the core and interface variables
\begin{equation}
\begin{gathered}
    \tilde{\bm \Sigma}_\mathrm{circ}(\omega) = \tilde {\bm Q}^\mathsf T \bm \Sigma_\mathrm{circ} (\omega) \tilde {\bm Q},\\
    \tilde {\bm Q} = 
    \begin{bmatrix}
        \bm Q & \bm 0 &
        \bm 0 & \bm 0 \\
        \bm 0 & \bm 1 &
        \bm 0 & \bm 0 \\
        \bm 0 & \bm 0 &
        \bm Q & \bm 0 \\
        \bm 0 & \bm 0 &
        \bm 0 & \bm 1
    \end{bmatrix}.
    \end{gathered}
\end{equation}
The sizes of the blocks of the matrix $\tilde{\bm Q}$ are from left to right equal to $N$, $K$, $N$, and $K$. The subscript in the blocks of self-energy ``$\mathrm{i}$'' corresponds to the interface degrees of freedom $\bm \psi'$ to be integrated out and the index ``$\mathrm c$'' corresponds to all the core degrees $\bm \psi$ of freedom and to the external fields $\bm v_i$ and $\bm \eta_\mathrm o$. See Fig.~\ref{fig:block-structure} on the block structure of the matrix $\tilde{\bm \Sigma}_\mathrm{circ}(\omega)$.

{The expression \eqref{eq:generatingfunc} is the first main result of this work. It provides a compact expression for the generating functional of the output fields and their correlators. More specifically, we recall that}
the core self-energy~\eqref{eq:core-self-energy} is temporally non-local, which makes the evaluation of the path integral for the generating functional challenging. However, as we show below, the non-local self-energy can be unraveled by introducing a discrete set of auxiliary bosonic modes, bilinearly coupled to the core degrees of freedom~\cite{QDMESS}. The frequencies of these bosonic modes correspond to the poles of $\bm \Sigma_\mathrm{core}(\omega)$ on a complex plane of frequencies or, in practice, the poles of a highly accurate rational approximation of $\bm \Sigma_\mathrm{core}(\omega)$. In the following subsections, we carry out this unraveling and derive temporally local equation of motion for the system extended by auxiliary degrees of freedom.

\subsection{Pole decomposition of self-energy}
In order to analyze the core self-energy as a function of the complex-valued frequency, we expand Eq.~\eqref{eq:core-self-energy} for retarded, advanced, and Keldysh components, separately:
\begin{equation}
    \bm \Sigma_\mathrm{core}^\mathrm R(\omega) =  \tilde{\bm \Sigma}_\mathrm{cc}^\mathrm R(\omega) \\ -\tilde{\bm \Sigma}_\mathrm{ci}^\mathrm R(\omega) \left[\tilde{\bm \Sigma}_\mathrm{ii}^\mathrm R(\omega)\right]^{-1}\tilde{\bm \Sigma}_\mathrm{ic}^\mathrm{R} (\omega),
\end{equation}
\begin{equation}
    \bm \Sigma_\mathrm{core}^\mathrm A(\omega) =  \tilde{\bm \Sigma}_\mathrm{cc}^\mathrm A(\omega) \\ -\tilde{\bm \Sigma}_\mathrm{ci}^\mathrm R(\omega) \left[\tilde{\bm \Sigma}_\mathrm{ii}^\mathrm R(\omega)\right]^{-1}\tilde{\bm \Sigma}_\mathrm{ic}^\mathrm{R} (\omega),
\end{equation}
\begin{multline}
    \bm \Sigma_\mathrm{core}^\mathrm K(\omega) =  \tilde{\bm \Sigma}_\mathrm{cc}^\mathrm K(\omega) -\tilde{\bm \Sigma}_\mathrm{ci}^\mathrm R(\omega) \left[\tilde{\bm \Sigma}_\mathrm{ii}^\mathrm R(\omega)\right]^{-1}\tilde{\bm \Sigma}_\mathrm{ic}^\mathrm K(\omega) \\
    +\tilde{\bm \Sigma}_\mathrm{ci}^\mathrm R(\omega)
    \left[\tilde{\bm \Sigma}_\mathrm{ii}^\mathrm R(\omega)\right]^{-1}
    \tilde{\bm \Sigma}_\mathrm{ii}^\mathrm K(\omega)
    \left[\tilde{\bm \Sigma}_\mathrm{ii}^\mathrm{A}(\omega)\right]^{-1}
    \tilde{\bm \Sigma}_\mathrm{ic}^\mathrm A(\omega)\\ -
    \tilde{\bm \Sigma}_\mathrm{ci}^\mathrm K(\omega)\left[\tilde{\bm \Sigma}_\mathrm{ii}^\mathrm A(\omega)\right]^{-1}\tilde{\bm \Sigma}_\mathrm{ic}^\mathrm A(\omega)
    .
\end{multline}
We notice, that for lumped-element circuits, the retarded and advanced components of the circuit self energy  $\tilde{\bm \Sigma}_\mathrm{circ}(\omega)$ are matrix-valued polynomial functions of frequency~$\omega$ with degree not higher than two, and the Keldysh component is proportional to~$\omega \coth\left[\hbar \omega / (2 k_\mathrm B T)\right]$. This implies that the retarded and advanced components of the core self-energy are rational matrix-valued functions of frequency. The poles of $\bm \Sigma_\mathrm{core}^\mathrm R(\omega)$ come as solutions of equation
\begin{equation}
\det \tilde{\bm \Sigma}_\mathrm{ii}^\mathrm R(\omega) = 0
\end{equation}
which we denote as $\omega_k^\mathrm d$. The poles of $\bm \Sigma_\mathrm{core}^\mathrm A(\omega)$ correspond to their complex conjugates $\left(\omega_k^\mathrm d\right)^\ast$. These poles have purely classical origin and correspond to frequencies where the impedance matrix of the interface is degenerate. We refer to these poles as \textit{dynamical poles}. The Keldysh component of the core self-energy can be expressed as
\begin{equation}
    \bm \Sigma_\mathrm{core}^\mathrm K(\omega) = \coth\left(\frac{\hbar \omega}{2 k_\mathrm B T}\right) \bm F(\omega),
\end{equation}
where $\bm F(\omega)$ is a matrix-valued rational function of frequency. This function $\bm F(\omega)$ also has poles at $\omega_k^\mathrm d$ and their complex conjugates, but the Keldysh component of the cores self-energy also has poles at the Matsubara frequencies $\imath \omega_n^\mathrm m = 2\pi \imath n k_\mathrm B T / \hbar$, $n = \pm 1, \pm 2, \ldots$ due to~$\coth\left[\hbar \omega / (2k_\mathrm B T)\right]$ factor. We employ Mittag-Leffler's theorem to express the self-energy in the form
\begin{multline}
    \bm \Sigma_\mathrm{core}(\omega) = \bm P(\omega) + \sum\limits_{k=1}^{M^\mathrm d} \left[
    \frac{\bm R_k^\mathrm d}{\omega - \omega_k^\mathrm d} + \frac{\bar{\bm R}_k^\mathrm d}{\omega - \left(\omega_k^\mathrm d\right)^\ast}
    \right] \\ + \sum\limits_{n\ne 0} \frac{\bm R_n^\mathrm m}{\omega + \imath \omega_n^\mathrm m},
    \label{eq:mittag-leffler}
\end{multline}
where $\bm P(\omega) = \bm P_0 + \bm P_1 \omega + \bm P_2\omega^2$ is a matrix-valued polynomial of degree two, $\bm R_k^\mathrm d$ and $\bar{\bm R}_k^\mathrm d$ are matrix-valued residues of core-self energy at the dynamical poles $\omega_k^\mathrm d$ and their complex conjugates, respectively, and $\bm R_n^\mathrm m$ are the matrix-valued residues at the Matsubara frequencies given by
\begin{equation}
    \bm R_n^\mathrm m = \frac{2 k_\mathrm B T}{\hbar \omega} \begin{bmatrix}
        \bm 0 & \bm 0 \\
        \bm 0 & \bm F(-\imath \omega_n^\mathrm m)
    \end{bmatrix}.
\end{equation}
Here, we assume that all the poles are simple. We notice, that the matrix residues at dynamical poles $\bm R_k^\mathrm d$ and $\bar{\bm R}_k^\mathrm d$ have unit rank, since these poles come from the inverse of $\tilde{\bm \Sigma}_\mathrm{ii}(\omega)$.

Strictly speaking, we have to introduce an infinite number of auxiliary modes, since we need to sum over all Matsubara frequencies in Eq.~\eqref{eq:mittag-leffler}, which is impossible in any non-trivial practical calculations. A naive way to approximate the series by a finite sum turns out to be impractical due to the poor convergence of the series. Alternatively, we may approximate the function $\coth\left[{\hbar \omega} /({2k_{\rm B} T})\right]$ over a range of real  frequencies \((- \Omega, + \Omega)\), where \(\Omega\) serves as the frequency cutoff, using a rational function with the minimal number of poles in the complex plane. We adopt the following decomposition 
\begin{equation}
    \coth\left(\frac{\hbar \omega}{2k_\mathrm B T}\right) \approx \frac{2k_\mathrm B T}{\hbar \omega} + \sum\limits_{n=1}^{M^\mathrm f} \frac{2\omega r_{n}^{\mathrm{f}}}{\omega^{2} + (\omega_{n}^{\mathrm{f}})^{2}},
    \label{eq:coth-approximation}
\end{equation}
where the residues and the poles are obtained using a modified symmetric version of the adaptive Antoulas–
Anderson (AAA) algorithm \cite{nakatsukasa2018aaa, xu2022taming} which turns out to be very accurate (see Appendix~\ref{sec:AAA}) in a finite range of frequencies even with very few poles taken into account.
We refer to the poles at frequencies $\pm \imath \omega_n^\mathrm{f}$ as \textit{fluctuation poles}, since the Keldysh component of the self-energy provides proper fluctuation-dissipation relation in the finite range of frequencies $(-\Omega, \Omega)$. Consequently, we find an approximation for the core self-energy as
\begin{multline}
    \bm \Sigma_\mathrm{core}(\omega) = \bm P(\omega) + \sum\limits_{k=1}^{M^\mathrm d} \left[
    \frac{\bm R_k^\mathrm d}{\omega - \omega_k^\mathrm d} + \frac{\bar{\bm R}_k^\mathrm d}{\omega - \left(\omega_k^\mathrm d\right)^\ast}
    \right] \\ + \sum\limits_{n=1}^{M^\mathrm f} \left(\frac{\bm R_n^\mathrm {f}}{\omega + \imath \omega_n^\mathrm {f}} + \frac{\bar{\bm R}_n^\mathrm {f}}{\omega - \imath \omega_n^\mathrm {f}}\right),
    \label{eq:mittag-leffler-2}
\end{multline}
where
\begin{equation}
    \begin{gathered}
    \bm R_n^\mathrm {f} = \begin{bmatrix}
        \bm 0 & \bm 0 \\
        \bm 0 & r_n^\mathrm{f} \bm F(-\imath \omega_n^\mathrm{f})
    \end{bmatrix},\\
    \bar {\bm R}_n^\mathrm {f} = \begin{bmatrix}
        \bm 0 & \bm 0 \\
        \bm 0 & r_n^\mathrm{f} \bm F(\imath \omega_n^\mathrm{f})
    \end{bmatrix}.
    \end{gathered}
\end{equation}

Next, we proceed to analyze the causality of the self-energy. In the time domain, each of the terms in Eq.~\eqref{eq:mittag-leffler-2} has a defined causality. The polynomial term $\bm P(\omega)$ is time local, the causality of the simple-pole terms is defined by the sign of the imaginary part of the respective poles: the poles in the lower (upper) complex half-plane yield terms with retarded (advanced) causality. In practice, it is convenient to work with poles in the lower half-plane, since the corresponding auxiliary modes decay in time. Note that since all the dynamical variables and source fields are real-valued in the time domain, the self-energy has the symmetry property $\bm \Sigma_\mathrm{core}(t) = \bm \Sigma^\mathsf{T}_\mathrm{core}(-t)$. This means that one can construct a modified self-energy by dropping terms with advanced causality
\begin{equation}
    \tilde{\bm \Sigma}_\mathrm{core}(\omega) = \frac{1}{2}{\bm P}(\omega) + \sum\limits_{k=1}^{M^\mathrm d} 
    \frac{\bm R_k^\mathrm d}{\omega - \omega_k^\mathrm d} + \sum\limits_{n=1}^{M^\mathrm f} \frac{\bm R_n^\mathrm {f}}{\omega + \imath \omega_n^\mathrm {f}},
    \label{eq:mittag-leffler-2}
\end{equation}
which satisfies
\begin{multline}
    \frac{1}{2} \int\limits_{-\infty}^{+\infty} \bm y^\dagger(\omega) \bm \Sigma_\mathrm{core}(\omega) \bm x(\omega)\;\frac{\mathrm d\omega}{2\pi} \\ = \int\limits_{-\infty}^{+\infty} \bm y^\dagger(\omega) \tilde{\bm \Sigma}_\mathrm{core}(\omega) \bm x(\omega)\;\frac{\mathrm d\omega}{2\pi},
\end{multline}
where $\bm x(\omega)$ and $\bm y(\omega)$ are Fourier images of arbitrary real-valued vector functions of time. Due to the causality structure of the core self-energy, the residues at the dynamical poles assume the form
\begin{equation}
    \bm R_k^\mathrm d = \begin{bmatrix}
        \bm 0 & \bm 0 \\
        \bm R_k^\mathrm {dR} & 
        \bm R_k^\mathrm {dK}
    \end{bmatrix},
\end{equation}
hence, we can consider that the modified self-energy $\tilde{\bm \Sigma}_\mathrm{core}(\omega)$ is rectangular. Introducing notations 
\begin{equation}
    \begin{gathered}
        \tilde{\bm R}_k^\mathrm d = \begin{bmatrix} \bm R_k^\mathrm {dR} & \bm R_k^\mathrm {dK} \end{bmatrix}, \\
        \tilde{\bm R}_n^\mathrm{f} = r_n^\mathrm{f} \bm F(-\imath \omega_n^\mathrm{f}),
    \end{gathered}
    \label{eq:matrix-residues}
\end{equation}
we can express the effective action of the core subsystem as
\begin{widetext}
    \begin{multline}
        S_\mathrm{core}[\bm \psi_\mathrm c, \bm v_\mathrm i, \bm \psi_\mathrm q, \bm \eta_\mathrm o] =S_\mathrm J[\bm \psi_\mathrm c,\bm \psi_\mathrm q]  - \int\limits_{-\infty}^{+\infty}
            \left\{
            \begin{bmatrix}
                    \dot{\bm \psi}_\mathrm q^\mathsf T(t) &
                    \dot{\bm \eta}_\mathrm o^\mathsf T(t)
                \end{bmatrix}
                \tilde{\bm P}_2 \partial_t +
                \begin{bmatrix}
                    \bm \psi_\mathrm q^\mathsf T(t) &
                    \bm \eta_\mathrm o^\mathsf T(t)
                \end{bmatrix}
                \left(
                \tilde{\bm P}_0
                +\imath
                \tilde{\bm P_1} \partial_t
                \right)
            \right\} 
            \begin{bmatrix}
                    \bm \psi_\mathrm c(t) \\
                    \bm v_\mathrm i(t) \\
                    \bm \psi_\mathrm q(t) \\
                    \bm \eta_\mathrm o(t)
                \end{bmatrix}
        \;\mathrm dt \\-  \int\limits_{-\infty}^{+\infty} 
        \begin{bmatrix}
                \bm \psi_\mathrm q^\dagger(\omega) &
                \bm \eta_\mathrm o^\dagger(\omega)
            \end{bmatrix}
        \left\{
            \sum\limits_{k=1}^{M^\mathrm d} 
            \frac{\tilde{\bm R}_k^\mathrm d}{\omega - \omega_k^\mathrm d}
            \begin{bmatrix}
                \bm \psi_\mathrm c(\omega) \\
                \bm v_\mathrm i(\omega) \\
                \bm \psi_\mathrm q(\omega) \\
                \bm \eta_\mathrm o(\omega)
            \end{bmatrix} +
            \sum\limits_{n=1}^{M^\mathrm f} 
            \frac{\tilde{\bm R}_n^\mathrm {f}}{\omega + i \omega_n^\mathrm {f}}
            \begin{bmatrix}
                \bm \psi_\mathrm q(\omega) \\
                \bm \eta_\mathrm o(\omega)
            \end{bmatrix}
            \right\}
        \frac{\mathrm d\omega}{2\pi}
        \label{eq:action-flux}.
    \end{multline}
\end{widetext}
The matrices $\tilde{\bm P}_k$ are defined as
\begin{equation}
    \tilde{\bm P}_k = \begin{bmatrix}
        \bm P_k^\mathrm R &
        \frac{1}{2}\bm P_k^\mathrm K
    \end{bmatrix},
\end{equation}
provided
\begin{equation}
    \bm P_k = \begin{bmatrix}
        \bm 0 & \bm P_k^\mathrm A \\
        \bm P_k^\mathrm R & \bm P_k^\mathrm K
    \end{bmatrix}.
\end{equation}

\subsection{Unraveling of temporally non-local terms in action}
We treat each non-local term in the action of Eq.~\eqref{eq:action-flux} separately by introducing auxiliary complex-valued vector degrees of freedom $\bm a$ via the Hubbard--Stratonovich transformation~\cite{hubbard1959calculation, stratonovich1957method, QDMESS}:
\begin{multline}\label{eq:hubbard-str}
    \exp\left[
    -\frac{\imath}{\hbar}\int\limits_{-\infty}^{+\infty} \bm y^\dagger(\omega) \frac{\bm R}{\omega - \omega_\ast}\bm x(\omega)\;\frac{\mathrm d\omega}{2\pi}
    \right] = \\
    \int \mathcal D\left[\bm a, \bm a^\dagger\right] \exp \left\{
    \frac{\imath}{\hbar} \int\limits_{-\infty}^{+\infty} \left[
    \hbar \bm a^\dagger(\omega) \left(\omega - \omega_\ast\right) \bm a(\omega) \right . \right . \\ \left. \phantom{\int\limits_{-\infty}^{+\infty}} \left. + \bm a^\dagger(\omega) \bm V^\mathsf T \bm x(\omega) + \bm y^\dagger (\omega) \bm U \bm a(\omega)
    \right]\frac{\mathrm d\omega}{2\pi}
    \right\}.
\end{multline}
where $\bm x(\omega)$ and $\bm y(\omega)$ are Fourier images of placeholder real valued-vector functions of time $\bm x(t)$ and $\bm y(t)$, respectively, $\bm R$ is a placeholder  constant complex-valued matrix, and $\omega_\ast$ is a placeholder complex-valued frequency with a negative imaginary part. The matrix $\bm R$ may be factorized as $\bm R = \hbar^{-1} \bm U \bm V^\mathsf T$, where the number of columns of matrices $\bm U$ and $\bm V$ is equal to the rank of $\bm R$. Such a factorization is not unique, it can be obtained, e.g., by singular value decomposition. The dimension of the auxiliary vector degree of freedom $\bm a$ is equal to the rank of $\bm R$. The validity of~\eqref{eq:hubbard-str} can be verified straightforwardly by evaluating the Gaussian path integral over $\bm a(\omega)$ and $\bm a^\dagger(\omega)$. Here, we assume that the normalization coefficient is included into the integration measure $\mathcal D\left[\bm a, \bm a^\dagger\right]$.

Note that the expression on the right hand side of Eq.~\eqref{eq:hubbard-str} is local in time and equals to
\begin{multline}
    \exp\left[
    -\frac{\imath}{\hbar}\int\limits_{-\infty}^{+\infty} \bm y^\dagger(\omega) \frac{\bm R}{\omega - \omega_\ast}\bm x(\omega)\;\frac{\mathrm d\omega}{2\pi}
    \right] = \\
    \int \mathcal D[\bm a, \bm a^\dagger] \exp\left\{
        \frac{\imath}{\hbar} \int\limits_{-\infty}^{+\infty} \left[
            \imath \hbar \bm a^\dagger(t) \dot{\bm a}(t) - \hbar \omega_\ast |\bm a(t)|^2 \right . \right. \\ \left.\phantom{\int\limits_{-\infty}^{+\infty}} \left.  +\bm a^\dagger(t) \bm V^\mathsf T \bm x(t) + \bm y^\mathsf T(t) \bm U \bm a(t)
        \right]\;\mathrm dt
    \right\}.
    \label{eq:action-aux-mode}
\end{multline}
Hence, we can apply this transformation to treat each of the simple pole terms in Eq.~\eqref{eq:action-flux}:
\begin{multline}
\exp\left\{
-\frac{\imath}{\hbar} \int\limits_{-\infty}^{+\infty} \begin{bmatrix}
    \bm \psi_\mathrm q^\dagger(\omega) &
    \bm \eta_\mathrm o^\dagger(\omega)
\end{bmatrix}
\frac{\tilde{\bm R}_k^\mathrm d}{\omega - \omega_k^\mathrm d}
\begin{bmatrix}
    \bm \psi_\mathrm c(\omega) \\
    \bm v_\mathrm i(\omega) \\
    \bm \psi_\mathrm q(\omega) \\
    \bm \eta_\mathrm o(\omega)
\end{bmatrix}\;
\frac{\mathrm d\omega}{2\pi}
\right\} \\
= \int \mathcal D[a_k, a_k^\ast] \exp\left\{
\frac{\imath}{\hbar} \int\limits_{-\infty}^{+\infty}
\left[
\imath \hbar a_k^\ast(t) \dot a_k(t) - \hbar \omega_k^\mathrm d |a_k(t)|^2 \right. \right. \\ \left .
\left.+ a_k^\ast(t) \bm v_k^{\mathrm d \mathsf T}
\begin{bmatrix}
    \bm \psi_\mathrm c(t) \\
    \bm v_\mathrm i(t) \\
    \bm \psi_\mathrm q(t) \\
    \bm \eta_\mathrm o(t)
\end{bmatrix} +
\begin{bmatrix}
    \bm \psi_\mathrm q(t) &
    \bm \eta_\mathrm o(t)
\end{bmatrix}
 \bm u_k^\mathrm d a_k(t)
 \right]\;\mathrm dt
\right\},
\label{eq:hubbard-stratonovich-dynamic}
\end{multline}
for every $k=1, \ldots, M^\mathrm d$ and
\begin{multline}
    \exp\left\{
    -\frac{\imath}{\hbar}\int\limits_{-\infty}^{+\infty}
    \begin{bmatrix}
    \bm \psi_\mathrm q^\dagger(\omega) &
    \bm \eta_\mathrm o^\dagger(\omega)
\end{bmatrix}
\frac{\tilde{\bm R}_n^\mathrm f}{\omega + \imath \omega_n^\mathrm f}
\begin{bmatrix}
    \bm \psi_\mathrm q(\omega) \\
    \bm \eta_\mathrm o(\omega)
\end{bmatrix}\;
\frac{\mathrm d\omega}{2\pi}
    \right\} \\
    =\int\mathcal D\left[\bm b_n, \bm b_n^\dagger\right]
    \exp \left\{
    \frac{\imath}{\hbar} \int\limits_{-\infty}^{+\infty}
    \sum\limits_{m=1}^{\varrho_n}
    \left[
    \imath \hbar b_{nm}^\ast(t) \dot b_{nm}(t) \right. \right. \\ \left. \left.  + \imath \hbar \omega_n^\mathrm f |b_{nm}(t)|^2 + b_{nm}^\ast(t) \bm v_{nm}^{\mathrm f \mathsf T} \begin{bmatrix}
        \bm \psi_\mathrm q(t) \\
        \bm \eta_\mathrm o(t)
    \end{bmatrix} \right. \right. \\ \left. \phantom{\int\limits_{-\infty}^{+\infty}}\left .\phantom{\begin{bmatrix}a\\b\end{bmatrix}} +
    \begin{bmatrix}
    \bm \psi_\mathrm q^\mathsf T(t) &
    \bm \eta_\mathrm o^\mathsf T(t)
    \end{bmatrix}
    \bm u_{nm}^\mathrm f b_{nm}(t)
    \right]\;\mathrm dt
    \right\},
    \label{eq:hubbard-stratonovich-fluctuation}
\end{multline}
for every $n = 1, \ldots, M^\mathrm f$.
Here we used that residues at dynamical poles  have unit rank $\tilde{\bm R}_k^\mathrm d = \hbar^{-1} \bm u_k^\mathrm d \bm v_k^{\mathrm f \mathsf T}$, where $\bm u_k^\mathrm d$ and $\bm v_k^{\mathrm f}$ are column vectors of dimensions $N_\mathrm{core} + K$ and $2 N_\mathrm{core} + 2K$, respectively. Therefore, to unravel temporally non-local contribution to the action~\eqref{eq:action-flux} coming from dynamical pole $k$ we introduce a single scalar auxiliary complex-valued degree of freedom corresponding $a_k$. However, this may not be the case for the fluctuation poles, and the factorization of the respective residues $\tilde{\bm R}_n^\mathrm f = \hbar^{-1} \bm U_n^\mathrm f \bm V_n^{\mathrm f \mathsf T}$ yields $(N_\mathrm{core} + K)\times \varrho_n$ matrices $\bm U_n^\mathrm f$ and $\bm V_n^\mathrm f$, where $\varrho_n = \operatorname{rank} \tilde{\bm R}_n^\mathrm f$. Hence, we for each fluctuation pole $n$ we have to introduce $\varrho_n$ auxiliary complex-valued degrees of freedom $b_{nm}$, $m=1,\ldots,\varrho_n$. In Eq.~\eqref{eq:hubbard-stratonovich-fluctuation}, we have also introduced $\bm u_{nm}^\mathrm f$ and $\bm v_{nm}^\mathrm f$ as $m$-th columns of factors $\bm U_n^\mathrm f$ and $\bm V_n^\mathrm f$, respectively.

The expressions under exponents in the path integrals in the right hand rides of~Eqs.~\eqref{eq:hubbard-stratonovich-dynamic} and~\eqref{eq:hubbard-stratonovich-fluctuation} correspond to actions of driven harmonic oscillators in the coherent-state formulation, therefore we refer to the auxiliary degrees of freedom by $a_k$ and $b_{nm}$ as the \textit{dynamical auxiliary modes} and \textit{fluctuation auxiliary modes}, respectively. Formally, these modes imitate memory effects of the non-Markovian bath for the core subsystem. The dynamical modes $a_k$ have a clear physical meaning, they correspond to the normal modes of the linear interface circuit. The physics of the fluctuation modes is more obscure, but they provide proper fluctuation-dissipation relation for the core circuit.

Consequently, we substitute Eqs.~\eqref{eq:hubbard-stratonovich-dynamic} and~\eqref{eq:hubbard-stratonovich-fluctuation} into the core action~\eqref{eq:action-flux} and express the generating functional as
\begin{widetext}
\begin{gather}
    \mathcal Z[\bm v_\mathrm i, \bm \eta_\mathrm o] = \int \mathcal D\left[\bm \psi_\mathrm c, \bm \psi_\mathrm q, \bm a, \bm a^\dagger, \bm b, \bm b^\dagger\right] \exp \left(
    \frac{\imath}{\hbar} 
    S_\text{QD-MESS}\left[\bm \psi_\mathrm c, \bm v_\mathrm i, \bm \psi_\mathrm q, \bm \eta_\mathrm o, \bm a, \bm a^\dagger, \bm b, \bm b^\dagger\right]
    \right),\\
    S_\text{QD-MESS}\left[\bm \psi_\mathrm c, \bm v_\mathrm i, \bm \psi_\mathrm q, \bm \eta_\mathrm o, \bm a, \bm a^\dagger, \bm b, \bm b^\dagger\right] = \int\limits_{-\infty}^{+\infty} \mathcal L_\text{QD-MESS}\left(\dot{\bm \psi}_\mathrm c, \dot{\bm v}_\mathrm i, \dot{\bm \psi}_\mathrm q, \dot{\bm \eta}_\mathrm o, \dot{\bm a}, \dot{\bm b},\bm \psi_\mathrm c, \bm v_\mathrm i, \bm \psi_\mathrm q, \bm \eta_\mathrm o, \bm a, \bm a^\dagger, \bm b, \bm b^\dagger\right)\;\mathrm dt,
    \label{eq:qd-mess-action}
\end{gather}
where
\begin{multline}
    \mathcal L_\text{QD-MESS}\left(\ldots\right) = \mathcal L_\mathrm J(\bm \psi_\mathrm c,\bm \psi_\mathrm q) 
    -
        \left\{
        \begin{bmatrix}
                \dot{\bm \psi}_\mathrm q^\mathsf T &
                \dot{\bm \eta}_\mathrm o^\mathsf T
            \end{bmatrix}
            \tilde{\bm P}_2
            \begin{bmatrix}
                \dot{\bm \psi}_\mathrm c \\
                \dot{\bm v}_\mathrm i \\
                \dot{\bm \psi}_\mathrm q \\
                \dot{\bm \eta}_\mathrm o
            \end{bmatrix}
             +
            \begin{bmatrix}
                \bm \psi_\mathrm q^\mathsf T &
                \bm \eta_\mathrm o^\mathsf T
            \end{bmatrix}
            \left(
            \tilde{\bm P}_0
            +\imath
            \tilde{\bm P}_1 \partial_t
            \right)\begin{bmatrix}
                \bm \psi_\mathrm c \\
                \bm v_\mathrm i \\
                \bm \psi_\mathrm q \\
                \bm \eta_\mathrm o
            \end{bmatrix}
        \right\} 
         \\+
        \sum\limits_{k=1}^{M^\mathrm d} \left(
        \imath\hbar a_k^\ast \dot a_k - \hbar\omega_k^\mathrm d |a_k|^2 +
        \begin{bmatrix}
            \bm \psi_\mathrm q^\mathsf T &
            \bm \eta_\mathrm o^\mathsf T
        \end{bmatrix}
        \bm u_k^\mathrm d a_k + 
        a_k^\ast \bm v_k^{\mathrm d \mathsf T}
        \begin{bmatrix}
            \bm \psi_\mathrm c \\
            \bm v_\mathrm i \\
            \bm \psi_\mathrm q \\
            \bm \eta_\mathrm o
        \end{bmatrix}
        \right) \\ +
        \sum\limits_{n=1}^{M^\mathrm f} \sum\limits_{m=1}^{\varrho_n} \left(
        \imath \hbar b_{nm}^\ast \dot{b}_{nm} + \imath \omega_n^\mathrm {f} | b_{nm}|^2+ \begin{bmatrix}
            \bm \psi_\mathrm q^\mathsf T &
            \bm \eta_\mathrm o^\mathsf T
        \end{bmatrix}
        \bm u_{nm}^\mathrm {f} b_{nm} +
        b_{nm}^\ast \bm v_{nm}^{\mathrm {f} \mathsf T} \begin{bmatrix}
            \bm \psi_\mathrm q\\
            \bm \eta_\mathrm o
        \end{bmatrix}
        \right).
    \label{eq:flux-qdmess-lagrangian}
\end{multline}
\end{widetext}
Here, we have introduced the Lagrangian and the time-local action of quantum dissipation in the minimally extended state space (QD-MESS) equation. 

In the Lagrangian~\eqref{eq:flux-qdmess-lagrangian} all the auxiliary modes are coupled to the flux degrees of freedom $\bm \psi_\mathrm c$ and $\bm \psi_\mathrm q$, and hence we refer to the QD-MESS equation produced by this Lagrangian as flux QD-MESS. However, this form is not unique. Instead of applying a Hubbard--Stratonovich transformation as in Eq.~\eqref{eq:hubbard-str}, we rewrite the integrand on its left hand side as
\begin{multline}
    \bm y^\dagger(\omega) \frac{\bm R}{\omega - \omega_\ast}\bm x(\omega)
     =
    -\bm y^\dagger(\omega) \bm R \left[\frac{\bm x(\omega)}{\omega_\ast} + \frac{\omega \bm x(\omega)}{\omega_\ast^2}\right] \\ + 
    \left[\frac{\imath \omega}{\omega_\ast} \bm y^\dagger(\omega)\right]\frac{\bm R}{\omega - \omega_\ast} \left[-\frac{\imath \omega}{\omega_\ast} \bm x(\omega)\right]
    .
\end{multline}
Multiplication by $\omega$ of a Fourier image corresponds to applying the $\imath \partial_t$ operator in the time domain. Hence, the first term is already local in time. We treat the last time-non-local term by applying the Hubbard--Stratonovich transformation as
\begin{multline}
    \exp\left[
    -\frac{\imath}{2\pi\hbar}\int\limits_{-\infty}^{+\infty} \bm y^\dagger(\omega) \frac{\bm R}{\omega - \omega_\ast}\bm x(\omega)\;\mathrm d\omega
    \right] \\ =
     \int \mathcal D[\bm a, \bm a^\dagger] \exp\left\{
         \frac{\imath}{\hbar} \int\limits_{-\infty}^{+\infty} \left[
             \imath \hbar \bm a^\dagger(t) \dot{\bm a}(t) \right . \right. \\ \left. \left. + \hbar \omega_\ast |\bm a(t)|^2  + \bm y^\mathsf T(t) \bm R\left(\frac{\bm x(t)}{\omega_\ast} + \frac{i \dot {\bm x}(t)}{\omega_\ast^2} \right) \right . \right. \\ \left.\phantom{\int\limits_{-\infty}^{+\infty}} \left.
             + \bm a^\dagger(t) \bm V^\mathsf T \frac{\dot{\bm x}(t)}{\omega_\ast} + \frac{\dot{\bm y}^\mathsf T(t)}{\omega_\ast} \bm U \bm a(t)
         \right]\;\mathrm dt
     \right\}.
    \label{eq:hubbard-str-voltage}
\end{multline}
This way, with a proper renormalization of the time-local terms in the action~\eqref{eq:action-flux}, one can obtain a QD-MESS Lagrangian where the auxiliary degrees of freedom couple to the voltages $\dot{\bm \psi}_\mathrm c$ and $\dot{\bm \psi}_\mathrm q$ for all or some of the auxiliary modes (see Appendix~\ref{sec:charge-qdmess} for details). In principle, it is even possible to derive QD-MESS equation with mixed coupling, where some degrees of freedom couple to auxiliary modes with flux and the others with charge. Such formulations may be beneficial for certain practical calculations, but for the sake of simplicity of the presentation, we proceed with the flux QD-MESS Lagrangian.

\subsection{QD-MESS equation}
To derive the quantum equation which corresponds to the Lagrangian~\eqref{eq:flux-qdmess-lagrangian}, we proceed with a standard transformation to the Schr\"odinger picture. First, we introduce the canonical conjugate variables to the fluxes, i.e., the charges
\begin{equation}
    \begin{gathered}
    \bm q_\mathrm c = \frac{\partial \mathcal L_\text{QD-MESS}}{\partial \dot{\bm \psi}_\mathrm q^\mathsf T}
    ,
    ~
    \bm q_\mathrm q^\mathsf T = \frac{\partial \mathcal L_\text{QD-MESS}}{\partial \dot{\bm \psi}_\mathrm c} 
    \end{gathered}
    \label{eq:legendre}
\end{equation}
Then, we apply the Legendre transformation associated with Eq.~\eqref{eq:legendre} and obtain the Liouvillian
\begin{multline}
    \mathfrak L = \bm q_\mathrm q^\mathsf T \dot{\bm \psi}_\mathrm c + \dot{\bm \psi}_\mathrm q^\mathsf T \bm q_\mathrm c + \imath \hbar \sum\limits_{k=1}^{M^\mathrm d} a_k^\ast \dot a_k\\ + \imath \hbar \sum\limits_{n=1}^{M^\mathrm f} \sum\limits_{m=1}^{\varrho_n} b_{nm}^\ast \dot{b}_{nm} - \mathcal L_\text{QD-MESS}.
    \label{eq:lagrangian-to-liouvillian}
\end{multline}
By necessity, the introduction of the charges conjugate to the core phase variables moves a few capacitive terms from the interface back into the core.

The resulting Liouvillian, written as a function of canonical variables, can be interpreted as a quantum Liouvillian 
by imposing the canonical commutation relations
\begin{equation}
    \begin{gathered}
        \left[\check \psi_{\mathrm c, j}, \check q_{\mathrm q,j'}\right] = \imath \hbar \delta_{jj'},~
        \left[\check \psi_{\mathrm q, j}, \check q_{\mathrm c,j'}\right] = \imath \hbar \delta_{jj'}, \\
        \left[\check a_k, \check a_{k'}^\dagger\right] = \delta_{kk'}, ~
        \left[\check b_{nm}, \check b_{n'm'}^\dagger\right] = \delta_{nn'} \delta_{mm'},
    \end{gathered}
    \label{eq:commutation-relations}
\end{equation}
with all other single-operator commutators vanishing.
Finally, the quantum Liouvillian, which governs the dynamics of the superconducting quantum circuit, reads
\begin{widetext}
\begin{multline}
    \check{\mathfrak L} = 
    \begin{bmatrix}
        \check{\bm q}_\mathrm q^\mathsf T &
        \check{\bm \psi}_\mathrm q^\mathsf T &
        \bm \eta_\mathrm o^\mathsf T &
        \dot{\bm \eta}_\mathrm o^\mathsf T
    \end{bmatrix}
    \left(
    \bm L^\mathrm R
    \begin{bmatrix}
        \check{\bm q}_\mathrm c \\
        \check{\bm \psi}_\mathrm c \\
        \bm v_\mathrm i \\
        \dot{\bm v}_\mathrm i
    \end{bmatrix}
    +\bm L^\mathrm K
    \begin{bmatrix}
        \check{\bm q}_\mathrm q \\
        \check{\bm \psi}_\mathrm q \\
        \bm \eta_\mathrm o \\
        \dot{\bm \eta}_\mathrm o
    \end{bmatrix}
    \right) - \mathcal L_\mathrm J\left(\check{\bm \psi}_\mathrm c, \check {\bm \psi}_\mathrm q\right)
    \\+
    \sum\limits_{k=1}^{M^\mathrm d} \left(
        \hbar \omega_k^\mathrm d \check a_k^\dagger \check a_k - \check a_k^\dagger \bm v_k^{\mathrm d \mathsf T}
        \begin{bmatrix}
            \check{\bm \psi}_\mathrm c \\
            \bm v_\mathrm i \\
            \check{\bm \psi}_\mathrm q \\
            \bm \eta_\mathrm o
        \end{bmatrix}
        -
        \begin{bmatrix}
            \check{\bm \psi}_\mathrm q^\mathsf T &
            \bm \eta_\mathrm o^\mathsf T
        \end{bmatrix}
        \bm u_k^\mathrm d
        \check a_k
    \right) \\ -
    \sum\limits_{n=1}^{M^\mathrm f}\sum\limits_{m=1}^{\varrho_n }\left(
    \imath \hbar \omega_n^\mathrm {f} \check{b}^\dagger_{nm} \check{b}_{nm} +
    \check{b}_{nm}^\dagger \bm v_{nm}^{\mathrm {f} \mathsf T} 
    \begin{bmatrix}
            \check{\bm \psi}_\mathrm q\\
            \bm \eta_\mathrm o
        \end{bmatrix}
    +
    \begin{bmatrix}
            \check{\bm \psi}_\mathrm q^\mathsf T &
            \bm \eta_\mathrm o^\mathsf T
        \end{bmatrix}
    \bm u_{nm}^\mathrm {f} \check{b}_{nm}
    \right) ,
    \label{eq:qdmess-liouvillian}
\end{multline}
\end{widetext}
where $\bm L^\mathrm{R}$ and $\bm L^\mathrm K$ are constant matrices, which describe the coupling between the quantum-classical and the quantum-quantum degrees of freedom of the core circuit and the source fields, respectively. Note that the ordering of the operators is such that the operators corresponding to the quantum degrees of freedom lie on the left, similarly to the creation operators of the auxiliary bosonic modes. The obtained Liouvillian by its form resembles a Hamiltonian of a bosonic impurity problem, where a nonlinear impurity is coupled to a continuum of harmonic modes. The most studied case in this context is the paradigmatic spin-boson problem~\cite{leggett1987dynamics}. An important difference to the situation is though that here the harmonic modes are discrete and have complex-valued frequencies and couplings.

The QD-MESS equation, which governs the dynamics of the dissipative quantum-electric circuit, reads as
\begin{equation}
    \imath \hbar \frac{\mathrm d}{\mathrm dt} |W\rrangle = \left.\check{\mathfrak L}\right|_{\bm \eta_\mathrm o = 0} |W\rrangle,
    \label{eq:qdmess}
\end{equation}
where $|W\rrangle$ is a multidimensional vector in an extended Liouville space, which completely describes the state of the core system and its correlations with the environment.
Equation~\eqref{eq:qdmess} has a conservation law 
\begin{equation}
     \llangle \hat 1 | \frac{\mathrm d}{\mathrm dt} | W\rrangle = 0,
     \label{eq:probability-conservation-law}
\end{equation}
where $\llangle \hat 1 |$ is a left zero eigenvector of the quantum operators of all dynamical variables $\check{\bm \psi}_\mathrm q$ and $\check{\bm q}_\mathrm q$, and of the creation operators $\check{\bm a}^\dagger$ and $\check{\bm b}^\dagger$ of all the auxiliary modes, i.e., it corresponds to the zero occupation number of each of these modes. Consequently, the proper ordering of the operators in the quantum Liouvillian~\eqref{eq:qdmess-liouvillian} results in
\begin{equation}
    \llangle \hat 1 | \left. \check{\mathfrak L}\right|_{\bm \eta_\mathrm o = 0} = 0.
\end{equation}
This left vector is equivalent to the vectorized identity operator in the Hilbert space of the core circuit and the product $\llangle \hat 1 | W \rrangle$ is equivalent to the trace operation of the full system-environment density operator. Therefore, Eq.~\eqref{eq:probability-conservation-law}
is equivalent to the conservation of probability for the standard Liouville--von Neumann equation. Thus, it is natural to choose a normalization of both $|W\rrangle$ and $\llangle \hat 1|$ such that $\llangle \hat 1 | W\rrangle = 1$. The thermal equilibrium state corresponds to a right zero eigenvector of the Liouvillian in the absence of classical and quantum sources fields
\begin{equation}
    \left.\check{\mathfrak L}\right|_{\bm \varphi_\mathrm i = 0,~\bm \eta_\mathrm o = 0} |W_\mathrm{eq}\rrangle = 0.
    \label{eq:qdmess-equilibrium}
\end{equation}

As we mention above, the QD-MESS Liouvillian equation~\eqref{eq:qdmess-liouvillian} has a formal resemblance to that of a many-body bosonic or continuous-variable Hamiltonians, which implies that it is not straightforward solve Eqs.~\eqref{eq:qdmess} and~\eqref{eq:qdmess-equilibrium}. However, note that in the Fock-state basis for auxiliary modes, this equation is equivalent to HEOM~\cite{QDMESS, xu2022taming}, where occupations numbers of the modes correspond to indices of auxiliary density operators. In practice, this means that all the numerical techniques which can be used for solving HEOM, such as tensor networks~\cite{shi2018efficient, borrelli2019density}, are also applicable for the QD-MESS framework.

The physical meaning of the components of $|W\rrangle$ may be non-trivial to identify, especially given that the choice of the degrees of freedom included in the system and in the environment may be arbitrary. This raises two issues: the preparation of the initial state for the simulations and the calculation of the probability distribution of observables. The first issue can be overcome by an explicit simulation of the initialization pulse, assuming that the state before the pulse is the equilibrium state $|W_\mathrm{eq}\rrangle$. Below, we introduce a method how to calculate mutual quasiprobability distributions of various observables.

\subsection{Observables}
Let us focus on a case where we apply an input field and measure the corresponding output field within a time interval $(t^\mathrm i, t^\mathrm f)$. Before $t^\mathrm i$, the open system is in the thermal state $|W_\mathrm{eq}\rrangle$. The result of a measurement is a time trace of the output field $\bm v_\mathrm o(t)$ sampled from a stochastic process. The observables $\bm o$ we are interested in are weighted averages of these fields over the above-mentioned time interval, and hence given by
\begin{equation}
    \bm o = \int\limits_{t^\mathrm i}^{t^\mathrm f} {\bm F}(t) \bm v_\mathrm o(t)\;\mathrm dt,
    \label{eq:observables}
\end{equation}
where $\bm F(t)$ is an arbitrary matrix-valued weight function. Since the output field is stochastic, these observables $\bm o$ are random variables. Their mutual quasiprobability distribution can be formally found as~\cite{kamenev2023field}
\begin{equation}
    p(\bm o) = \left.\left\{
    \delta \left[-\imath \hbar \int\limits_{t^\mathrm i}^{t^\mathrm f} \bm F(t) \frac{\delta}{\delta \bm \eta_\mathrm o^\mathsf T(t)}\;\mathrm dt - \bm o\right] \mathcal Z[\bm v_\mathrm i, \bm \eta_\mathrm o]
    \right\}\right|_{\bm \eta_\mathrm o = 0},
\end{equation}
where the argument of the delta function is a linear functional acting on the generating functional $\mathcal Z[\bm v_\mathrm i, \bm \eta_\mathrm o]$. The generating functional can be calculated by solving the QD-MESS equation~\eqref{eq:qdmess} with a non-zero quantum source
\begin{equation}
    \mathcal Z[\bm v_\mathrm i, \bm \eta_\mathrm o] = \llangle \hat 1 |\mathbb T \exp \left[-\frac{\imath}{\hbar} \int\limits_{t^\mathrm i}^{t^\mathrm f} \check{\mathfrak L}(t)\;\mathrm dt\right] |W_\mathrm{eq}\rrangle ,
\end{equation}
where $\mathbb T$ is a time-ordering symbol.
By employing a Fourier transformation of a multidimensional delta function,
we find that the quasiprobability density can be calculated as
\begin{equation}
    p(\bm o) = \int \mathrm e^{\imath \bm s^\mathsf T \bm o} \mathcal Z[\bm v_\mathrm i, \hbar \bm F^\mathsf T(t) \bm s]
    \;\frac{\mathrm d\bm s}{(2\pi)^{\operatorname{dim}(\bm o)}}.
    \label{eq:probability-density}
\end{equation}

Since $\mathcal Z[\bm v_\mathrm i, 0] = 1$ for an arbitrary drive field $\bm v_\mathrm i$, the quasiprobability density $p(\bm o)$ integrates to unity. However, it is not clear at this point whether it satisfies the non-negativity requirement for a probability density. For example, the Wigner quasiprobability function, which may have negative values, falls into this class of distribution functions if one chooses $\bm q_\mathrm c$ and $\bm \psi_\mathrm c$ as observables instead of the time-averaged output fields.

\section{Dispersive readout of a transmon qubit}

\label{sec:transmon-readout}

\begin{figure}[t]
    \centering
    \includegraphics[width=\linewidth]{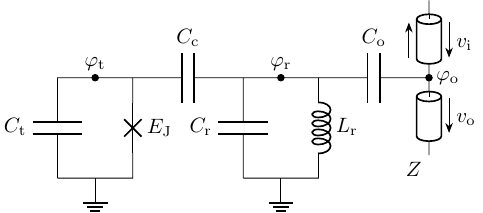}
    \caption{Circuit diagram of a dispersive-readout setup for a transmon qubit. The circuit consists of the qubit, capacitively coupled to a resonator, which is in turn capacitively coupled to two semi-infinite transmission lines.}
    \label{fig:transmon}
\end{figure}

\begin{table}[t]
    \caption{Parameters of the circuit shown in Fig.~\ref{fig:transmon} used for numerical calculations.}
    \label{tab:parameters}
    \begin{tabularx}{\linewidth}{c|c|c|c|c|c|c}
        \hline\hline
        $C_\mathrm t$\.(fF) &
        $E_\mathrm J/h$\.(GHz) &
        $C_\mathrm r$\;(fF)&
        $L_\mathrm r$\;(nH)&
        $C_\mathrm c$\;(fF)&
        $C_\mathrm o$\;(fF)&
        $Z$\;($\Omega$)\\
        \hline
        80 &
        12.5&
        570 &
        0.7 &
        40 &
        15 &
        50\\
        \hline\hline
    \end{tabularx}
\end{table}

\begin{table}[t]
    \caption{Frequencies of the dynamical auxiliary modes of the circuit of Fig.~\ref{fig:transmon} given the parameter values shown in Tab.~\ref{tab:parameters}.}
    \begin{tabularx}{\linewidth}{C|C|C}
        \hline\hline
        $\omega_1^\mathrm d/(2\pi)$ (GHz)&
        $\omega_2^\mathrm d/(2\pi)$  (GHz)&
        $\omega_3^\mathrm d/(2\pi)$  (GHz)\\
        \hline
        $7.61 - 1.63\imath \cdot 10^{-3}$ &
        $-7.61 -1.63\imath \cdot 10^{-3}$ &
        $-434\imath$ \\
        \hline \hline
    \end{tabularx}
    \label{tab:dyn-poles}
\end{table}

Here, we demonstrate our theory in an experimentally feasible scenario by applying it to a model of the dispersive readout of a superconducting qubit~\cite{blais2021circuit}. The purpose here is, first, to validate the new framework by reproducing results that can be obtained with conventional weak coupling approaches, second, to show that even in this limit the new approach provides new insight, and, third, to provide new results that cannot be obtained from a standard IO theory. 

As a specific system, we employ a typical transmon qubit~\cite{charge2007koch}, shown in Fig.~\ref{fig:transmon}, which is capacitively coupled to an off-resonant readout resonator. The resonator is capacitively coupled to two semi-infinite transmission lines, one of which is used to drive the readout mode and the other to measure the output field. No input is applied on the transmission line for the measurement and the output field to the drive line is disregarded. 

The retarded, advanced, and Keldysh components of self energy of the circuit~\eqref{eq:inv-greens-function} read as
\begin{multline}
    \bm \Sigma_\mathrm{circ}^\mathrm R(\omega) \\= 
        \begin{bmatrix}
            \omega^2 \tilde C_\mathrm t &
            -\omega^2 C_\mathrm c &
            0 &
            0 \\
            -\omega^2 C_\mathrm c &
            \omega^2 \tilde C_\mathrm r - \frac{1}{L_\mathrm r} &
            -\omega^2 C_\mathrm o & 0 \\
            0 &
            -\omega^2 C_\mathrm o &
            \omega^2 C_\mathrm o + \frac{2\imath}{Z}\omega &
            \frac{2}{Z} \\
            0 & 0 & -\imath \omega & 0
        \end{bmatrix},
\end{multline}
\begin{equation}
    \bm \Sigma_\mathrm{circ}^\mathrm A(\omega) = \left[\bm \Sigma^\mathrm R(\omega^\ast)\right]^\dagger,
\end{equation}
\begin{equation}
    \bm \Sigma_\mathrm{circ}^\mathrm K(\omega)= 
        \begin{bmatrix}
            0 &
            0 &
            0 &
            0 \\
            0 &
            0 &
            0 & 0 \\
            0 &
            0 &
            \frac{2\imath}{Z}\omega &
            -\frac{\imath}{2} \omega \\
            0 & 0 & -\frac{\imath}{2} \omega & \frac{\imath}{4} Z\omega
        \end{bmatrix}  \coth\left(\frac{\hbar \omega}{2 k_\mathrm B T}\right),
\end{equation}
where $\tilde C_\mathrm t = C_\mathrm t + C_\mathrm c$ and $\tilde C_\mathrm r = C_\mathrm r + C_\mathrm c + C_\mathrm o$, and the parameters of the circuit are introduced in Fig.~\ref{fig:transmon}. The first $3\!\times\! 3$ block describes intrinsic dynamics of the circuit degrees of freedom $\varphi_{\mathrm t,\mathrm c/\mathrm q}$, $\varphi_{\mathrm r,\mathrm c/\mathrm q}$, and $\varphi_{\mathrm o,\mathrm c/\mathrm q}$, see Fig.~\ref{fig:transmon}. The rest of the matrix provides coupling to the external field. Since we use only one line for drive and one line for measurement, there is only one input field $v_\mathrm i(t)$ and one counting field $\eta_\mathrm o(t)$.
The dynamical degree of freedom of the core circuit is represented by transmon flux $\psi_{\mathrm c/ \mathrm q} = \varphi_{\mathrm t, \mathrm c/\mathrm q}$, the interface dynamical degrees of freedom $\bm \psi'_{\mathrm c/\mathrm q} = \begin{bmatrix} \varphi_{\mathrm r, \mathrm c/ \mathrm q} & \varphi_{\mathrm o, \mathrm c/\mathrm q}\end{bmatrix}^\mathsf T$ are to be eliminated, i.e., the transformation matrix $\bm Q$ of Eq.~\eqref{eq:core-interface-splitting} is an identity matrix. 
The poles of the retarded component of the core self-energy $\omega_k^\mathrm d$ satisfy the following equation
\begin{equation}
    \left[\left(\omega^\mathrm d_k\right)^2 \tilde C_\mathrm r - \frac{1}{L_\mathrm r}\right] \left(\omega^\mathrm d_k C_\mathrm o + \frac{2\imath}{Z}\right) - \left(\omega^\mathrm d_k\right)^3 C_\mathrm o^2 = 0.
\end{equation}
Two of the resulting dynamical auxiliary modes have frequencies close to $\pm 1 / \sqrt{L_\mathrm r\tilde C_\mathrm r}$ with small negative imaginary parts and the third one has a purely imaginary frequency around $-2 \imath / (Z C_\mathrm o)$. For our calculations we use the parameters of the circuit given in Tab.~\ref{tab:parameters}. Table~\ref{tab:dyn-poles} shows the frequencies of the dynamical auxiliary modes obtained for these parameters.

It appears to be more convenient to use coupling to the auxiliary modes through voltage $\dot \psi$ instead of flux coupling shown in Eq.~\eqref{eq:flux-qdmess-lagrangian}. This leads to the coupling through the charge of the transmon island to the auxiliary modes in the Liouvillian. More details can be found in Appendix~\ref{sec:charge-qdmess}.
Since the coupling is weak, we may apply the dynamical Schrieffer--Wolf approach presented in Appendix~\ref{sec:schrieffer-wolf} for all of the fluctuation auxiliary modes and the strongly damped dynamical auxiliary mode. We restrict the linear space of $\check{\mathfrak L}^{(0)}(t)$ and the coupling operators in the Liouvillian~\eqref{eq:qdmess-liouvillian-general} to the four lowest levels for the isolated transmon and to six levels for the weakly dissipative dynamical auxiliary modes. See more technical details on the construction of the Liouvillian in Appendix~\ref{sec:qdmess-approximation}.

\subsection{Transmission coefficient}

\begin{figure*}[t]
    \includegraphics[width=\linewidth]{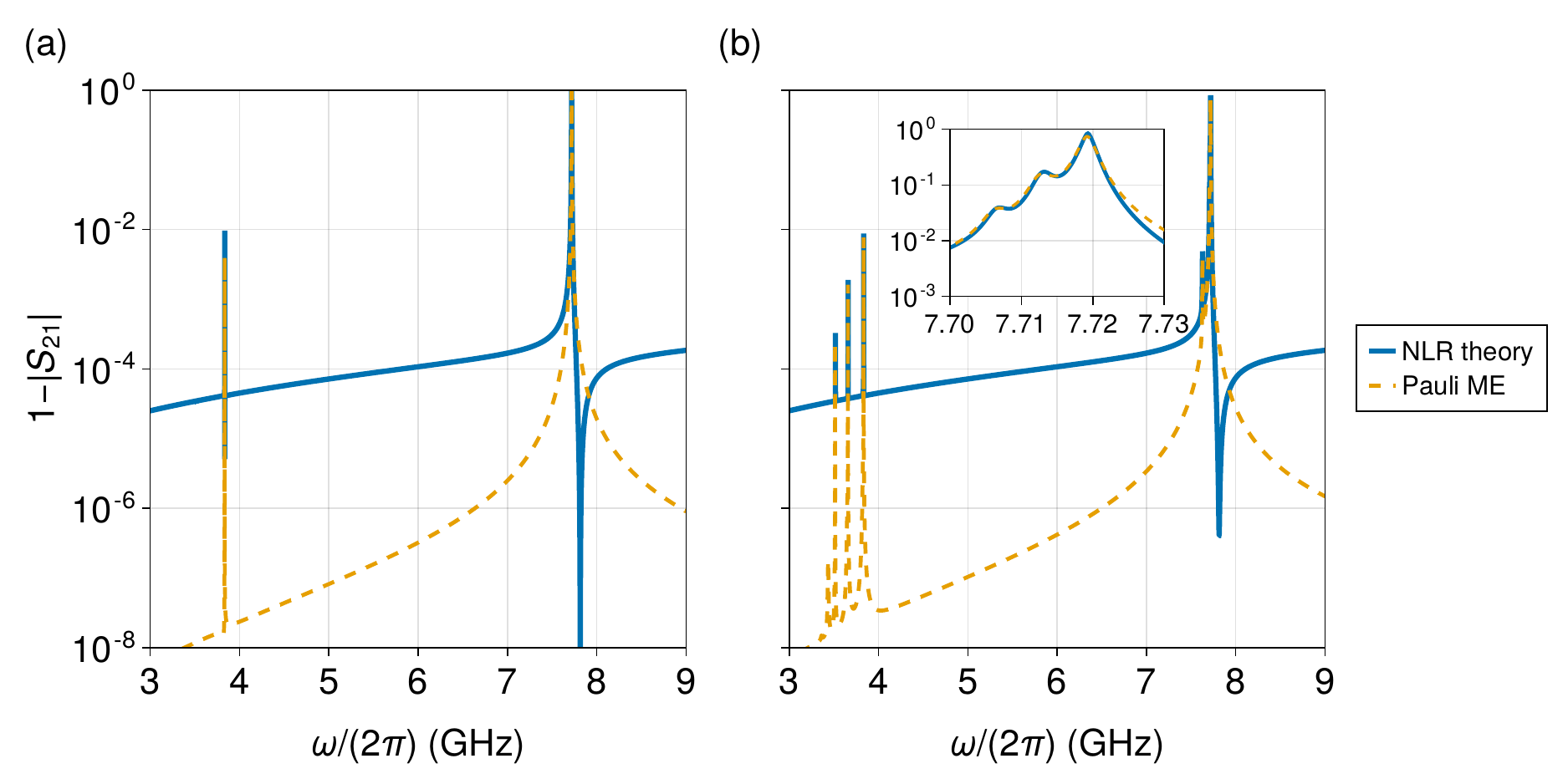}
    \caption{Absolute value of the transmission coefficient $S_{21}(\omega)$ calculated using NLR theory (blue solid line) and Pauli master equation (orange dashed line) for (a) cold transmon at temperature $T = 0.01$~K and (b) hot transmon at temperature $T = 0.1$~K. The parameters of the circuit are given in Tab.~\ref{tab:parameters}. The frequencies and linewidths of the qubit and the resonator, qubit anharmonicity, and dispersive shift are shown in Tab.~\ref{tab:spectrum}. The local Hilbert spaces of the transmon and the readout resonator were truncated to four and six lowest levels, respectively.}
    \label{fig:transmon-S21}
\end{figure*}

\begin{table}[t]
\caption{Frequency $\omega_\mathrm t$, linewidth $\gamma_\mathrm t$, and anharmonicity $\alpha$ of the transmon, frequency $\omega_\mathrm r$ and linewidth $\gamma_\mathrm r$ of the resonator, and the dispersive shift $\chi$, obtained from the diagonalization of the system Liouvillian after the Schrieffer--Wolf transformation. The qubit linewidth is given at low (top cell) and high (bottom cell) temperatures. 
}
\label{tab:spectrum}
\begin{tabularx}{\linewidth}{C|C|C}
    \hline
    \hline
    $\omega_\mathrm t / (2\pi)$ (GHz)&
    $\gamma_\mathrm t / (2\pi)$ (kHz)&
    $\alpha / (2\pi) $ (MHz) \\
    \hline
    \multirow{2}{*}[0em]{3.834} &
    $1$ (0.01~K) &
    \multirow{2}{*}[0em]{$-175$}
     \\
     \cline{2-2}
    &
    $69$ (0.1~K)&
    \\
    \hline\hline
    $\omega_\mathrm r / (2\pi)$ (GHz)&
    $\gamma_\mathrm r / (2\pi)$ (MHz)&
    $\chi / (2\pi)$ (MHz) \\
    \hline
    {7.719}&
    {$1.71$} &
    {$-6.4$} \\
    \hline \hline
\end{tabularx}
\end{table}

As the first step we probe the spectrum of the device by calculating the transmission coefficient $S_{21}(\omega)$.
It is defined as $\langle v_\mathrm o(\omega)\rangle = S_{21}(\omega)  v_\mathrm i(\omega)$ for infinitesimal drive $v_\mathrm i$, where the average is taken over the realizations of the output field. It can be evaluated as a second variation of the generating functional~\eqref{eq:generating-functional} with respect to the input and the counting fields~\cite{kamenev2023field}
\begin{equation}
    S_{21}(\omega) = \left.-\imath \hbar \frac{\delta^2 \mathcal Z[v_\mathrm i, \eta_\mathrm o]}{\delta v_\mathrm i(\omega) \delta \eta_\mathrm o^\ast(\omega)} \right|_{v_\mathrm i = 0,~\eta_\mathrm o=0}.
\end{equation}
In order to calculate it, we apply the Schrieffer--Wolf transformation for $v_\mathrm i = 0$ and $\eta_\mathrm o = 0$ to the static part of Liouvillian and to the terms which provide coupling to the input and counting fields $v_\mathrm i$ and $\eta_\mathrm o$. We find the steady state of the resulting Liouvillian and use the first-order perturbation theory with respect to the input field $v_\mathrm i$ to evaluate the tramsmission coefficient. The details of the calculations are presented in Appendix~\ref{sec:transmon-S21-calculation}.

The frequency dependence of the absolute value of $S_{21}(\omega)$ is shown in Fig.~\ref{fig:transmon-S21} at (a) low temperature of $T = 0.01$~K and (b) high temperature of $T = 0.1$~K. At the low temperature, the circuit is close to its ground state and there are only two transitions visible. There is a strong feature at the resonator frequency $\omega_\mathrm r$ and a weak signature at the transmon frequency $\omega_\mathrm t$. At the high temperature, both the qubit and the resonator become thermally populated, and consequently more transitions become visible. In Fig.~\ref{fig:transmon-S21}(b), we can clearly observe lines which correspond to transitions between different transmon levels as well as the splitting of the resonator feature caused by the dispersive shift (see the inset). The spurious feature at the frequency slightly below the main resonator peak arises due to the hard cut-off of the transmon levels which results in the increased dispersive shift for the highest considered transmon level. The positions of the peaks and their widths correspond to the real and imaginary parts of the eigenvalues of the Liouvillian $\check{\mathfrak L}|_{v_\mathrm i = 0,\eta_\mathrm o = 0}$ in the absence of the drive. The results are shown in Tab.~\ref{tab:spectrum}. Note that the linewidth of the qubit transition is increased at high temperature which would result in the lower $T_2$ time of the qubit.

\subsection{Comparison to Markovian input-output theory}
As already mentioned, one purpose of the calculation of the transmission coefficient in the transmon readout setup is to explicitly demonstrate that the developed NLR theory provides valid results also in the domain of weak coupling. Therefore, in this subsection, we provide a comparison between NLR and the standard Markovian input-output theory. We consider $\varphi_\mathrm r$ and $\varphi_\mathrm t$ to be the system degrees of freedom, and the flux of the output node $\varphi_\mathrm o$ we include in the bath. The Hamiltonian of the system reads as
\begin{multline}
    \hat H_\mathrm{tr} = \frac{\tilde C_\mathrm r\hat q_\mathrm t^2  + 2 C_\mathrm c \hat q_\mathrm t \hat q_\mathrm r + \tilde C_\mathrm t\hat q_\mathrm r^2 }{2\left(\tilde C_\mathrm r \tilde C_\mathrm t - C_\mathrm c^2\right)}\\ - E_\mathrm J \cos \left(\frac{2\pi \hat \varphi_\mathrm t}{\Phi_0}\right) + \frac{\hat \varphi_\mathrm r^2}{2L_\mathrm r} + C_\mathrm o \dot v_\mathrm i(t) \hat \varphi_\mathrm r.
\end{multline}
The system is coupled to the bath with a super-Ohmic spectral density
\begin{equation}
    J(\omega) = \frac{1}{\pi}\frac{\omega^3 \tau_{RC} C_\mathrm o}{1 + \left(\omega \tau_{RC}\right)^2}
\end{equation}
via operator $\hat \varphi_\mathrm r$. Here, $\tau_{RC} = C_\mathrm o Z / 2$ is a time constant which provides a crossover to an Ohmic spectral density at high frequencies. The average output field in the Fourier domain is given by
\begin{equation}
    \left \langle v_\mathrm o (\omega) \right \rangle = -\tau_{RC}\omega^2 \left\langle \hat \varphi_\mathrm r(\omega)\right\rangle  + v_\mathrm i(\omega),
\end{equation}
where
\begin{equation}
    \begin{gathered}
    \langle \varphi_\mathrm r(t)\rangle = \operatorname{Tr}\left[ \hat \rho(t) \hat \varphi_\mathrm r\right],\\
    \langle \hat \varphi_\mathrm r(\omega) \rangle = \int\limits_{-\infty}^{+\infty} \langle \varphi_\mathrm r(t)\rangle e^{\imath \omega t}\;\mathrm dt,
    \end{gathered}
\end{equation}
and $\hat \rho(t)$ is the density operator of the system at time $t$.
Since we consider infinitesimally weak drive, we evaluate dissipators for stationary Hamiltonian.

We employ the standard Born--Markov and rotating-wave approximations~\cite{blais2021circuit}, which results in a Pauli master equation for the occupations and relaxation equations for coherences~\cite{gardiner2004quantum} and use first order perturbation theory with respect to the drive (see Appendix~\ref{sec:pauli} for the details).
The transmission coefficient calculated using Eq.~\eqref{eq:s21-lindblad} is shown in Fig.~\ref{fig:transmon-S21}. We indeed observe  a quantitative agreement between the two approaches in the \emph{vicinity} of the resonances. However, there is a clear discrepancy in the behavior \emph{away} from the resonances, where the NLR theory predicts lower transmission coefficients. Agreement is also seen for the frequencies of the qubit resonances and for features corresponding to the dispersively shifted resonator mode. To provide the most direct comparison, we have truncated the local Hilbert space of the transmon to four levels. Practically, the observed discrepancy away from the resonances can be eliminated  by introducing  in the Markovian treatment \emph{ad hoc} an additional $RC$-filter with a time constant $\tau_{RC}$ to the input and output voltages. In the presented Markovian approach this filtering is not taken into account, since $\tau_{RC}$ is supposed to be the shortest time scale in the system. However, within the framework of the NLR theory such a filtering appears as an integral part of the formulation and needs not be added by hand. We conclude that for weakly coupled superconducting circuits the presented NLR theory based on the QD-MESS reproduces results of the Markovian IO theory close to resonances  but provides a more consistent treatment over the entire frequency range.


\subsection{Readout of a hot transmon}

\label{sec:single-shot-ro}

\begin{figure*}[t]
    \includegraphics[width=\linewidth]{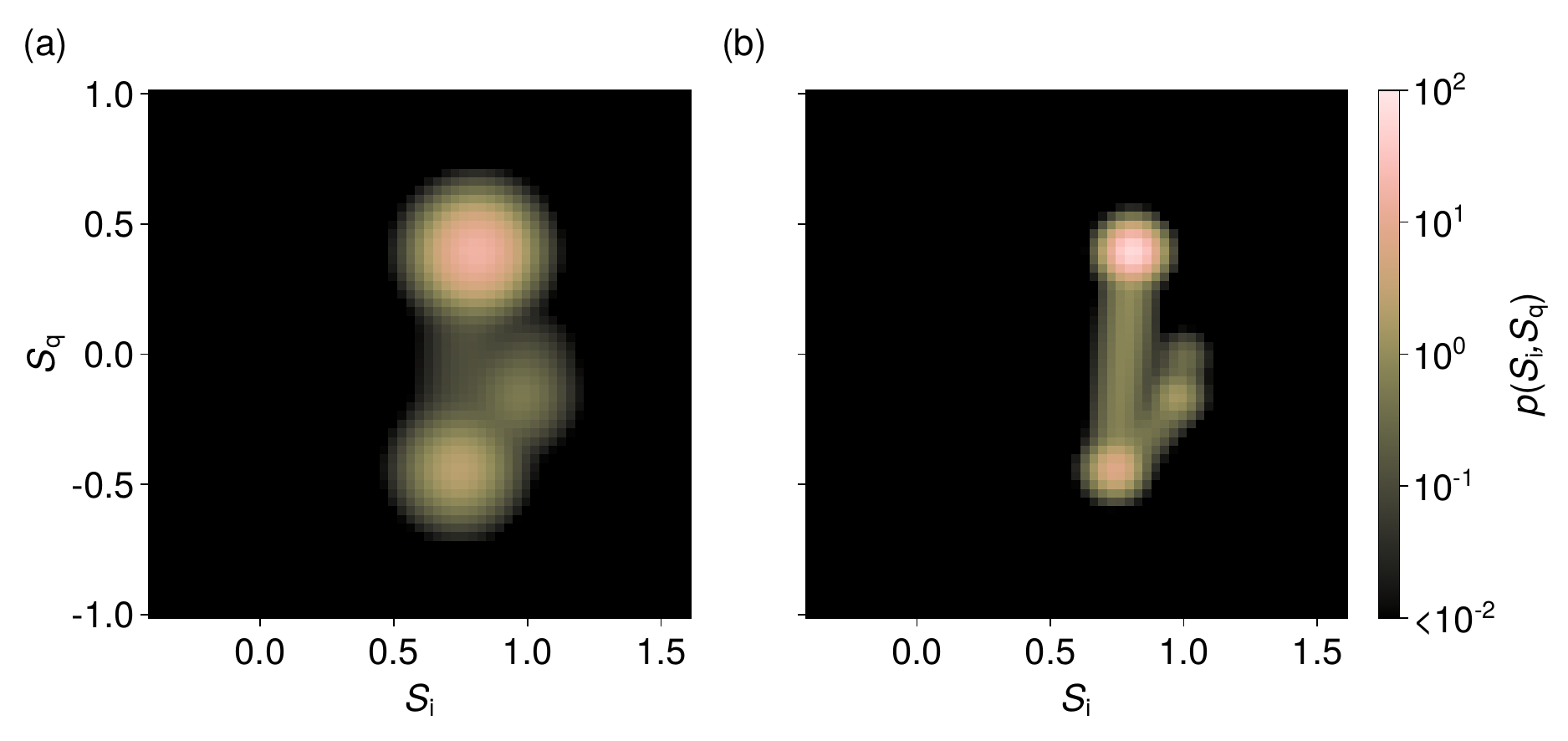}
    \caption{Quasiprobability density of the normalized in-phase and quadrature components of the output field. The temperature of the qubit environment is $T=0.1$~K, the resonator drive frequency is $\omega_\mathrm d = 2\pi \times 7.7146$~GHz, the drive amplitude is $V_\mathrm d = 0.2$~$\mu$V, and the integration time defined in Eq.~\eqref{eq:s21-integ} is (a) $t_\mathrm{int} = 5.18$~$\mu$s and (b) $t_\mathrm{int} = 20.75$~$\mu$s.}
    \label{fig:transmon-single-shot}
\end{figure*}

Next, we proceed to the simulation of the state of the output field during transmon readout. For the sake of simplicity, we do not consider the preparation of the qubit state and the simulation of its subsequent dynamics with respect to the readout pulse. Instead, we consider a continuous monochromatic drive $v_\mathrm i(t) = V_\mathrm d \cos \left(\omega_\mathrm d t\right)$ and the qubit-resonator system at temperature $T$ relaxed to the steady state with respect to the drive. The details of the calculation can be found in Appendix~\ref{sec:qdmess-periodic}. Note that such a readout is not an instantaneous von Neumann-type measurement and that the standard Born approximation for the calculation of probabilities cannot be applied.

The quantity of interest is the homodyne demodulated signal averaged over a finite time interval of length $t_\mathrm{int}$, given by
\begin{equation}
    S_\mathrm i + \imath S_\mathrm q = \frac{2}{t_\mathrm{int} V_\mathrm d} \int\limits_0^{t_\mathrm{int}} v_\mathrm o(t) \mathrm e^{\imath \omega_\mathrm d t}\;\mathrm dt.\label{eq:s21-integ}
\end{equation}
The real and imaginary components $S_\mathrm i$ and $S_\mathrm q$ correspond to in-phase and quadrature components of the output field, respectively,  normalized by the drive amplitude. These components are measured also in the single-shot readout experiments although typically using a weighted time average over the duration of the readout pulse. 
The quantities $S_\mathrm i$ and $S_\mathrm q$ belong to the class of observables defined in Eq.~\eqref{eq:observables} with
\begin{equation}
\bm o = \begin{bmatrix} S_\mathrm i \\ S_\mathrm q \end{bmatrix}
\end{equation}
and
\begin{equation}
    \bm F(t) = \frac{2 \Theta(t) \Theta(t_\mathrm{int} - t)}{t_\mathrm{int} V_\mathrm d}\begin{bmatrix}
        \cos (\omega_\mathrm d t) \\
        \sin (\omega_\mathrm d t)
    \end{bmatrix}
    ,
    \label{eq:weight-homodyne}
\end{equation}
where $\Theta(t)$ is the Heaviside step function.

We calculate the quasiprobability density $p(S_\mathrm i, S_\mathrm q)$ using Eqs.~\eqref{eq:weight-homodyne} and~\eqref{eq:probability-density} and show the results and simulation parameters in Fig.~\ref{fig:transmon-single-shot}. We can clearly distinguish four disk-shaped clouds which correspond to the different transmon states. The intensities of the clouds are proportional to the thermal populations of the qubit states. In addition there are faint lines which connect different clouds. These lines correspond to the spontaneous transitions between the qubit states which may occur during the integration time~\cite{gambetta2007protocols}. Similar connecting lines have indeed been observed in experiments~\cite{gunyho2023singleshot}.

The value of $t_\mathrm{int}$ is pivotal for the accurate readout protocol. For long integration times, the spontaneous transitions become likely and in the limit $t_\mathrm{int} \to \infty$, the quasi-probability distribution collapses into a delta function centered at a single point where $S_{\mathrm i}+\imath S_{\mathrm q} = S_{21}(\omega_\mathrm d)$. In the opposite limit, with the decrease of $t_\mathrm{int}$, the clouds begin to grow and, at some point, overlap. Therefore, there is an optimal integration time, not too short such that the clouds can still be distinguishable, and not too long such that probability of spontaneous transition during the measurement is small.

The choice of the qubit readout problem, in addition to the purely illustrative purpose, is also motivated by the problem of the quantum measurement. The qubit readout falls into a class of continuous inefficient measurements~\cite{jacobs2014quantum} since the state of the qubit is not probed directly, but through averaging the output field over a finite time interval.
In a contrast to the standard presentation of  measurements~\cite{jacobs2014quantum}, we do not consider an initial product state of the system and the environment or the measurement device, but a thermal initial state. Moreover, we do not separate the system and the environment but study the open quantum system as a whole. This approach may be favorable outside the weak coupling regime, where the product initial states gives rise to non-exponential short-time decoherence~\cite{braun2001universality, tuorila2019system}.

\section{Resistively shunted junction}

\label{sec:rsj}

\begin{figure}[t]
    \includegraphics[width=\linewidth]{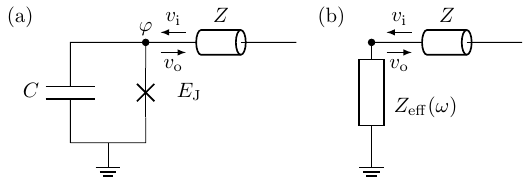}
    \caption{(a) Circuit diagram of a capacitively and resistively shunted junction. The resistor is replaced by a transmission line of characteristic impedance $Z$. (b) Equivalent linear circuit to that in (a) where the junction and its capacitor are replaced by a general frequency-dependent effective impedance such that the reflection coefficient $S_{11}(\omega)$ remains unchanged.}
    \label{fig:rsj}
\end{figure}

\begin{figure*}[t]
    \includegraphics[width=\linewidth]{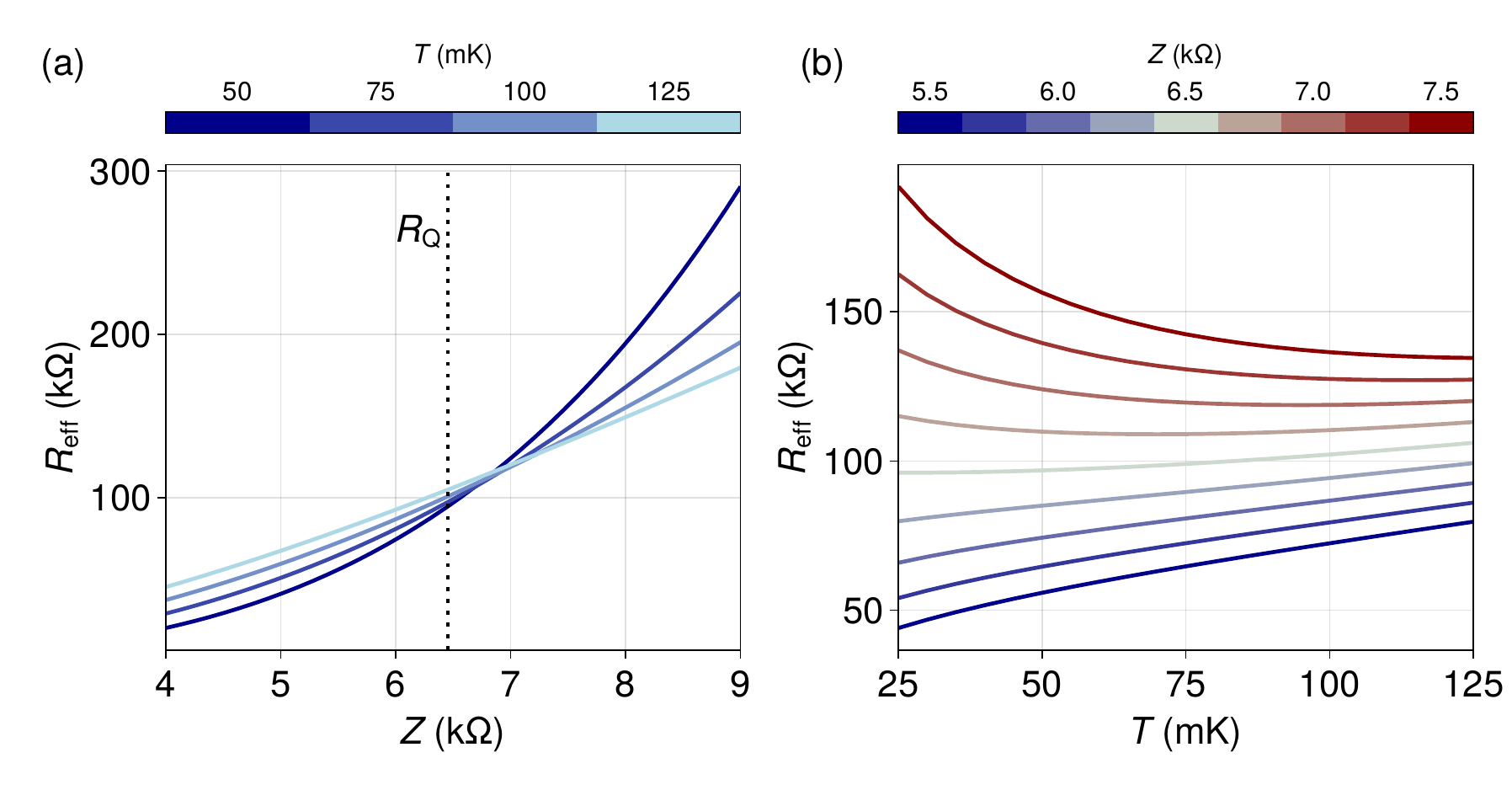}
    \caption{(a) Effective resistance of the junction as function of impedance of the transmission line calculated using Eq.~\eqref{eq:effective-resistance} for different temperatures. (b) Effective resistance as function of temperature calculated for different impedances of the transmission line. The parameters of the circuit in Fig.~\ref{fig:rsj} are $C = 1.43$~fF and $E_\mathrm J / h = 2.5$~GHz, which correspond to experimental values from Ref.~\cite{Murani2020}.}
    \label{fig:rsj-results}
\end{figure*}

To illustrate the applicability of the NLR theory in a highly non-trivial situation beyond the weak-coupling regime, we consider a capacitively and resistively shunted Josephson junction~\cite{schon1990quantum}, where the role of the resistor is played by a transmission line, see the circuit diagram in Fig.~\ref{fig:rsj}(a). A quantum phase transition between superconducting and insulating phases of the junction at $Z<R_\mathrm Q:= h / (2e)^2$ and $Z>R_\mathrm Q$, respectively, was predicted for this system by Schmid~\cite{Schmid1983} and Bulgadaev~\cite{Bulgadaev1984phase}. 
This quantum phase transition has been experimentally observed~\cite{Penttilae1999, Yagi1997, Penttilae2001, Liu2009}. However, a recent microwave experiment hints towards the absence of the insulating phase~\cite{Murani2020}. This result has triggered a lively discussion in the community~\cite{Hakonen2021, Murani2021} and new  theoretical and experimental studies of the resistively shunted junction~\cite{Houzet2024, Burshtein2024, Yamamoto2024a, Giacomelli2024, Leger2023, Kuzmin2023, Subero2023}.

We employ the developed NLR theory to calculate the reflection coefficient $S_{11}(\omega)$ which relates the average reflected field to the infinitesimal input field by
\begin{equation}
    \langle v_\mathrm o(\omega)\rangle = S_{11}(\omega) v_\mathrm i(\omega).
\end{equation}
The scattering coefficient $S_{11}(\omega)$ is identical to that of a linear circuit shown Fig.~\ref{fig:rsj}(b), given the proper choice of effective impedance
\begin{equation}
    Z_\mathrm{eff}(\omega) = Z \frac{1 + S_{11}(\omega)}{1 - S_{11}(\omega)}.
    \label{eq:effective-resistance}
\end{equation}
We restrict our analysis to the low frequency limit $\omega \to 0$ where this effective impedance transforms into an active effective resistance $R_\mathrm{eff} = Z_\mathrm{eff}(\omega)|_{\omega = 0}$. This resistance provides the relation between the dc voltage across the junction and the current injected through the transmission line. Since the Josephson junction presents a nonlinear element, its effective resistance depends on the parameters of the environment, namely its impedance and temperature. The aim of this section is to calculate the effective resistance $R_\mathrm{eff}$ and discuss its behavior in the context of the previous studies on this system.

From the point of view of our nonlinear-response theory, the core subsystem is given by the junction itself and there are no explicit dynamical modes. 
The quantum Liovillian governing the dynamics of the lossy Josephson junction reads as
\begin{widetext}
\begin{multline}
     \check{\mathfrak L} = \left(\check q_\mathrm q + \frac{\check \psi_\mathrm q}{Z} - \eta_\mathrm o\right)\frac{\check q_\mathrm c}{C} + 2 E_\mathrm J \sin \left(\frac{\pi \check \psi_\mathrm q}{\Phi_0}\right) \sin \left(\frac{2 \pi \check \psi_\mathrm c}{\Phi_0}\right)
     - \frac{2 \check \psi_\mathrm q v_\mathrm i}{Z} -\frac{\imath k_\mathrm B T}{\hbar Z} \left(\check \psi_\mathrm q - \frac{Z \eta_\mathrm o}{2}\right)^2 + v_\mathrm i \eta_\mathrm o\\- \imath \hbar \sum\limits_{n=1}^{M^\mathrm f} \omega_n^\mathrm{f}\left[
    \check b_n^\dagger - \sqrt{\frac{r_n^\mathrm{f}}{\hbar \omega_n^\mathrm{f} Z}} \left(\check \psi_\mathrm q - \frac{Z \eta_\mathrm o}{2}\right)\right] \left[
    \check b_n - \sqrt{\frac{r_n^\mathrm{f}}{\hbar \omega_n^\mathrm{f} Z}} \left(\check \psi_\mathrm q - \frac{Z \eta_\mathrm o}{2}\right)\right].
    \label{eq:rsj-liouvillian}
\end{multline}
\end{widetext}
The details of the mathematical derivation are shown in Appendix~\ref{sec:rsj-derivation-qdmess}.
Note that the Liouvillian is periodic with respect to $\check \psi_\mathrm c$, resulting in
\begin{equation}
    \exp \left(\frac{\imath}{\hbar} \Phi_0 \check q_\mathrm q\right) \check{\mathfrak L} \exp \left(-\frac{\imath}{\hbar} \Phi_0 \check q_\mathrm q\right) = \check{\mathfrak L}.
\end{equation}
Therefore, we treat this variable as compact and the canonically conjugated quantum charge $\check q_\mathrm q$ as discrete. The classical charge $\check q_\mathrm c$ operator has a continuous spectrum, since the Liouvillian is not periodic with respect to its conjugate $\check \psi_\mathrm q$. The expectation value of the output field in the state $|W\rrangle$ can be found as
\begin{equation}
    \begin{gathered}
        \langle v_\mathrm o\rangle = \llangle \hat 1 | \check V_\mathrm c  |W\rrangle - v_\mathrm i,~
        \check V_\mathrm c = -\left.\frac{\partial \check{\mathfrak L}}{\partial \eta_\mathrm o}\right|_{v_\mathrm i = 0, \eta_\mathrm o = 0},\\
        \llangle \hat 1 | \check V_\mathrm c = \llangle \hat 1 |\left[\frac{\check q_\mathrm c}{C} +\frac{\imath}{2} \sum\limits_{m=1}^M \sqrt{\hbar \omega_n^\mathrm{f} r_n^\mathrm{f} Z} \check b_n\right].
    \end{gathered}
\end{equation}
Here, we used the fact that $\llangle \hat 1|$ is a common left eigenvector of $\check \psi_\mathrm q$ and $\check b_n^\dagger$, corresponding to zero eigenvalue.

First, we find the equilibrium state $|W_\mathrm{eq}\rrangle$ which is solution of Eq.~\eqref{eq:qdmess-equilibrium}. Then, using the first-order perturbation theory, we can evaluate reflection coefficient as
\begin{equation}
    S_{11}(\omega) = -1 + \llangle \hat 1 | \check V_\mathrm c \left(\omega - \left. \check{\mathfrak L}\right|_{v_\mathrm i = 0, \eta_\mathrm o = 0}\right)^{-1} \check V_\mathrm q|W_\mathrm{eq}\rrangle,
    \label{eq:rsj-reflection}
\end{equation}
where
\begin{equation}
    \check V_\mathrm q = \left.\frac{\partial \check{\mathfrak L}}{\partial v_\mathrm i}\right|_{v_\mathrm i = 0, \eta_\mathrm o = 0} = -\frac{2 \check \psi_\mathrm q}{Z}.
\end{equation}

Solving the stationary QD-MESS equation~\eqref{eq:qdmess-equilibrium} with the Liouvillian~\eqref{eq:rsj-liouvillian} is quite challenging even numerically due to the high dimension of the extended Liouville space in the presence of multiple auxiliary modes. Here, advanced methods based on matrix product states could potentially by adapted \cite{xu2022taming}. To make progress here, we first split the full Liouvillian into a sum of the Josephson term and a remainder as
\begin{equation}
    \begin{gathered}
        \left. \check{\mathfrak L}\right|_{v_\mathrm i, \eta_\mathrm o = 0} = \check{\mathfrak L}_0 + \check{\mathfrak L}_\mathrm J, \\
        \check{\mathfrak L}_\mathrm J = 2 E_\mathrm J \sin \left(\frac{\pi \check \psi_\mathrm q}{\Phi_0}\right) \sin \left(\frac{2\pi \check \psi_\mathrm c}{\Phi_0}\right).
    \end{gathered}
    \label{eq:rsj-liouvillian-perturb}
\end{equation}
Then, we consider the case of low Josephson energies and use the second-order perturbation theory with respect to $E_\mathrm J$ to evaluate the reflection coefficient. In contrast to the perturbation theory used in the previous section, this perturbative approach fully accounts the strong dissipation induced by the resistive shunt. In the stationary limit introduced above, we obtain
\begin{widetext}
\begin{multline}
    \left.S_{11}(\omega)\right|_{\omega=0} = 1 - \frac{4\pi \zeta \tau^2_{RC} E_\mathrm J^2}{\hbar^2}  \int\limits_{0}^{+\infty} \tau \sin\left[\pi \zeta \left(1 - \mathrm e^{-\tau}\right)\right] \\ \times 
    \exp\left\{\frac{2 \pi k_\mathrm B T \zeta \tau_{RC}}{\hbar } \left(1 - \tau - \mathrm e^{-\tau}\right)+
    \sum\limits_{n=1}^{M^\mathrm f} 
    \frac{2 \pi r_n^\mathrm f \zeta \tau_{RC} \left[\left(\mathrm e^{-\tau} - 1\right) - \left(\nu_n^{\mathrm f}\right)^{- 1} \left(\mathrm e^{-\nu_n^{\mathrm{f}} \tau} - 1\right)\right]}{\left(\nu_n^\mathrm{f}\right)^2 - 1}
    \right\} \;\mathrm d\tau,
    \label{eq:rsj-static-reflection}
\end{multline}
\end{widetext}
where $\zeta = Z / R_\mathrm Q$, $\tau_{RC} = ZC$, and $\nu_n^\mathrm f = \tau_{RC} \omega_n^\mathrm f$. This expression is valid provided $|1 - S_{11}(\omega)|_{\omega=0} \ll 1$ which is equivalent to $R_\mathrm{eff} \gg Z$. The details of the derivation are presented in Appendix~\ref{sec:rsj-solving-qdmess}.

We numerically evaluate the integral in Eq.~\eqref{eq:rsj-static-reflection} and obtain the effective resistance of the junction using Eq.~\eqref{eq:effective-resistance}. We study the effective resistance as function of the impedance of the environment and of the temperature, with our results summarized in Fig.~\ref{fig:rsj-results}. The validity of the employed perturbation theory for the considered parameters is justified by the results, since effective resistance $R_\mathrm{eff}$ varies from tens to hundreds of $\mathrm k\Omega$ while the impedance of the shunt $Z$ varies within the range of few $\mathrm k\Omega$. At low values of the shunt resistance $Z$, the junction shows metallic behavior where its effective resistance decreases with the decreasing temperature. At high $Z$, the effective resistance increases with decreasing temperature, which corresponds to insulating behavior. The crossover point which separates these two regimes turns out to be temperature dependent and lie at a somewhat higher impedance than theoretically predicted value of $R_\mathrm Q \approx 6.453$~k$\Omega$, which we expect to be a finite temperature effect. Therefore, we conclude that these results provide a signature of the Schmid--Bulgadaev transition at non-zero temperatures.

\section{Conclusions}\label{sec:conclusions}

We presented an exact nonlinear-response theory, free of the limiting requirements of weak coupling, low temperature, and weak drive. The dissipation and the driving field introduced by the coupled transmission lines is reduced to the coupling of the non-linear system degrees of freedom to a discrete set of auxiliary bosonic modes. We split the modes into two classes: dynamical and fluctuation auxiliary modes. The first ones provide the coupling to the classical field and the latter are responsible for the thermal and quantum fluctuations in the circuit. This results in a time-local QD-MESS equation which fully describes the open-quantum-system dynamics of the circuit. This introduced equation has great potential in accurately and feasibly modeling quantum systems beyond the reach of the previous methods. 

For practical purposes, we introduced observables as weighted time averages of the output fields emitted to the transmission lines and derived an analytical expression of their mutual quasiprobability distribution based on the Fourier transformation of the generating functional. This formalism allows to calculate statistics of homo- and heterodyne demodulated signals without the need for stochastic sampling of the output field as well as to calculate broadband characteristics of the output field.

Importantly, we illustrated the consistency of our findings with those of the known Markovian methods by studying the dispersive readout of a superconducting transmon qubit. Convincingly, we found good agreement between our theory and the standard Markovian IO theory. The technical challenge of solving QD-MESS was avoided by applying a dynamical Schrieffer--Wolf transformation to lower the dimension of the extended Liouville space of the system. 

Furthermore, we thoroughly analyzed capacitively and resistively shunted Josephson junctions for different values and temperatures of the resistance. Interestingly, we observed signatures of the Schmid--Bulgadaev phase transition at non-zero temperatures, a topic that has been under an intense debate recently. Our method free of questionable approximations in this case provides direct instructions for achieving corresponding experimental observations and may resolve the debate. This example also shows that our theory is useful also in the regime where the Markovian methods fail.

The developed NLR theory provides a powerful framework for the theoretical analysis of open quantum systems such as superconducting quantum circuits, including extraction of experimentally observable quantities. Going beyond the weak-coupling regime can in many cases be achievable by solving the QD-MESS equation with advanced numerical techniques such as tensor networks. In principle, the developed formalism can be applied not only to lumped-element circuits, but also to distributed elements. The necessary self-energies can be computed explicitly using high-frequency simulation software for linear components of the distributed-element circuits. This procedure paves the way for exact quantum simulations of practical devices.

In the future, the presented formalism can also be modified for the calculation of the thermodynamic properties of the circuit. A non-equilibrium case can be simulated by introducing fluctuation auxiliary modes corresponding to different temperatures. Here, the counting field corresponding to the heat flow couples to the classical expression for Joule losses which are quadratic in terms of the microwave field in the circuit. Such a generalization may allow to study heat transport in non-equilibrium systems and implement accurate simulations of quantum heat engines.

Our treatment of open quantum systems is free from explicit system-environment separation and from an initial product state condition. Thus it may turn invaluable for the development of the measurement theory in the regime of intermediate and strong coupling between the system and the measurement device. Deeper understanding of the quantum measurement in general is of key importance to the whole description and interpretation of natural phenomena in general.

\acknowledgments
V.V. thanks Kalle Kansanen and Riya Baruah for fruitful discussions. This work has been financially supported by the Academy of Finland Centre of Excellence program (project no.~336810) and THEPOW (project no.~349594), the European Research Council under Advanced Grant no.~101053801 (ConceptQ), by Horizon Europe programme HORIZON-CL4-2022-QUANTUM-01-SGA via the project 101113946 OpenSuperQPlus100, the German Science Foundation (DFG) under AN336/12-1 (For2724), the State of Baden-W{\"u}ttemberg under KQCBW/SiQuRe, and the BMBF within the QSolid project. We acknowledge the computational resources provided by the Aalto Science-IT project.

\appendix

\section{Initial density matrix}
\label{sec:initial-condition}

\begin{figure}[ht]
    \includegraphics[width=\linewidth]{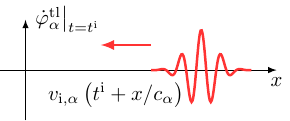}
    \caption{Average electromagnetic field in the initial state of transmission line $\alpha$ corresponding to a wave package propagating to the coupling node $n_\alpha$.}
    \label{fig:initial-condition}
\end{figure}

In this Appendix, we specify formal initial state density matrix of the transmission lines which we use in the definition of the generating functional~\eqref{eq:generating-functional-0}. We assume that at time $t^\mathrm i$ in a distant past, the state of the extended system, comprising the circuit and the transmission lines, was in a product state
\begin{multline}
    W\left[\bm \phi_+, \bm \phi_+^{\mathrm{tl}}, \bm \phi_-, \bm \phi_-^{\mathrm{tl}}, \bm v_\mathrm i\right] = \rho\left(\bm \phi_+, \bm \phi_-\right) \\ \times \prod\limits_{\alpha=1}^K \tilde W_{T,\alpha}\left[\phi_{+, \alpha}^\mathrm{tl}, \phi_{-, \alpha}^\mathrm{tl}, v_{\mathrm i,\alpha}\right].
    \label{eq:initial-condition}
\end{multline}
Here, density matrix of transmission line $\alpha$ corresponds to the thermal coherent state with temperature $T$:
\begin{multline}
    \tilde W_{T,\alpha}\left[\phi_{+, \alpha}^\mathrm{tl}, \phi_{-, \alpha}^\mathrm{tl}, v_\mathrm i\right] \\= 
    \exp\left\{\frac{\imath}{\hbar} \int\limits_0^{+\infty} C_{\ell,\alpha} \dot{\phi}^\mathrm{tl}_{0, \alpha} \left[\phi^\mathrm{tl}_{+,\alpha} - \phi^\mathrm{tl}_{-,\alpha}\right]\;\mathrm dx \right\} \\ \times W_{T, \alpha}\left[\phi^\mathrm{tl}_{+,\alpha} - \phi^{\mathrm{tl}}_{0, \alpha}, \phi^\mathrm{tl}_{-,\alpha} - \phi^{\mathrm{tl}}_{0, \alpha}\right].
\end{multline}
The coherent shifts, in accordance with~\eqref{eq:coherent-displacement}, are given by
\begin{equation}
    \begin{gathered}
        \dot \phi^\mathrm{tl}_{0, \alpha}(x) = v_{\mathrm i,\alpha}\left(t^\mathrm i + \frac{x}{c_\alpha}\right), \\
        \phi^\mathrm{tl}_{0, \alpha}(x) = \int\limits_{0}^x v_{\mathrm i, \alpha}\left(t^\mathrm i + \frac{x'}{c_\alpha}\right) \;\frac{\mathrm dx'}{c_\alpha}.
    \end{gathered}
\end{equation}
They describe the initial conditions for the electromagnetic wave, propagating towards the node $n_\alpha$, see Fig.~\ref{fig:initial-condition}. Such initial condition results in the average voltage of the input field at the node $\langle \hat v_{\mathrm i,\alpha}(x, t)\rangle |_{x=0}$ to be equal to $v_{\mathrm i, \alpha}(t)$. The thermal density matrix $W_{T,\alpha}\left[\phi_{+, \alpha}^\mathrm{tl}, \phi_{-, \alpha}^\mathrm{tl}\right]$ can be expressed through the path integral of the exponential of the Euclidean action as
\begin{multline}
    W_{T,\alpha}\left[\phi_{+, \alpha}^\mathrm{tl}, \phi_{-, \alpha}^\mathrm{tl}\right] \\= \int\mathcal D\left[\varphi_\alpha^\mathrm{tl}\right] \exp \left\{-\frac{1}{\hbar}S_{\mathrm E,\alpha}\left[\varphi_\alpha^\mathrm{tl}\right]\right\}
    \label{eq:thermal}
\end{multline}
\begin{multline}
    S_{\mathrm E,\alpha}\left[\varphi_\alpha^\mathrm{tl}\right] = \int\limits_0^{\hbar \beta} \left\{\left.\frac{\left(\varphi_\alpha^\mathrm{tl}\right)^2}{2L_\varepsilon}\right|_{x=0} \right . \\ \left .+ \int\limits_{0}^{+\infty}
    \left[\frac{C_{\ell,\alpha} \left(\partial_\tau \varphi_\alpha^\mathrm{tl}\right)^2}{2} + \frac{\left(\partial_x \varphi_\alpha^\mathrm{tl}\right)^2}{2L_{\ell, \alpha}}\right]\;\mathrm dx\right\}\;\mathrm d\tau,
\end{multline}
where $\beta = 1 / (k_\mathrm B T)$ and $T$ is temperature of the transmission line.
The path integration in Eq.~\eqref{eq:thermal} is carried out over the fixed point trajectories $\bm \varphi^{\mathrm{tl}}(x, \tau)|_{\tau = 0} = \bm \phi^\mathrm{tl}_-(x)$, and $\bm \varphi^{\mathrm{tl}}(x, \tau)|_{\tau = \hbar \beta} = \bm \phi^\mathrm{tl}_+(x)$. We assume that the partition function is included into the integration measure of the path integral~\eqref{eq:thermal}. This path integral is Gaussian, and hence the influence functional can be evaluated as
\begin{equation}
    W_{T,\alpha}\left[\phi_{+, \alpha}^\mathrm{tl}, \phi_{-, \alpha}^\mathrm{tl}\right] = N_{T,\alpha} \exp \left\{-\frac{1}{\hbar}S_{\mathrm E,\alpha}\left[\tilde \varphi_\alpha^\mathrm{tl}\right]\right\},
    \label{eq:thermal-result}
\end{equation}
where $\tilde \varphi^\mathrm{tl}_\alpha$ is the extreme trajectory which satisfies the Laplace equation with the boundary conditions
\begin{gather}
        \partial_\tau^2 \tilde \varphi_\alpha^\mathrm{tl} + c_\alpha^2 \partial_x^2 \tilde \varphi_\alpha^\mathrm{tl} = 0, \\
        \left.\left(\tilde \varphi_\alpha^\mathrm{tl} - \frac{L_\varepsilon}{L_{\ell,\alpha}} \partial_x \tilde \varphi_\alpha^\mathrm{tl}\right)\right|_{x=0} = 0,    \label{eq:laplace-boundary}\\
        \left.\tilde \varphi_\alpha^\mathrm{tl}\right|_{\tau = 0} = \phi_{-,\alpha}^{\mathrm{tl}},~
        \left.\tilde \varphi_\alpha^\mathrm{tl}\right|_{\tau = \hbar\beta} = \phi_{+,\alpha}^{\mathrm{tl}},
\end{gather}
and $N_{T,\alpha}$ is a normalization factor.

To solve these equations, we find the eigenfunctions of the operator $-\partial_x^2$ with the boundary condition~\eqref{eq:laplace-boundary} as
\begin{equation}
    \begin{gathered}
    u_\alpha (k, x) = \frac{\sin(kx) + \left(k L_\varepsilon / L_{\ell, \alpha}\right) \cos(kx)}{\sqrt{1 + \left(k L_\varepsilon / L_{\ell, \alpha}\right)^2}}.
    %
    \end{gathered}
\end{equation}
Then we define a generalized Fourier transform as
\begin{equation}
    \begin{gathered}
    \varphi(k, \tau) = \int\limits_0^{+\infty} \varphi(x, \tau) u_{\alpha}(k, x)\;\mathrm dx, \\
    \varphi(x, \tau) = \frac{2}{\pi}\int\limits_0^{+\infty}  \varphi(k, \tau)u_\alpha(k, x)\;\mathrm dk.
    \end{gathered}
    \label{eq:gen-fourier}
\end{equation}
The Fourier image of the extreme trajectory reads as
\begin{multline}
    \tilde \varphi_\alpha^\mathrm{tl}(k, \tau)  = \varphi_{+,\alpha}^{\mathrm{tl},\mathrm i}(k) \frac{\ \sinh(c_\alpha k \tau) }{\sinh(\hbar c_\alpha k \beta)} \\ +\varphi_{-,\alpha}^{\mathrm{tl},\mathrm i}(k) \frac{\sinh \left[c_\alpha k (\hbar \beta - \tau)\right]}{\sinh (\hbar c_\alpha k \beta)},
\end{multline}
and the thermal density matrix is given by
\begin{widetext}
\begin{equation}
    W_{T, \alpha}\left[\phi_{+,\alpha}^{\mathrm{tl}}, \phi_{-,\alpha}^{\mathrm{tl}}\right]  = N_{T,\alpha} \exp \left\{
    \int\limits_0^{+\infty} k \frac{2 \phi_{+,\alpha}^{\mathrm{tl}}(k) \phi_{-,\alpha}^{\mathrm{tl}}(k) -
    \left[\left(\phi_{+,\alpha}^{\mathrm{tl}}(k)\right)^2  +
    \left(\phi_{-,\alpha}^{\mathrm{tl}}(k)\right)^2\right]\cosh(\hbar c_\alpha k \beta)}{\pi \hbar Z_\alpha \sinh(\hbar c_\alpha k \beta)}
    \;\mathrm dk\right\}.
    \label{eq:thermal-density-matrix}
\end{equation}
\end{widetext}

\section{Influence functional for transmission~lines}
\label{sec:influence-functional}

In this Appendix we present the derivation of the influence functional~\eqref{eq:tl-nonlocal} derived from the definition of the generating functional~\eqref{eq:generating-functional-0}.
Since the initial state~\eqref{eq:initial-condition} is a product state, we can evaluate Gaussian integrals in~\eqref{eq:generating-functional-0} over each $\bm \varphi_{\pm, \alpha}^\mathrm{tl}$ separately for each $\alpha$. For each line $\alpha$, we introduce an influence functional
\begin{widetext}
\begin{multline}
    \mathcal Z_\alpha^\mathrm{tl} \left[\varphi_{n_\alpha,+}, \varphi_{n_\alpha,-}, v_{\mathrm i,\alpha}, \eta_{\mathrm o,\alpha}\right] = \int\mathcal D\left [\varphi_{+,\alpha}^\mathrm{tl}, \varphi_{-,\alpha}^\mathrm{tl}\right] \tilde W_{T,\alpha}\left[\varphi_{\alpha,+}^{\mathrm{tl}, {\mathrm i, \alpha}}, \varphi_{\alpha,-}^{\mathrm{tl}, \mathrm i}, v_\mathrm i\right]  
    \exp\left\{\frac{\imath}{\hbar} \int\limits_{t^\mathrm i}^{t^\mathrm f} \left[
    \mathcal L_\alpha^\mathrm{tl}\left[\dot \varphi_{+,\alpha}^\mathrm{tl}, \varphi_{+,\alpha}^\mathrm{tl}, \varphi_{+,n_\alpha}\right] \phantom{\frac{1}{4}}\right. \right.\\ \left. 
    \phantom{\int\limits_{t^\mathrm i}^{t^\mathrm f}} \left.-
        \mathcal L_\alpha^\mathrm{tl}\left[\dot \varphi_{-,\alpha}^\mathrm{tl}, \varphi_{-,\alpha}^\mathrm{tl}, \varphi_{-,n_\alpha}\right]
    +\frac{1}{4}
    \eta_{\mathrm o, \alpha} \left. {\left(\partial_t - c_\alpha \partial_x\right)} \left(\varphi^\mathrm{tl}_{+, \alpha} + \varphi^\mathrm{tl}_{-,\alpha}\right)\right|_{x=x_0}
    \right]\;\mathrm dt
    \right\}.
    \label{eq:influence-functional-qwe}
\end{multline}\end{widetext}
Now, we evaluate fixed end path integrals over real time
\begin{multline}
    K_\alpha[\tilde \phi^{\mathrm{tl}}_\alpha,
    \phi^{\mathrm{tl}}_\alpha, \varphi_{n_\alpha}, \eta_{\mathrm o, \alpha}] \\
    = \int\mathcal D\left[\varphi^{\mathrm{tl}}_\alpha\right] \exp\left\{\frac{\imath}{\hbar}
    \int\limits_{t^\mathrm i}^{t^\mathrm f}
    \left[
    \mathcal L_\alpha^\mathrm{tl}\left[
    \dot \varphi_\alpha^\mathrm{tl}, \varphi_\alpha^\mathrm{tl}, \varphi_{n_\alpha}
    \right] \phantom{\frac{1}{4}}\right. \right. \\ \left. \phantom{\int\limits_{t^\mathrm i}^{t^\mathrm f}}\left.+
    \frac{1}{4} \eta_{\mathrm o,\alpha} \left(\partial_t - c_\alpha \partial_x\right) \left.\varphi_{\alpha}^\mathrm{tl}\right|_{x=x_0}\right]\;\mathrm dt
    \right\},
    \label{eq:kernel}
\end{multline}
where the boundary conditions are given by $\phi_\alpha^{\mathrm{tl}}(x) = \varphi_\alpha^\mathrm{tl}\left(x, t^{\mathrm{i}}\right)$ and $\tilde \phi_\alpha^{\mathrm{tl}}(x) = \varphi_\alpha^\mathrm{tl}\left(x, t^{\mathrm{f}}\right)$. This integral coincides with integral over $\varphi_{+, \alpha}^\mathrm{tl}$ and after flipping the sign of $\eta_{\mathrm o,\alpha}$ to complex conjugation of the integral over $\varphi_{-,\alpha}^\mathrm{tl}$. Equation~\eqref{eq:kernel} can be evaluated in the same way, by employing the generalized Fourier transformation~\eqref{eq:gen-fourier} and solving the equation for the saddle-point trajectory. The integrand in the argument of the exponent in Eq.~\eqref{eq:kernel} after the Fourier transformation reads
\begin{widetext}
\begin{multline}
    \mathcal L_\alpha^\mathrm{tl}\left[\dot \varphi^\mathrm{tl}_\alpha, \varphi_\alpha^\mathrm{tl}, \varphi_{n_\alpha}\right] + \frac{1}{4} \eta_{\mathrm o, \alpha} \left(\partial_t - c_\alpha \partial_x\right) \left.\varphi_\alpha^\mathrm{tl}\right|_x =
    \frac{2}{\pi} \int\limits_{0}^{+\infty}\left\{
    \frac{C_{\ell, \alpha} \left[\dot \varphi_\alpha^\mathrm{tl}(k, t)\right]^2}{2} -
    \frac{k^2 \left[\varphi_\alpha^\mathrm{tl}(k, t)\right]^2}{2 L_{\ell,\alpha}} \right. \\ 
    +\frac{\eta_{\mathrm o,\alpha}(t)}{2} \left[ \dot \varphi_\alpha^\mathrm{tl}(k, t) \frac{\sin(kx_0) + (k L_\varepsilon / L_{\ell, \alpha}) \cos(kx_0)}{\sqrt{1 + (k L_\varepsilon / L_{\ell, \alpha})^2}}
    - c k \varphi_\alpha^\mathrm{tl}(k, t) \frac{\cos(kx_0) - (k L_\varepsilon / L_{\ell, \alpha}) \sin(kx_0)}{\sqrt{1 + (k L_\varepsilon / L_{\ell, \alpha})^2}}
    \right] \\
     \left.+ \frac{k \varphi_\alpha^\mathrm{tl}(k, t) \varphi_{n_\alpha}(t)}{L_{\ell,\alpha}\sqrt{1 + \left(k L_\varepsilon / L_{\ell,\alpha}\right)^2}}
    \right\}\;\mathrm dk - \frac{\varphi_{n_\alpha}^2(t)}{2L_\varepsilon}.
\end{multline}
\end{widetext}
The Euler--Lagrange equation for the saddle point trajectory $\tilde \varphi_\alpha^\mathrm{tl}(k, t)$ is expressed as
\begin{equation}
\begin{gathered}
    \ddot {\tilde \varphi}_\alpha^\mathrm{tl}(k, t) + c_\alpha^2 k^2 \tilde \varphi_\alpha^\mathrm{tl}(k, t) = \chi_\alpha(k, t),\\
    \tilde \varphi_\alpha^\mathrm{tl}\left(k, t^{\mathrm i}\right) = \phi_\alpha^{\mathrm{tl}}(k),~\tilde \varphi_\alpha^\mathrm{tl}\left(k, t^{\mathrm f}\right) = \tilde \phi_\alpha^{\mathrm{tl}}(k)
\end{gathered}
\end{equation}
where
\begin{multline}
    \chi_\alpha(k, t) = \frac{c_\alpha^2 k \varphi_{n_\alpha}(t)}{ \sqrt{1 + \left(k L_\varepsilon / L_{\ell,\alpha}\right)^2}} \\
    -\frac{1}{4 C_{\ell, \alpha}} \left[ \dot \eta_{\mathrm o,\alpha}(t) \frac{\sin(kx_0) + (k L_\varepsilon / L_{\ell, \alpha}) \cos(kx_0)}{\sqrt{1 + (k L_\varepsilon / L_{\ell, \alpha})^2}} \right. \\
    \left.
    + c k \eta_{\mathrm o,\alpha}(t) \frac{\cos(kx_0) - (k L_\varepsilon / L_{\ell, \alpha}) \sin(kx_0)}{\sqrt{1 + (k L_\varepsilon / L_{\ell, \alpha})^2}}
    \right].
\end{multline}
The solution of this equation is given by
\begin{widetext}
\begin{equation}
    \tilde \varphi_\alpha^\mathrm{tl}(k, t) =  \int\limits_{t^\mathrm i}^{t^\mathrm f}G_\alpha(k, t, t') \chi_\alpha(k, t')\;\mathrm dt'  + \frac{\phi_\alpha^{\mathrm{tl}}(k) \sin\left[c_\alpha k \left(t^\mathrm f - t\right)\right] + \tilde \phi_\alpha^{\mathrm{tl}}(k)\sin\left[c_\alpha k \left(t - t^\mathrm i\right)\right]}{\sin \left[c_\alpha k \left(t^\mathrm f - t^\mathrm i\right)\right]},
\end{equation}
where the zero-boundary-condition Green's function $G_\alpha(k, t, t')$ reads
\begin{equation}
    G_\alpha(k, t, t') 
    = -\frac{\Theta(t - t')\sin\left[c_\alpha k \left(t^\mathrm f - t\right)\right]\sin\left[c_\alpha k \left(t' - t^\mathrm i\right)\right]}{c_\alpha k \sin \left[c_\alpha k \left(t^\mathrm f - t^\mathrm i\right)\right]} 
    -\frac{\Theta(t' - t)\sin\left[c_\alpha k \left(t^\mathrm f - t'\right)\right]\sin\left[c_\alpha k \left(t - t^\mathrm i\right)\right]}{c_\alpha k \sin \left[c_\alpha k \left(t^\mathrm f - t^\mathrm i\right)\right]}.
\end{equation}
The path integral~\eqref{eq:kernel} is equal to

\begin{multline}
    K_\alpha\left[\tilde \phi_\alpha^{\mathrm{tl}}, \phi_\alpha^{\mathrm{tl}}, \varphi_{n_\alpha}, \eta_{\mathrm o,\alpha}\right] = \exp\left\{
    \frac{2\imath}{\pi \hbar}\int\limits_{0}^{+\infty}
    \left[
    k \frac{\left[\left(\tilde \phi_\alpha^{\mathrm{tl}}(k)\right)^2   + 
    \left(\phi_\alpha^{\mathrm{tl}}(k)\right)^2\right]\cos\left[c_\alpha k \left(t^\mathrm f - t^\mathrm i\right)\right]-2\tilde \phi_\alpha^{\mathrm{tl}}(k) \phi_\alpha^{\mathrm{tl}}(k)}{2 Z_\alpha \sin \left[c_\alpha k \left(t^\mathrm f - t^\mathrm i\right)\right]}\right. \right. \left. \left.    \right. \right. \\ 
    \left. \left. +
    C_{\ell,\alpha}\phi_\alpha^{\mathrm{tl}}(k) \int\limits_{t^\mathrm i}^{t^\mathrm f}\chi_\alpha (k, t) \frac{ \sin\left[c_\alpha k \left(t^\mathrm f - t\right)\right]
    }{\sin\left[c_\alpha k \left(t^\mathrm f - t^\mathrm i\right)\right]}\;\mathrm dt \right. \right. 
    \left. \left .+
    C_{\ell,\alpha} \tilde\phi_\alpha^{\mathrm{tl}}(k) \int\limits_{t^\mathrm i}^{t^\mathrm f}\chi_\alpha (k, t) \frac{\sin\left[c_\alpha k \left(t - t^\mathrm i\right)\right]
    }{\sin\left[c_\alpha k \left(t^\mathrm f - t^\mathrm i\right)\right]}\;\mathrm dt \right. \right. \\
    \left. \left. + \frac{C_{\ell,\alpha}}{2}\iint\limits_{t^\mathrm i}^{t^\mathrm f} \chi_\alpha(k, t) G_\alpha(k, t, t') \chi_\alpha(k, t')\;\mathrm dt\;\mathrm dt'
    \right]\;\mathrm dk
    \frac{\imath}{\hbar} \int\limits_{t^\mathrm i}^{t^\mathrm f} \frac{\varphi_{n_\alpha}^2}{2 L_\varepsilon}\;\mathrm dt
    \right\}.
\end{multline}

The influence functional~\eqref{eq:influence-functional-qwe} can be expressed as a path integral over the boundaries as
\begin{multline}
    \mathcal Z_\alpha^\mathrm{tl}\left[\varphi_{+,n_\alpha}, \varphi_{-,n_\alpha}, v_{\mathrm i,\alpha}, \eta_{\mathrm o, \alpha}\right]  =\int \mathcal D\left[\phi_{+,\alpha}^{\mathrm{tl}}, \phi_{-,\alpha}^{\mathrm{tl}}, \tilde \phi_{\alpha}^{\mathrm{tl}} \right]
    K\left[\tilde \phi_\alpha^{\mathrm{tl}}, \phi_{+,\alpha}^{\mathrm{tl}}, \varphi_{+, n_\alpha}, \eta_{\mathrm o,\alpha}\right] \\ \times
    \tilde W_{T,\alpha} \left[\phi_{+,\alpha}^{\mathrm{tl}}, \phi_{-,\alpha}^{\mathrm{tl}}, v_{\mathrm i,\alpha}\right]
    K^\ast \left[\tilde \phi_\alpha^{\mathrm{tl}}, \phi_{-,\alpha}^{\mathrm{tl}}, \varphi_{-, n_\alpha}, -\eta_{\mathrm o,\alpha}\right].
    \label{eq:influence-tl}
\end{multline}

We carry out a Keldysh rotation
\begin{equation}
    \begin{gathered}
        \phi_{\pm, \alpha}^{\mathrm{tl}} = \phi_{\mathrm c,\alpha}^{\mathrm{tl}} \pm \frac{\phi_{\mathrm q, \alpha}^{\mathrm{tl}}}{2},~
        \varphi_{\pm,n_\alpha} = \varphi_{\mathrm c,n_\alpha} \pm \frac{\varphi_{\mathrm q, n_\alpha}}{2}
    \end{gathered}
\end{equation}
and evaluate the path integral~\eqref{eq:influence-tl} as
\begin{multline}
    \mathcal Z_\alpha^\mathrm{tl}[\varphi_{\mathrm c, n_\alpha}, v_{\mathrm i,\alpha}, \varphi_{\mathrm q, n_\alpha}, \eta_{\mathrm o, \alpha}] = \exp\left\{
    \frac{2 \imath}{\pi \hbar}\int\limits_0^{+\infty}\left[
        \frac{C_{\ell,\alpha}}{c_\alpha k} \int\limits_{t^\mathrm i}^{t^\mathrm f}\mathrm dt \int\limits_{t^\mathrm i}^t \chi_{\mathrm q, \alpha}(k, t) {\sin[c_\alpha k (t - t')]} \chi_{\mathrm c, \alpha}(k, t')\;\mathrm dt' + \right. \right. \\
        \left. \left .
        + C_{\ell,\alpha} \varphi_{0, \alpha}(k) \int\limits_{t^\mathrm i}^{t^\mathrm f} \cos\left[c_\alpha k \left(t - t^\mathrm i\right)\right] \chi_{\mathrm q, \alpha}(k, t)\;\mathrm dt + \frac{C_{\ell,\alpha}}{c_\alpha k} \dot \varphi_{0,\alpha}(k) \int\limits_{t^\mathrm i}^{t^\mathrm f} \sin\left[c_\alpha k \left(t - t^\mathrm i\right)\right]\chi_{\mathrm q, \alpha}(k, t)\;\mathrm dt \right . \right .
        \\ \left. \left .
        +\frac{\imath C_{\ell, \alpha}}{4 c_\alpha k} \coth \left(\frac{\hbar c_\alpha k}{2 k_\mathrm B T}\right) \iint\limits_{t^\mathrm i}^{t^\mathrm f} \chi_{\mathrm q, \alpha}(k, t) \cos\left[c_\alpha k\left(t - t'\right)\right] \chi_{\mathrm q, \alpha}(k, t')\;\mathrm dt\;\mathrm dt'
    \right]\;\mathrm dk - \frac{\imath}{\hbar} \int\limits_{t^\mathrm i}^{t^\mathrm f} \frac{\varphi_{\mathrm q, n_\alpha}(t) \varphi_{\mathrm c, n_\alpha}(t)}{L_\varepsilon}\;\mathrm dt
    \right\},
    \label{eq:influence-fun-another}
\end{multline}
where
\begin{equation}
    \chi_{\mathrm c, \alpha} (k, t)= \frac{c_\alpha^2 k \varphi_{\mathrm c, n_\alpha}(t)}{ \sqrt{1 + \left(k L_\varepsilon / L_{\ell,\alpha}\right)^2}},
    \label{eq:chi-class}
\end{equation}
\begin{multline}
    \chi_{\mathrm q, \alpha}(k, t) = \frac{c_\alpha^2 k \varphi_{\mathrm q, n_\alpha}(t)}{\sqrt{1 + \left(k L_\varepsilon / L_{\ell,\alpha}\right)^2}} 
    -\frac{1}{2 C_{\ell, \alpha}} \left[ \dot \eta_{\mathrm o,\alpha}(t) \frac{\sin(kx_0) + (k L_\varepsilon / L_{\ell, \alpha}) \cos(kx_0)}{\sqrt{1 + (k L_\varepsilon / L_{\ell, \alpha})^2}} \right. \\
    \left.
    + c k \eta_{\mathrm o,\alpha}(t) \frac{\cos(kx_0) - (k L_\varepsilon / L_{\ell, \alpha}) \sin(kx_0)}{\sqrt{1 + (k L_\varepsilon / L_{\ell, \alpha})^2}}
    \right].
    \label{eq:chi-quant}
\end{multline}
\end{widetext}
We substitute~\eqref{eq:chi-class} and~\eqref{eq:chi-quant} into~\eqref{eq:influence-fun-another}, preform integration over $k$, and consider limits $L_\varepsilon \to +0$, $x_0 \to +0$, $t^\mathrm i \to -\infty$, and $t^\mathrm f \to +\infty$. We also carry out the Fourier transformation in the time domain. The final result reads as
\begin{widetext}
\begin{multline}
    \mathcal Z_\alpha^\mathrm{tl}[\varphi_{\mathrm c, n_\alpha}, v_{\mathrm i,\alpha}, \varphi_{\mathrm q, n_\alpha}, \eta_{\mathrm o, \alpha}] = \exp\left\{
    \frac{\imath}{2\pi \hbar}\int\limits_{-\infty}^{+\infty}\left[
    \frac{i\omega}{Z_\alpha} \varphi_{\mathrm q, n_\alpha}^\ast (\omega) \varphi_{\mathrm c, n_\alpha}(\omega) 
    + i \omega \eta_{\mathrm o, \alpha}^\ast(\omega) \varphi_{\mathrm c, n_\alpha}(\omega) \right. \right . \\+ 
    \left. \left. \left(\frac{2 \varphi_{\mathrm q,n_\alpha}^\ast(\omega)}{Z_\alpha} - \eta_{\mathrm o, \alpha}(\omega)\right)  v_{\mathrm i,\alpha}(\omega) +
    \frac{\imath \omega}{2 Z} \coth\left(\frac{\hbar \omega}{2 k_\mathrm B T}\right)
    \left|\varphi_{\mathrm q,n_\alpha}(\omega) - \frac{Z_\alpha\eta_{\mathrm o, \alpha}(\omega)}{2} \right|^2
    \right]\;\mathrm d\omega
    \right\}.
\end{multline}
\end{widetext}
The expression under the exponent is equivalent to the combination of~\eqref{eq:tl-nonlocal} and~\eqref{eq:tl-self-energy}.

\section{Rational approximation of Bose--Einstein distribution function}
\label{sec:AAA}
    \begin{table}[ht]
        \caption{Parameters of the rational approximation~\eqref{eq:AAA-approx} of $\coth(z)$.}
        \begin{tabularx}{\linewidth}{C|C}
            \hline \hline
            $z_n$ & $r_n$ \\
            \hline 
            3.141592655445434 & 1.0000000055441378\\ \hline
            6.283189771187831 & 1.0000092259581892\\ \hline
            9.425879949143997 & 1.001614741279204\\ \hline
            12.61536493615531 & 1.0472247085376767\\ \hline
            16.243736719976432 & 1.3143116676767104\\ \hline
            21.179619301069728 & 1.8699539196161084\\ \hline
            28.257562489430427 & 2.684608268078368\\ \hline
            38.445551248863126 & 3.8854577814112417\\ \hline
            53.41937113027605 & 5.816887481357558\\ \hline
            76.65204493098419 & 9.385726603683633\\ \hline
            117.04233355146623 & 17.659527331995577\\ \hline
            207.166009024319 & 46.985668625964905\\ \hline
            640.9532317009297 & 411.2142895506761\\ \hline
\hline
        \end{tabularx}
        \label{tab:AAA-results}
    \end{table}
    In this Appendix, we discuss construction and accuracy of decomposition~\eqref{eq:coth-approximation}. Using the AAA algorithm~\cite{nakatsukasa2018aaa, xu2022taming}, we find a rational approximation of $\coth(z)$ function in the form
    \begin{equation}
        \coth(z) \approx \frac{1}{z} + \sum\limits_{n=1}^M \frac{2 z r_n}{z_n^2 + z^2}.
        \label{eq:AAA-approx}
    \end{equation}
    The poles at $\pm \imath z_n$ have respective residues $r_n$. The obtained values $z_n$ and $r_n$ are shown in Tab.~\ref{tab:AAA-results}. These parameters provide an approximation for $\coth(z)$ in the range $z \in [-100, 100]$ within a deviation of $2 \times 10^{-13}$ with only 13 poles in the lower and upper half-planes each. Using the approximation~\eqref{eq:AAA-approx}, one can obtain the parameters for Eq.~\eqref{eq:coth-approximation} as
    \begin{equation}
        \begin{gathered}
            \omega_n^\mathrm {f} = \frac{2 k_\mathrm B T z_n}{\hbar},~
            r_n^\mathrm{f} = \frac{2 k_\mathrm B T r_n}{\hbar}.
        \end{gathered}
    \end{equation}
\section{Charge QD-MESS equation}
\label{sec:charge-qdmess}
In this Appendix, we provide the charge QD-MESS equation in contrast to the flux version considered in the main article. We begin with the classical QD-MESS Lagrangian with the voltage coupling to the auxiliary degrees of freedom, given by
\begin{widetext}
\begin{multline}
    \mathcal L_\text{QD-MESS}'\left(\ldots\right) = \mathcal L_\mathrm J(\bm \psi_\mathrm c,\bm \psi_\mathrm q) 
    -
        \left\{
        \begin{bmatrix}
                \dot{\bm \psi}_\mathrm q^\mathsf T &
                \dot{\bm \eta}_\mathrm o^\mathsf T
            \end{bmatrix}
            \tilde{\bm P}_2
            \begin{bmatrix}
                \dot{\bm \psi}_\mathrm c \\
                \dot{\bm v}_\mathrm i \\
                \dot{\bm \psi}_\mathrm q \\
                \dot{\bm \eta}_\mathrm o
            \end{bmatrix}
             +
            \begin{bmatrix}
                \bm \psi_\mathrm q^\mathsf T &
                \bm \eta_\mathrm o^\mathsf T
            \end{bmatrix}
            \left(
            \tilde{\bm P}_0'
            +\imath
            \tilde{\bm P}_1' \partial_t
            \right)\begin{bmatrix}
                \bm \psi_\mathrm c \\
                \bm v_\mathrm i \\
                \bm \psi_\mathrm q \\
                \bm \eta_\mathrm o
            \end{bmatrix}
        \right\} 
         \\+
        \sum\limits_{k=1}^{M^\mathrm d} \left(
        \imath\hbar a_k^\ast \dot a_k - \hbar\omega_k^\mathrm d |a_k|^2 +
        \begin{bmatrix}
            \dot{\bm \psi}_\mathrm q^\mathsf T &
            \bm \eta_\mathrm o^\mathsf T
        \end{bmatrix}
        \bm u_k^\mathrm d a_k + 
        a_k^\ast \bm v_k^{\mathrm d \mathsf T}
        \begin{bmatrix}
            \dot{\bm \psi}_\mathrm c \\
            \bm v_\mathrm i \\
            \dot{\bm \psi}_\mathrm q \\
            \bm \eta_\mathrm o
        \end{bmatrix}
        \right) \\ +
        \sum\limits_{n=1}^{M^\mathrm f} \sum\limits_{m=1}^{\varrho_n}\left(
        \imath \hbar b_{nm}^\ast \dot{b}_{nm} + \imath \omega_n^\mathrm {f} |b_{nm}|^2+ \begin{bmatrix}
            \dot{\bm \psi}_\mathrm q^\mathsf T &
            \bm \eta_\mathrm o^\mathsf T
        \end{bmatrix}
        \bm u_{nm}^\mathrm {f} b_{nm} +
        b_{nm}^\ast \bm v_{nm}^{\mathrm {f} \mathsf T} \begin{bmatrix}
            \dot{\bm \psi}_\mathrm q\\
            \bm \eta_\mathrm o
        \end{bmatrix}
        \right),
    \label{eq:charge-qdmess-lagrangian}
\end{multline}
\end{widetext}
where the Lagrangian $\mathcal L_\mathrm{QD-MESS}'(\ldots)$ shares the same arguments as the Lagrangian $\mathcal L_\mathrm{QD-MESS}(\ldots)$ introduced in Eqs.~\eqref{eq:qd-mess-action} and~\eqref{eq:flux-qdmess-lagrangian},
\begin{equation}
    \begin{gathered}
        \tilde{\bm P}_0' = \tilde{\bm P}_0 + \sum\limits_k \bm I_1 \frac{\tilde{\bm R}_k^\mathrm d}{\omega_k^\mathrm d} \bm I_2
        + \sum\limits_{n} \bm I_3 \frac{\tilde{\bm R}_n^\mathrm{f}}{\omega_n^\mathrm{f}} \bm I_1,\\
        \tilde{\bm P}_1' = \tilde{\bm P}_1 + \sum\limits_k \bm I_1 \frac{\imath \tilde{\bm R}_k^\mathrm d}{\left(\omega_k^\mathrm d\right)^2} \bm I_2
        + \sum\limits_{n} \bm I_3 \frac{\imath \tilde{\bm R}_n^\mathrm{f}}{\left(\omega_n^\mathrm{f}\right)^2} \bm I_1,
    \end{gathered}
\end{equation}
and
\begin{equation}
    \begin{gathered}
        \bm I_1 = \operatorname{diag}(\overbrace{1, \ldots, 1}^{N_\mathrm{core}}, \overbrace{0, \ldots, 0}^{K}), \\
        \bm I_2 = \operatorname{diag}(\overbrace{1, \ldots, 1}^{N_\mathrm{core}}, \overbrace{0, \ldots, 0}^{K},\overbrace{1, \ldots, 1}^{N_\mathrm{core}}, \overbrace{0, \ldots, 0}^{K}), \\
        \bm I_3 = \begin{bmatrix}
            \bm 0 & \bm I_1
        \end{bmatrix}.
    \end{gathered}
\end{equation}
The expressions for the classical and quantum charges~\eqref{eq:legendre}, the Liouvillian~\eqref{eq:lagrangian-to-liouvillian}, and the commutation relations~\eqref{eq:commutation-relations} remain valid but the part of the Liouvillian which describes the dynamics of the auxiliary modes becomes non-diagonal. The form of Eq.~\eqref{eq:qdmess-liouvillian-general} can be obtained by applying the Bogoliubov transformation to the operators of the auxiliary modes. The difference from the flux QD-MESS Liouvillian is that the operators of the auxiliary modes become coupled to the charge degrees of freedom.

The main advantage of the charge QD-MESS equation for the transmon readout circuit shown in Fig.~\ref{fig:transmon} is that it evades spurious inductive terms in the Liouvillian which break the quasi-charge conservation of the transmon island.

\section{Weak-coupling regime}
\label{sec:schrieffer-wolf}
In this Appendix, we introduce weak-coupling approach to solving QD-MESS equation which we employ in our analysis of the qubit readout in Sec.~\ref{sec:transmon-readout}.
Solving QD-MESS equation remains challenging due to the exponentially high dimension of the state vector $|W\rrangle$. This issue is pronounced at ultra-low temperatures, where the fluctuation poles are densely clustered near zero frequency. Here, we suggest a systematic approach to eliminate weakly coupled auxiliary modes, which results in a substantial reduction of dimensionality required for the Liouville space of linear degrees of freedom. The general form of the QD-MESS Liouvillian reads as
\begin{multline}
    \check{\mathfrak L}(t) = \check{\mathfrak L}^{(0)}(t) + \sum\limits_{k=1}^M \left\{
    \left[\check x_k' + \alpha_k'(t)\right] \check a_k \phantom{\check a_k^\dagger}\right. \\ \left. + 
    \left[\check x_k'' + \alpha_k''(t)\right] \check a_k^\dagger
    + \hbar \omega_k \check a_k^\dagger \check a_k
    \right\},
    \label{eq:qdmess-liouvillian-general}
\end{multline}
where we denote the Liouvillian of all nonlinear degrees of freedom as $\check{\mathfrak L}^{(0)}(t)$, unify the dynamical and fluctuation auxiliary modes denoted by the operators $\check a_k$ and $\check a_k^\dagger$, denote all the coupling operators as $\check x_k'$ and $\check x_k''$, and introduce the drive variables $\alpha_k'(t)$ and $\alpha_k''(t)$ which may originate from both, classical $\bm v_\mathrm i$ and quantum $\bm \eta_\mathrm o$ sources. We assume both $\alpha_k'(t)$ and $\alpha_k''(t)$ to be bounded. Note that this form is not unique since one can incorporate some of the auxiliary modes to $\check{\mathfrak L}^{(0)}(t)$.

First, we eliminate time-dependent source terms by carrying out a shift transformation
\begin{equation}
    |W\rrangle = e^{\check T(t)}|W_\mathrm{shift}\rrangle,
\end{equation}
where $\check T(t) = \sum_{k=1}^M \left[\beta_k'(t) \check a_k + \beta_k''(t) \check a_k^\dagger\right]$ and the time-dependent shifts $\beta_k'(t)$ and $\beta_k''(t)$ are given by the bounded solutions of the equations
\begin{equation}
    \begin{gathered}
        \imath \hbar \dot \beta_k'(t) = - \hbar \omega_k \beta_k'(t) + \alpha_k'(t),\\
        \imath \hbar \dot \beta_k''(t) = \hbar \omega_k \beta_k''(t) + \alpha_k''(t).\\
    \end{gathered}
\end{equation}
Thus, the dynamics of $|W_\mathrm{shift}\rrangle$ is governed by the following Liouvillian
\begin{multline}
        \check{\mathfrak L}_\mathrm{shift}(t) = \check{\mathfrak L}^{(0)}_\mathrm{shift}(t) +  \\ \sum\limits_{k=1}^M \left(
    \check x_k' \check a_k + 
    \check x_k'' \check a_k^\dagger
    + \hbar \omega_k \check a_k^\dagger \check a_k
    \right),
\end{multline}
\begin{multline}
    \check{\mathfrak L}^{(0)}_\mathrm{shift}(t) = \check{\mathfrak L}^{(0)}(t) + \sum\limits_{k=1}^M \left[
    \check x_k'(t) \beta_k''(t) - \check x_k''(t) \beta_k'(t) \right] \\  + \frac{1}{2}\sum\limits_k\left[\alpha_k'(t) \beta_k''(t) - \alpha_k''(t) \beta_k'(t)
    \right].
\end{multline}

Up to this point, we have not done any approximations to arrive at the above Liouvillian since the shift transformation is exact. Next, we proceed with eliminating the transverse coupling to the auxiliary modes by a dynamical Schrieffer--Wolf transformation~\cite{schrieffer1966relation, bravyi2011schrieffer}
\begin{equation}
    |W_\mathrm{shift}\rrangle = e^{\check S(t)}  |W_\mathrm{SW}\rrangle,
\end{equation}
where $\check S(t) = \sum_{k=1}^M \left[\check p_k'(t) \check a_k + \check p_k''(t) \check a_k^\dagger\right]$ and the operators $\check p_k'(t)$ and $\check p_k''(t)$ act in the space of the Liouvillian $\check{\mathfrak L}_\mathrm{shift}^{(0)}(t)$. We emphasize that this Schrieffer--Wolf transformation is carried out at the Liouvillian level and is non-unitary. The first-order expansion of the exponents with respect to the operators $\check p_k'(t)$ and $\check p_k''(t)$ reveals that the transverse coupling may be eliminated if $\check p_k'(t)$ and $\check p_k''(t)$ are the bounded solutions of the following equations:
\begin{equation}
    \begin{gathered}
        \imath\hbar \frac{\mathrm d\check p_k'(t)}{\mathrm dt} = \left[\check{\mathfrak L}_\mathrm{shift}^{(0)}(t), \check p_k'(t)\right] - \hbar \omega_k \check p_k'(t) + \check x_k', \\
        \imath\hbar \frac{\mathrm d\check p_k''(t)}{\mathrm dt} = \left[\check{\mathfrak L}_\mathrm{shift}^{(0)}(t), \check p_k''(t)\right] + \hbar \omega_k \check p_k''(t) + \check x_k''.
    \end{gathered}
    \label{eq:sw-liouville}
\end{equation}
The second-order expansion yields the Schrieffer--Wolf correction to the Liouvillian as
\begin{multline}
    \check{\mathfrak L}_\mathrm{SW} \approx \check{\mathfrak L}^{(0)}_\mathrm{shift}(t) + \\ \sum\limits_{k=1}^M \left\{
    \hbar \omega_k \check a_k^\dagger \check a_k + \frac{1}{2}\left[
    \check x_k' \check a_k + \check x_k'' \check a_k^\dagger, \check S(t)
    \right]\right\}.
\end{multline}
Next, we apply a partial rotating-wave approximation for the auxiliary degrees of freedom, keeping only the diagonal terms $\check a_k^\dagger \check a_k$. This results in
\begin{multline}
    \check{\mathfrak L}_\mathrm{SW}(t) \approx \check {\mathfrak L}_\mathrm{SW}^{(0)}(t) + \sum\limits_{k=1}^M \left\{ \hbar \omega_k + \left[\check x_k', \check p_k''(t)\right]\right. \\ \left. + \left[\check x_k'', \check p_k'(t)\right]\right\} \check a_k^\dagger \check a_k,
\end{multline}
\begin{equation}
    \check{\mathfrak L}_\mathrm{SW}^{(0)}(t) = \check{\mathfrak L}_\mathrm{shift}^{(0)}(t) + \frac{1}{2}\sum\limits_{k=1}^M \left[\check x_k' \check p_k''(t) - \check p_k'(t) \check x_k''\right].
\end{equation}
The last step is valid only if there are no pairs $(k, k')$, for which $\omega_k \pm \omega_{k'}$ is small. Such a delicate case appears, for example, for two weakly decaying dynamical auxiliary modes with frequencies $\omega_k = -\omega_{k'}^\ast$. These pairs should be treated carefully: one may either keep the off-diagonal terms $\check a_k \check a_{k'}$ and $\check a_k^\dagger \check a_{k'}^\dagger$ in the transformed Liouvillian, or take these modes into account exactly by incorporating them into $\check{\mathfrak L}^{(0)}(t)$ in the original Liouvillian~\eqref{eq:qdmess-liouvillian-general}.

The dynamical Schrieffer--Wolf transformation is perturbative, hence it can be applied only to systems weakly coupled to the auxiliary modes. In the case of the stationary Liouvillian $\check{\mathfrak L}_\mathrm{shift}$, it is valid if the matrix elements of the coupling operators $\check x_k'$ and $\check x_k''$ fulfil the conditions
\begin{equation}
    \begin{gathered}
        \left|\llangle n | \check x_k' | n'\rrangle \right| \ll \hbar |\Omega_{n} - \Omega_{n'} + \omega_k|,\\
        \left|\llangle n | \check x_k'' | n'\rrangle \right| \ll \hbar |\Omega_{n} - \Omega_{n'} - \omega_k|,
    \end{gathered}
\end{equation}
where $\llangle n|\check{\mathfrak L}_\mathrm{shift} = \hbar \Omega_n \llangle n|$ and $\check{\mathfrak L}_\mathrm{shift} | n'\rrangle = \hbar \Omega_{n'} |n'\rrangle$.
This perturbative Schrieffer--Wolf approach applied to QD-MESS equation is similar to the standard Born--Markov approximations widely used for open quantum systems in the weak-coupling regime. 
However, the presented approach provides a systematic way to derive the dissipative terms in the Liouvillian in case of complicated circuits, for which employing the standard IO theory~\cite{collett1984squeezing, gardiner1985input, yurke1984quantum} becomes hindered. Imporantly, the weakly decaying or strongly coupled auxiliary modes can be treated more accurately by incorporating them into the Liouvillian $\check{\mathfrak L}^{(0)}(t)$.

In addition to the weak-coupling requirement, one of the most restrictive limitations of this approach is that it can be used only for sufficiently small systems, for which Eq.~\eqref{eq:sw-liouville} can be solved exactly. This difficulty is not related to the dissipation in the circuit, but arises from the high number of degrees of freedom in the core subsystem or auxiliary modes to be taken into account non-perturbatively. This results in a prohibitively high dimension of the Liouvillian $\check{\mathfrak L}_0(t)$ required for accurate analysis, which is a typical problem for large electrical quantum circuit. However, accurate solutions of even small circuits may be of great interest, for example, from the point of view of single-qubit operation fidelities. Note also that it is unlikely that an efficient open-quantum-system solver is found in a case where there is no efficient solver for the closed system.

\section{Discrete approximation of QD-MESS for the transmon}
\label{sec:qdmess-approximation}

Direct numerical simulation of the QD-MESS equation is impossible, since even the Liouvillian $\check{\mathfrak L}^{(0)}(t)$ in Eq.~\eqref{eq:qdmess-liouvillian-general} is infinite-dimensional and one needs to come up with its finite-dimensional approximation in some suitable basis. For a general circuit with intermediate or strong dissipation, it may be a challenging problem. For the transmon readout circuit we use, the coupling of the qubit to the resonator and to the transmission lines is weak, and hence we can use a finite number of bare-transmon basis states for our calculations. The dissipation-free part of the transmon Liouvillian reads
\begin{equation}
    \check{\mathfrak L}_\mathrm{t} = \frac{\check q_\mathrm q \check q_\mathrm c}{C_\mathrm{eff}} + 2 E_\mathrm j \sin \left(\frac{2\pi \check \psi_\mathrm c}{\Phi_0}\right) \sin\left(\frac{\pi\check \psi_\mathrm q}{\Phi_0}\right).
\end{equation}
Next, we make a substitution
\begin{equation}
\begin{gathered}
    \check \psi_\mathrm c = \frac{\check \psi_+ + \check \psi_-}{2},~
    \check q_\mathrm c = \frac{\check q_+ + \check q_-}{2}, \\
    \check \psi_\mathrm q = \check \psi_+ - \check \psi_-,~\check q_\mathrm q = \check q_+ - \check q_-,
    \end{gathered}
\end{equation}
which is a transformation from the classical and quantum variables to the forward and backward-path variables. In terms of these variables, the Liouvillian assumes the form
\begin{equation}
    \begin{gathered}
    \check{\mathfrak L}_\mathrm t = \check H_+ - \check H_-,\\
    \check H_\pm = \frac{\check q_\pm^2}{2C_\mathrm{eff}} - E_\mathrm j \cos \left(\frac{2\pi \check \psi_\pm}{\Phi_0}\right).
    \end{gathered}
\end{equation}
This Liouvillian is nothing but a commutator with a standard transmon Hamiltonian
\begin{equation}
\begin{gathered}
\check{\mathfrak L}_\mathrm t \operatorname{vec} \left(\hat \rho\right)
= \operatorname{vec} \left(\left[\hat H, \hat \rho\right]\right),\\
\hat H = \frac{\hat q^2}{2C_\mathrm{eff}} - E_\mathrm j \cos \left(\frac{2\pi \hat \psi}{\Phi_0}\right),
\end{gathered}
\end{equation}
where $\operatorname{vec}\left(\hat \rho\right)$ stands for the vectorization of the density operator.
For the latter, we can construct a finite-difference approximation, diagonalize it numerically and use a few lowest transmon levels $|n\rangle$ with energies $E_n$ as a basis. Consequently, an arbitrary operator $\check A_\pm$ acting in the Liouville space can be expressed through its counterpart $\hat A$ acting in the Hilbert space as
\begin{equation}
    \check A_+ \tilde{\rightarrow} \hat A \otimes \hat 1,~ \check A_- \tilde{\rightarrow} \hat 1 \otimes \hat A^\mathsf T,
\end{equation}
where $\tilde{\rightarrow}$ stands for the isomorphism relation, $\otimes$ denotes the Kronecker product, and $\hat 1$ is the identity operator in the Hilbert space. For the bosonic operators of the auxiliary modes, which we do not eliminate with the dynamical Schrieffer--Wolf transformation, we cut the Liouville space to a few lowest levels. For our calculations, we use four levels of the transmon and six levels of the auxiliary modes.

\section{Calculation of scattering coefficient}

\label{sec:transmon-S21-calculation}

We apply the dynamical Schrieffer--Wolf transformation to the static part of Liouvillian and drive terms. The resulting Liouvillian assumes the form
\begin{multline}
    \check{\mathfrak L}_\mathrm{SW} = 
    \check{\mathfrak L}_\mathrm{SW}^{(\mathrm{st})} \\ + \begin{bmatrix}
        \eta_\mathrm o & \dot \eta_\mathrm o & 1
    \end{bmatrix}
    \begin{bmatrix}
        L^{(00)}_\mathrm{oi} & L^{(01)}_\mathrm{oi} & \check V_\mathrm c^{(0)} \\
        L^{(10)}_\mathrm{oi} & L^{(11)}_\mathrm{oi} & \check V_\mathrm c^{(1)} \\
        \check V_\mathrm q^{(0)} & \check V_\mathrm q^{(1)} & 0
    \end{bmatrix}
    \begin{bmatrix}
        v_\mathrm i\\
        \dot v_\mathrm i \\
        1
    \end{bmatrix},
\end{multline}
where
\begin{equation}
    \check{\mathfrak L}_\mathrm{SW}^{(\mathrm{st})} = \left.\check{\mathfrak L}\right|_{v_\mathrm i = 0, \eta_\mathrm o = 0} + \frac{1}{2}\left[\left.\check{\mathfrak L}\right|_{v_\mathrm i = 0, \eta_\mathrm o = 0}, \check S\right],
\end{equation}
and $\check S$ is the generator of the Schrieffer--Wolf transformation calculated for vanishing drive with $v_\mathrm i = 0$ and $\eta_\mathrm 0 = 0$.
The operators which couple the system degrees of freedom with the input and output field are given by
\begin{equation}
\begin{gathered}
    \check V_\mathrm c^{(0)} = \left.\left(\frac{\partial \check{\mathfrak L}}{\partial \eta_\mathrm o} + \left[\frac{\partial \check{\mathfrak L}}{\partial \eta_\mathrm o}, \check S\right]\right)\right|_{v_\mathrm i = 0, \eta_\mathrm o = 0}, \\
    \check V_\mathrm c^{(1)} = \left.\left(\frac{\partial \check{\mathfrak L}}{\partial \dot \eta_\mathrm o} + \left[\frac{\partial \check{\mathfrak L}}{\partial \dot \eta_\mathrm o}, \check S\right]\right)\right|_{v_\mathrm i = 0, \eta_\mathrm o = 0},\\
    \check V_\mathrm q^{(0)} = \left.\left(\frac{\partial \check{\mathfrak L}}{\partial v_\mathrm i} + \left[\frac{\partial \check{\mathfrak L}}{\partial v_\mathrm i}, \check S\right]\right)\right|_{v_\mathrm i = 0, \eta_\mathrm o = 0}, \\
    \check V_\mathrm q^{(1)} = \left.\left(\frac{\partial \check{\mathfrak L}}{\partial \dot v_\mathrm i} + \left[\frac{\partial \check{\mathfrak L}}{\partial \dot v_\mathrm i}, \check S\right]\right)\right|_{v_\mathrm i = 0, \eta_\mathrm o = 0},
    \end{gathered}
\end{equation}
and the direct coupling between the input and output field is provided by
\begin{equation}
\begin{gathered}
    L_\mathrm{oi}^{(00)} = \left.\frac{\partial^2 \check{\mathfrak L}}{\partial \eta_\mathrm o \partial v_\mathrm i}\right|_{v_\mathrm i = 0, \eta_\mathrm o = 0},~ L_\mathrm{oi}^{(01)} =\left.\frac{\partial^2 \check{\mathfrak L}}{\partial \eta_\mathrm o \partial \dot v_\mathrm i}\right|_{v_\mathrm i = 0, \eta_\mathrm o = 0},\\
    L_\mathrm{oi}^{(10)} = \left.\frac{\partial^2 \check{\mathfrak L}}{\partial \dot \eta_\mathrm o \partial v_\mathrm i}\right|_{v_\mathrm i = 0, \eta_\mathrm o = 0},~ L_\mathrm{oi}^{(11)} =\left.\frac{\partial^2 \check{\mathfrak L}}{\partial \dot \eta_\mathrm o \partial \dot v_\mathrm i}\right|_{v_\mathrm i = 0, \eta_\mathrm o = 0}.
    \end{gathered}
\end{equation}
The latter are complex-valued numbers since the both the classical and quantum fields couple linearly to the system and auxiliary degrees of freedom in the Liouvillian~\eqref{eq:qdmess-liouvillian}. Here, we drop the terms quadratic in $\eta_\mathrm o$, since they contribute only to the fluctuations.

 The transmission coefficient can be expressed as
\begin{multline}
    S_{21}(\omega) = \sum\limits_{jj'=0}^{1}
    \imath^{j-j'} \omega^{j+j'}
    \llangle \hat 1 | \check V_\mathrm c^{(j)}\check {\mathfrak G}(\omega)
    \check V_\mathrm q^{(j')}|W_\mathrm{eq}\rrangle \\+
    \sum\limits_{jj'=0}^1 \imath^{j-j'} \omega^{j+j'} L_\mathrm{oi}^{(jj')},
\end{multline}
where $\check{\mathfrak G}(\omega) = \left[\omega - \check{\mathfrak L}_\mathrm{SW}^{(\mathrm{st})}\right]^{-1}$ which can be efficiently calculated using a numerical diagonalization of the static Liouvillian.

\section{QD-MESS equation for the periodically driven system}
\label{sec:qdmess-periodic}

In this Appendix we discussed evaluation of the generating functional for non-vanishing drive $v_\mathrm i(t)$ and counting field $\eta_\mathrm o(t)$ in application to the transmon readout presented in Sec.~\ref{sec:single-shot-ro}. We solve the QD-MESS equation with the following classical and quantum sources:
\begin{equation}
    \begin{gathered}
    v_\mathrm i(t) = V_\mathrm d \cos(\omega_\mathrm d t), \\
    \eta_\mathrm o(s_1, s_2, t) =  s_1 \cos(\omega_\mathrm d t) + s_2 \sin(\omega_\mathrm d t),
    \end{gathered}
\end{equation}
for different values of the parameters $s_1$ and $s_2$. The total Liouvillian is periodic in time, and hence Floquet theory can be applied. We consider the evolution operator over a single period as
\begin{equation}
    \check U(s_1, s_2) = \mathbb T \exp \left[-\frac{\imath}{\hbar}\int\limits_0^{{2\pi}/{\omega_\mathrm d}} \check{\mathfrak L}(t)\;\mathrm dt\right].
\end{equation}
The steady state of the system corresponds to the unit eigenstate of this operator in the absence of the quantum drive
\begin{equation}
    \check U(0)|W_\mathrm{ss}\rrangle = |W_\mathrm{ss}\rrangle.
\end{equation}
We consider that the integration time $t_\mathrm{int}$ is an integer multiple $N$ of the drive period $2\pi / \omega_\mathrm d$. Thus the characteristic function can be evaluated as
\begin{equation}
    \mathcal Z(s_1, s_2) = \llangle \hat 1 | \left[\check U\left(\frac{s_1}{N}, \frac{s_2}{N}\right)\right]^N |W_\mathrm{ss}\rrangle.
\end{equation}
These quantities can be efficiently evaluated using the dynamical Schrieffer--Wolf approach for weakly coupled auxiliary modes.

\section{Pauli master equation}
\label{sec:pauli}

Pauli master equation which governs dynamics of a system of coupled transmon and resonator reads as
\begin{multline}
        \frac{\mathrm d  \rho_{nn}}{\mathrm dt} = \sum\limits_m \left[\frac{\imath }{\hbar} C_\mathrm o \dot v_\mathrm i(t) \left(\rho_{nm} \varphi_{\mathrm r, mn} - \varphi_{\mathrm r, nm} \rho_{mn}\right) \right .\\ \left .+ \Gamma_{nm} \rho_{mm} - \Gamma_{mn} \rho_{nn}\right] ,
\end{multline}
\begin{multline}
        \frac{\mathrm d \rho_{nm}}{\mathrm dt} = \left(-\imath \omega_{nm}  - \Gamma^\phi_{nm}\right)\rho_{nm} \\ +  \frac{\imath}{\hbar} C_\mathrm o \dot v_\mathrm i(t) \sum\limits_k \left(\rho_{nk} \varphi_{\mathrm r, km} - \varphi_{\mathrm r, nk} \rho_{km} \right),~n\ne m,
\end{multline}
where $\omega_{nm} = (E_n - E_m) / \hbar$, $\rho_{nm} = \langle n | \hat \rho |m\rangle$ are the elements of the density matrix of the system, $\varphi_{\mathrm r, nm} = \langle n | \hat \varphi_\mathrm r | m\rangle$ is a matrix element of the resonator flux operator, $|n\rangle$ is the $n$-th eigenstate of the non-driven Hamiltonian corresponding to the energy $E_n$ so that
\begin{equation}
    \left.\hat H_{\mathrm{tr}}\right|_{v_\mathrm{i}(t) = 0} |n\rangle = E_n |n\rangle,
\end{equation}
the transition rates are given by
\begin{equation}
    \Gamma_{nm} = \frac{2\pi}{\hbar} \frac{|\langle n | \hat \varphi_\mathrm r |m\rangle|^2 J\left(\omega_{mn}\right)}{1 - \exp\left(-\frac{\hbar \omega_{mn}}{k_\mathrm B T}\right)},
\end{equation}
and the dephasing rates read
\begin{equation}
    \Gamma^\phi_{nm} = \frac{1}{2}\sum\limits_{k}\left(\Gamma_{kn} + \Gamma_{km}\right).
\end{equation}
In the absence of the drive, $v_\mathrm i(t) = 0$, the stationary solution of the above equations corresponds to a thermal density matrix
\begin{equation}
    \rho_{nm}^{(0)} = \delta_{nm} \frac{\exp\left(-\frac{E_n}{k_\mathrm B T}\right)}{\sum_{n'} \exp\left(-\frac{E_{n'}}{k_\mathrm B T}\right)}.
\end{equation}
For weak but finite drive, we utilize the first-order perturbation theory with respect to the drive, which results in the following expression for the transmission coefficient
\begin{multline}
    S_{21}(\omega) = 1  + \frac{C_\mathrm o \tau_{RC} \omega^3}{\hbar} \\  \times \sum\limits_{n\ne m} \frac{\left(\rho_{mm}^{(0)} - \rho_{nn}^{(0)}\right) \left| \varphi_{\mathrm r, nm}\right|^2}{\omega - \omega_{nm} + \imath \Gamma_{nm}^\phi}.
    \label{eq:s21-lindblad}
\end{multline}

\section{Derivation of the QD-MESS for resistively shunted junction}

\label{sec:rsj-derivation-qdmess}

In this Appendix, we provide the key steps for the derivation of the Liouvillian~\eqref{eq:rsj-liouvillian}. The core self-energy components~\eqref{eq:effective-action} read as
\begin{equation}
    \begin{gathered}
        \bm \Sigma_\mathrm{core}^\mathrm R(\omega) =
        \begin{bmatrix}
            -C \omega ^ 2 - \imath \omega Z^{-1} & -2 Z^{-1} \\
            \imath \omega & 1
        \end{bmatrix}, \\
        \bm \Sigma_\mathrm{core}^\mathrm A(\omega) =
        \begin{bmatrix}
            -C \omega ^ 2 + \imath \omega Z^{-1} & -\imath \omega \\
            -2 Z^{-1} & 1
        \end{bmatrix}, \\
        \bm \Sigma_\mathrm{core}^\mathrm K(\omega) =
        -\frac{\imath \omega}{4} \coth \left(\frac{\hbar \omega}{2 k_\mathrm B T}\right)
        \begin{bmatrix}
            4Z^{-1} & -2 \\
            -2 & Z
        \end{bmatrix}.
    \end{gathered}
\end{equation}
The matrix-valued residues~\eqref{eq:matrix-residues} corresponding to the fluctuation poles and arising from rational approximation~\eqref{eq:coth-approximation} of $\coth[\hbar \omega/(2 k_\mathrm B T)]$~can be evaluated analytically and are given by
\begin{multline}
    \tilde{\bm R}_n^\mathrm{f} = -\frac{\omega_n^\mathrm{f} r_n^\mathrm{f}}{Z} \begin{bmatrix}
        1 & -\frac{Z}{2} \\
        -\frac{Z}{2} & \frac{Z^2}{4}
    \end{bmatrix} \\ = 
    \frac{1}{\hbar} \left(\imath \sqrt{\frac{\hbar \omega_n^\mathrm{f} r_n^\mathrm{f}}{Z}}
    \begin{bmatrix}
        1 \\
        -\frac{Z}{2}
    \end{bmatrix}\right)
    \left(\imath \sqrt{\frac{\hbar \omega_n^\mathrm{f} r_n^\mathrm{f}}{Z}}
    \begin{bmatrix}
        1 &
        -\frac{Z}{2}
    \end{bmatrix}\right).
\end{multline}
Each of the poles gives rise to an auxiliary mode with a purely imaginary frequency. All the residues have unit rank, therefore there is only a single fluctuation auxiliary mode per pole. Finally, we can express the Lagrangian of QD-MESS as
\begin{widetext}
\begin{multline}
    \mathcal L_\text{QD-MESS}\left(\dot \psi_\mathrm c, \dot \psi_\mathrm q, \psi_\mathrm c, \psi_\mathrm q, \dot{\bm b}, \bm b^\ast, \bm b\right) = C \dot \psi_\mathrm q \dot \psi_\mathrm c - \frac{\psi_\mathrm q \dot \psi_\mathrm c}{Z}  - 2 E_\mathrm J \sin \left(\frac{\pi \psi_\mathrm q}{\Phi_0}\right) \sin \left(\frac{2\pi \psi_\mathrm c}{\Phi_0}\right) + \frac{\imath k_\mathrm B T}{\hbar Z} \left(\psi_\mathrm q - \frac{Z \eta_\mathrm o}{2}\right)^2 \\
     + \imath \hbar \sum\limits_{n=1}^M \left[
     b_n^\ast \dot b_n + \omega_n^\mathrm{f} \left|
    b_n - \sqrt{\frac{r_n^\mathrm{f}}{\hbar \omega_n^\mathrm{f} Z}} \left(\psi_\mathrm q - \frac{Z \eta_\mathrm o}{2}\right)\right|^2
    \right]  + \eta_\mathrm o \left(\dot \psi_\mathrm c - v_\mathrm i\right) + \frac{2 \psi_\mathrm q v_\mathrm i}{Z}.
\end{multline}
\end{widetext}
Canonical charges~\eqref{eq:legendre} are given by
\begin{equation}
    \begin{gathered}
        q_\mathrm c = C \dot \psi_\mathrm c,~
        q_\mathrm q =  C \dot \psi_\mathrm q - \frac{\psi_\mathrm q}{Z} + \eta_\mathrm o.
    \end{gathered}
\end{equation}
Finally, the Legendre transformation~\eqref{eq:lagrangian-to-liouvillian} and the quantization with proper ordering of the operators results in the Liouvillian~\eqref{eq:rsj-liouvillian}.

\section{Solving the QD-MESS for resistively shunted junction}

\label{sec:rsj-solving-qdmess}
In this Appendix, we present key steps in derivation of Eq.~\eqref{eq:rsj-static-reflection}.
We begin with the perturbative treatment of the stationary QD-MESS equation~\eqref{eq:qdmess-equilibrium} for the Liouvillian~\eqref{eq:rsj-liouvillian} as
\begin{gather}
    \left(\check{\mathfrak L}_0 + \check{\mathfrak L}_\mathrm J\right)|W_\mathrm{eq}\rrangle = 0,\\
    |W_\mathrm{eq}\rrangle \approx \left[1 - \check{\mathfrak L}_0^{-1} \check{\mathfrak L}_\mathrm J  + \left(\check{\mathfrak L}_0^{-1} \check{\mathfrak L}_\mathrm J\right)^2\right] |W_\mathrm{eq}^{(0)}\rrangle,
    \label{eq:rsj-eq-perturb}
\end{gather}
where $|W_\mathrm{eq}^{(0)}\rrangle$ is a right zero eigenvector of the Liouvillian $\check{\mathfrak L}_0$. Since the Liouvillian $\check{\mathfrak L}_0$ is singular, we interpret $\check{\mathfrak L}_0^{-1}|W\rrangle$ as a solution of
\begin{equation}
    \check{\mathfrak L}_0^{-1}|X\rrangle = |W\rrangle
\end{equation}
which satisfies the condition $\llangle \hat 1 | X \rrangle = 0$. Such a solution always exists provided $\llangle \hat 1 | W \rrangle = 0$.
First, we analyze the Liouvillian $\check{\mathfrak L}_0$ from Eq.~\eqref{eq:rsj-liouvillian-perturb}. Since $\check {\mathfrak L}_0$ commutes with $\check q_\mathrm q$, we can express it as
\begin{equation}
    \begin{gathered}
        \check {\mathfrak L}_0 = \sum\limits_{k=-\infty}^{\infty} 
        | k \rrangle \llangle k | \otimes \check{\mathfrak L}_{0,k},
    \end{gathered}
\end{equation}
where
\begin{multline}
    \check{\mathfrak L}_{0, k} = \frac{2e \check q_\mathrm c}{C} k + \frac{\check \psi_\mathrm q \check q_\mathrm c}{ZC} - \frac{\imath k_\mathrm B T}{\hbar Z} \check \psi_\mathrm q ^ 2\\ - \imath \hbar \sum\limits_{n=1}^{M^\mathrm f} \omega_n^\mathrm{f} \left(\check b_n^\dagger - \gamma_n \check \psi_\mathrm q\right)\left(\check b_n - \gamma_n \check \psi_\mathrm q\right),
\end{multline}
where we have introduced
\begin{equation}
    \gamma_n = \sqrt{\frac{r_n^\mathrm{f}}{\hbar \omega_n^\mathrm{f} Z}},
\end{equation}
and $|k\rrangle$ and $\llangle k |$ are the left and right eigenvectors of $\check q_\mathrm q$, respectively:
\begin{equation}
    \begin{gathered}
        \check q_\mathrm q|k\rrangle = 2 e k |k \rrangle,\\
        \llangle k|\check q_\mathrm q = 2 e k \llangle k |,
    \end{gathered}
\end{equation}
normalized as $\llangle k | k \rrangle = 1$.
The eigenvalues of $\check q_\mathrm q$ are integer multiples of the double electron charge $2e$ since its conjugated phase $\check \varphi_\mathrm c$ is a periodic compact variable.
The Liouvillian term, associated with the Josephson energy, reads as
\begin{multline}
    \check{\mathfrak L}_\mathrm J = \imath E_\mathrm J \sum\limits_{k=-\infty}^\infty \left(|k\rrangle \llangle k + 1| - |k+1\rrangle \llangle k|
    \right) \\ \otimes \sin \left(\frac{\pi \check \psi_\mathrm q}{\Phi_0}\right) 
\end{multline}
in this picture.
The block Liouvillians $\check{\mathfrak  L}_{0,k}$ are quadratic in terms of the operators $\check \psi_\mathrm q$, $\check q_\mathrm c$, $\check b_n$, and $\check b_n^\dagger$, therefore they can be diagonalized by a Bogoliubov transformation:
\begin{equation}
        \check \psi_\mathrm q = \frac{\hbar}{q_T} \check d_0^\dagger,
\end{equation}
\begin{multline}
    \check q_\mathrm c = \imath q_T \left\{
            \check d_0^\dagger -
            \check d_0 + \sum\limits_{n=1}^{M^\mathrm f} \frac{\hbar \omega_n^\mathrm{f} \gamma_n}{q_T}\left[
                \frac{\check d_n}{\omega_n^\mathrm{f} - \omega_0}
                \right. \right. \\ \left. \left.
                -\frac{\check d_n^\dagger}{\omega_n^\mathrm{f} + \omega_0} - 
                \frac{\hbar \omega_0 \gamma_n \check d_0^\dagger}{q_T \left(\left(\omega_n^\mathrm{f}\right)^2 - \omega_0^2\right)}
            \right]
        \right\}
\end{multline}
\begin{gather}
    \check b_n = \check d_n + \frac{\hbar \omega_n^\mathrm{f} \gamma_n}{q_T \left(\omega_n^\mathrm{f} - \omega_0\right)} \check d_0^\dagger,\\
    \check b_n^\dagger = \check d_n^\dagger + \frac{\hbar \omega_n \gamma_n}{q_T \left(\omega_n^\mathrm{f} - \omega_0\right)} \check d_0^\dagger,
\end{gather}
where $\omega_0 = (ZC)^{-1}$, $q_T = \sqrt{k_\mathrm B T  C}$, and operators $\check d_n$ and $\check d_n^\dagger$ obey the standard bosonic commutation relations
\begin{equation}
    \begin{gathered}
        \left[\check d_n, \check d_{n'}^\dagger\right] = \delta_{nn'},\\
        \left[\check d_n, \check d_{n'}\right] = 0,\\
        \left[\check d_n^\dagger, \check d_{n'}^\dagger\right] = 0,
        \end{gathered}
\end{equation}
for $n, n' = 0,\ldots,M$. Note that this transformation simultaneously diagonalizes the block Liouvillians $\check L_{0,k}$ for all $k$. We also emphasize, that since the Liouvillians $\check{\mathfrak L}_{0,k}$ are non-Hermitian, the operators $\check d_n$ and $\check d_n^\dagger$ are not related by Hermitian conjugation. The block Liouvillians expressed through the operators $\check d_n$ and $\check d_n^\dagger$ assume the form
\begin{multline}
    \check{\mathfrak L}_{0,k} = - \imath k_\mathrm B T \frac{ 2 \pi Z}{R_\mathrm Q} k^2\\ 
    -\imath \hbar \sum\limits_{n=0}^{M^\mathrm f} \omega_n
    \left(\check d_n^\dagger - k \beta_n\right)\left(\check d_n - k \alpha_n\right),
\end{multline}
where we use a short-hand notation $\omega_n = \omega_n^\mathrm{f}$, and the shifts $\alpha_n$ and $\beta_n$, $n=0,\ldots, M$ are given by
\begin{equation}
    \begin{gathered}
        \beta_0 = -\frac{2e q_T }{\hbar \omega_0 C},\\
        \alpha_0 = -\frac{2e q_T }{\hbar \omega_0 C} \left[
        \sum\limits_{n=1}^{M^\mathrm f} \frac{\hbar^2 \gamma^2_n \omega_0 \omega_n}{q_T^2 \left(\omega_n^2 - \omega_0^2\right)} - 1
        \right],\\
        \beta_n = \frac{2e \gamma_n }{C \left(\omega_n - \omega_0\right)},~n=1,\ldots,{M^\mathrm f},\\
        \alpha_n = -\frac{2e \gamma_n }{C\left(\omega_n + \omega_0\right)},~n=1,\ldots,{M^\mathrm f}.
    \end{gathered}
\end{equation}
The left and right eigenstates are of $\check{\mathfrak L}_{0,k}$ are specified by the integer non-negative occupation numbers $m_n$ of the shifted auxiliary bosonic modes $\check d_n$ as
\begin{multline}
    \check{\mathfrak L}_{0,k} |k; m_0, \ldots, m_{M^\mathrm f}\rrangle = -\imath \left(k_\mathrm B T \frac{2\pi Z}{R_\mathrm Q} k^2\right. \\ \left. + \hbar \sum\limits_{n=0}^{M^\mathrm f} m_n \omega_n\right) |k; m_0,\ldots, m_{M^\mathrm f}\rrangle, 
\end{multline}
\begin{multline}
    \llangle k; m_0, \ldots, m_{M^\mathrm f}|\check{\mathfrak L}_{0,k}  = -\imath \left(k_\mathrm B T \frac{2\pi Z}{R_\mathrm Q} k^2\right. \\ \left. + \hbar \sum\limits_{n=0}^{M^\mathrm f} m_n \omega_n\right) \llangle k; m_0,\ldots, m_{M^\mathrm f}|, 
\end{multline}
\begin{multline}
    |k; m_0, \ldots, m_{M^\mathrm f}\rrangle =
    \prod\limits_{n=0}^{M^\mathrm f} \frac{\left(\check d_n^\dagger - k \beta_n\right)^{m_n}}{\sqrt{m_n!}} \\ \times e^{-\frac{k^2}{2}\beta_n \alpha_n + k \alpha_n \check d_n^\dagger} |0_\mathrm{aux}\rrangle,
\end{multline}
\begin{multline}
    \llangle k; m_0, \ldots, m_{M^\mathrm f}| = \llangle 0_\mathrm{aux} | \prod\limits_{n=0}^M e^{-\frac{k^2}{2} \beta_n \alpha_n + k \beta_n \check d_n}\\ \times \frac{\left(\check d_n - k \alpha_n\right)^{m_n}}{\sqrt{m_n!}},
\end{multline}
and the states $|0_\mathrm{aux}\rrangle$ and $\llangle 0_\mathrm{aux}|$ correspond to the state where the occupation numbers of all the bosonic modes vanish:
\begin{equation}
    \begin{gathered}
        \check d_n|0_\mathrm{aux}\rrangle = 0,~n=0,\ldots,{M^\mathrm f},\\
        \llangle 0_\mathrm{aux} | \check d_n^\dagger = 0, n=0,\ldots,{M^\mathrm f}.
    \end{gathered}
\end{equation}
We impose a unit normalization as $\llangle 0_\mathrm{aux} | 0_\mathrm{aux}\rrangle = 1$. Next, we express the non-perturbed equilibrium state $|W_\mathrm{eq}^{(0)}\rrangle$ and the trace operation $\llangle \hat 1 |$ as
\begin{equation}
    \begin{gathered}
    |W_\mathrm{eq}^{(0)}\rrangle = |0\rrangle \otimes |0_\mathrm{aux}\rrangle,\\
    \llangle \hat 1 | = \llangle 0 | \otimes \llangle 0_\mathrm{aux}|.
    \end{gathered}
\end{equation}

Subsequently, we substitute the perturbative expansion~\eqref{eq:rsj-eq-perturb} into~\eqref{eq:rsj-reflection} and in the stationary limit, we obtain
\begin{multline}
    S_{11}(\omega)|_{\omega=0} = -1 - \llangle \hat 1| \check V_\mathrm c \left(
        \check{\mathfrak L}_0^{-1} \check V_\mathrm q \right. \\ \left. +
        \check{\mathfrak L}_0^{-1} \check V_\mathrm q
        \check{\mathfrak L}_0^{-1} \check{\mathfrak L}_\mathrm J
        \check{\mathfrak L}_0^{-1} \check{\mathfrak L}_\mathrm J \right. \\ \left. +
        \check{\mathfrak L}_0^{-1} \check{\mathfrak L}_\mathrm J 
        \check{\mathfrak L}_0^{-1} \check V_\mathrm q
        \check{\mathfrak L}_0^{-1} \check{\mathfrak L}_\mathrm J
        \right. \\ \left. + 
        \check{\mathfrak L}_0^{-1} \check{\mathfrak L}_\mathrm J
        \check{\mathfrak L}_0^{-1} \check{\mathfrak L}_\mathrm J 
        \check{\mathfrak L}_0^{-1} \check V_\mathrm q
    \right)|W_\mathrm{eq}^{(0)}\rrangle,
\end{multline}
where we keep all the terms up to the second power with respect to $\check{\mathfrak L}_\mathrm J$. We express the operators $\check V_\mathrm q$, $\check V_\mathrm c$, and $\check{\mathfrak L}_\mathrm J$ in terms of the operators $\check d_n$ and $\check d_n^\dagger$ as
\begin{equation}
    \check V_\mathrm q = -\frac{2 \hbar}{q_T Z} \check d_0^\dagger,
\end{equation}
\begin{multline}
    \llangle \hat 1 | \check V_\mathrm c = \frac{\imath}{C}\llangle \hat 1 | \left\{
     \sum\limits_{n=1}^{M^\mathrm f}
    {\hbar \omega_n \gamma_n}
    \left[
    \frac{1}{\omega_n - \omega_0} \right. \right. \\ \left. \left. + \frac{1}{2 \omega_0}
    \right] \check d_n
    - {q_T} \check d_0
    \right\},
\end{multline}
\begin{multline}
    \check{\mathfrak L}_\mathrm J = \imath E_\mathrm J \sum\limits_{k=-\infty}^\infty \left(|k\rrangle \llangle k + 1| - |k+1\rrangle \llangle k|
    \right) \\ \otimes \sin \left(\frac{\pi \hbar \check d_0^\dagger}{q_T \Phi_0}\right).
\end{multline}
To evaluate the reflection coefficient, we use the identities
\begin{equation}
    \begin{gathered}
    \check{\mathfrak L}_{0,k}^{-1} = \frac{\imath}{\hbar} \int\limits_0^{+\infty}\exp\left(-\frac{\imath}{\hbar} \check{\mathfrak L}_{0,k} t\right)\;\mathrm dt,\\
    \end{gathered}
\end{equation}
\begin{multline}
    \exp\left(-\frac{\imath}{\hbar} \check{\mathfrak L}_{0,k} t\right) \check d_n^\dagger   \exp\left(\frac{\imath}{\hbar} \check{\mathfrak L}_{0,k} t\right) = k \beta_n \\ + \left(\check d_n^\dagger - k \beta_n\right) 
        e^{-\omega_n t},
\end{multline}
\begin{multline}
    \exp\left(\frac{\imath}{\hbar} \check{\mathfrak L}_{0,k} t\right) \check d_n \exp\left(-\frac{\imath}{\hbar} \check{\mathfrak L}_{0,k} t\right) = k \alpha_n \\ + \left(\check d_n - k \alpha_n\right) 
        e^{-\omega_n t},
\end{multline}
\begin{multline}
    \exp\left(-\frac{\imath}{\hbar} \check{\mathfrak L}_{0,k} t\right) \prod_{n=0}^M e^{k \alpha_n \check d_n^\dagger}|0_\mathrm{aux}\rrangle \\= \exp\left(-\frac{2\pi k_\mathrm B T}{\hbar} \frac{Z}{R_\mathrm Q} k^2 t\right) \prod_{n=0}^M e^{k \alpha_n \check d_n^\dagger} |0_\mathrm{aux}\rrangle,
\end{multline}
\begin{multline}
    \llangle 0_\mathrm{aux}| \prod_{n=0}^M e^{k \beta_n \check d_n} \exp\left(-\frac{\imath}{\hbar} \check{\mathfrak L}_{0,k} t\right) \\= \llangle 0_\mathrm{aux}| \prod_{n=0}^M e^{k \beta_n \check d_n} \exp\left(-\frac{2\pi k_\mathrm B T}{\hbar} \frac{Z}{R_\mathrm Q} k^2 t\right),
\end{multline}
and obtain the expression for the reflection coefficient in the low frequency limit
\begin{multline}
    \left.S_{11}(\omega)\right|_{\omega=0} = 1 - \frac{4 \pi  Z}{R_\mathrm Q} \frac{E_\mathrm J^2}{\hbar^2} \int\limits_{0}^{+\infty} t \exp\left[-\frac{2\pi k_\mathrm B T Z t}{\hbar R_\mathrm Q} \right] \\ \times
    \prod\limits_{n=0}^{M^f} \exp\left[
    \alpha_n \beta_n \left(e^{-\omega_n t} - 1\right)
    \right]
    \\ \times \sin\left[\frac{4\pi Z}{R_\mathrm Q} \left(1 - e^{-\omega_0 t}\right)\right]  \;\mathrm dt,
\end{multline}
which up to notations coincides with~\eqref{eq:rsj-reflection}.

\bibliography{bibliography}

\end{document}